\documentclass[12pt]{article}
\usepackage{enumerate}
\usepackage{amsfonts}
\usepackage{amsmath}
\usepackage{amssymb}
\usepackage{amsthm}

\def\H{\mathcal{H}}

\def\S{\mathfrak{S}}

\def\T{\mathfrak{T}}
\def\B{\mathfrak{B}}

\newcommand{\rank}{\mathrm{rank}}
\newcommand{\id}{\mathrm{Id}}
\newcommand{\Tr}{\mathrm{Tr}}

\newcommand{\shs}{\hspace{1pt}}

\newcounter{defin}  \newcounter{lemma}  \newcounter{theorem}
\newcounter{property} \newcounter{corol}  \newcounter{remark} \newcounter{example}

\newenvironment{lemma}{\par\refstepcounter{lemma}
     \textbf{Lemma \thelemma.} }{\rm\par}
\newenvironment{theorem}{\par\refstepcounter{theorem}
     \textbf{Theorem \thetheorem.}\ }{\rm\par}
\newenvironment{property}{\par\refstepcounter{property}
     \textbf{Proposition \theproperty.}\ }{\rm\par}
\newenvironment{corollary}{\par\refstepcounter{corol}
     \textbf{Corollary \thecorol.} }{\rm\par}
\newenvironment{definition}{\par\refstepcounter{defin}
     \textbf{Definition \thedefin.}\ }{\rm\par}
\newenvironment{remark}{\par\refstepcounter{remark}
     \textbf{Remark \theremark.}}{\rm\par}
\newenvironment{example}{\par\refstepcounter{example}
     \textbf{Example \theexample.}}{\rm\par}

\begin{document}

\title{Measures of quantum correlations in infinite-dimensional systems}
\author{M.E. Shirokov\footnote{Steklov Mathematical Institute, RAS, Moscow, email:msh@mi.ras.ru}}
\date{}
\maketitle

\begin{abstract}
Several important measures of quantum correlations of a state of a
finite-dimensional composite system are defined as linear
combinations of marginal entropies of this state. This paper is
devoted to the infinite-dimensional generalizations of such
quantities and to the analysis of their properties.

We introduce the notion of faithful extension of a linear
combination of marginal entropies and consider several concrete
examples starting with the quantum mutual information and the
quantum conditional entropy.

Then we show that the conditional mutual information can be uniquely
defined as a lower semicontinuous function on the set of all states
of a tripartite infinite-dimensional system possessing all the basic
properties valid in finite dimensions.

Infinite-dimensional generalizations of some other measures of
quantum correlations in multipartite quantum systems are also
considered. It is shown that almost all of  these generalized
measures are globally lower semicontinuous and possess local
continuity properties which essentially simplify their use in
analysis of quantum systems.

In the second part of the paper we consider  applications of the
general results of its first part, in particular,  to the theory of
infinite-dimensional quantum channels and their capacities. We also
show the existence of the Fawzi-Renner recovery channel reproducing
marginal states for all tripartite states (including states with
infinite marginal entropies) starting with the corresponding
finite-dimensional result.
\end{abstract}\pagebreak

\tableofcontents

\pagebreak

\section{Introduction}

Characterization of quantum correlations in composite quantum
systems is one of the main problems in quantum information theory
which attracted attention from the middle of 20-th century. A
notable progress in this direction had been achieved during the last
two decades when several quantities characterizing special forms of
quantum correlations had been found and explored. It is natural that
initially all these characteristics were studied in
finite-dimensional settings to avoid analytical problems arising in
dealing with infinite-dimensional quantum systems (such as
discontinuity and infinite values of the von Neumann entropy,
noncompactness of a state space, etc.) Nevertheless, keeping in mind
the physical applications it seems reasonable to construct
infinite-dimensional generalizations of the commonly used
information quantities and to study their properties.

One of the main problems in infinite-dimensional generalization of
information quantities is the appearance of the uncertainty
$"\infty-\infty"$ in their original definitions. For example, many
important characteristics of a state $\omega$ of a multipartite
finite-dimensional system $A_{1}...A_{n}$ are defined as a real
linear combination of the marginal entropies
\begin{equation}\label{g-form+}
  \sum_k \alpha_k H(\omega_{X_k}),
\end{equation}
where $\omega_{X_k}$ is a partial state of $\omega$ corresponding to
a subsystem $X_k$ of $A_{1}...A_{n}$. Linear combination
(\ref{g-form+}) correctly determines a value in $[-\infty,+\infty]$
only for states $\omega$ of an infinite-dimensional system
$A_{1}...A_{n}$ for which  all the summands in (\ref{g-form+}) are
either $>-\infty$ or $<+\infty$. Since the states with finite von
Neumann entropy form a first category subset within the set of all
states \cite{W}, a direct translation of definition (\ref{g-form+})
to the case of infinite-dimensional multipartite systems makes the
corresponding quantity undefined for "almost all" states.

Fortunately, the above problem can be (partially or completely)
solved by using alternative expressions for (\ref{g-form+})
consisting of terms which are more stable under passage to infinite
dimensions. The simplest example is using the expression
\begin{equation}\label{mi-re}
H(\omega_{AB}\shs\Vert\shs\omega_{A}\otimes \omega_{B}),
\end{equation}
where $H(\cdot\|\cdot)$ is the quantum relative entropy, instead of
the linear combination
\begin{equation}\label{mi-d}
H(\omega_{A})+H(\omega_{B})-H(\omega_{AB})
\end{equation} defining
quantum mutual information of a state $\omega_{AB}$ of a
finite-dimensional bipartite system. Properties of the relative
entropy show that (\ref{mi-re}) gives adequate definition of the
quantum mutual information for any state $\omega_{AB}$ of
infinite-dimensional bipartite system inheriting all basic
properties of this quantity (nonnegativity, monotonicity, etc, see
Sec. 4.1).

Motivated by the coincidence of (\ref{mi-re}) and (\ref{mi-d}) the
quantum conditional entropy
$\,H(A|B)_{\omega}=H(\omega_{AB})-H(\omega_B)\,$ is extended in
\cite{Kuz} to the convex set
$\{\shs\omega_{AB}\,|\,H(\omega_{A})<+\infty\shs\}$ (containing
states with $H(\omega_{AB})=H(\omega_B)=+\infty$) by the formula
\begin{equation}\label{c-ent-def+}
H(A|B)_{\omega}=H(\omega_{A})-H(\omega_{AB}\shs\Vert\shs\omega_{A}\otimes
\omega_{B})
\end{equation}
preserving all basic properties of the quantum conditional entropy
(monotonicity, concavity, subadditivity). This extension has several
important applications (see Section 5).\smallskip

In the first part of this paper we begin with introducing the notion
of a faithful extension ($\mathfrak{F}$-extension) of linear
combination (\ref{g-form+}) as an extension satisfying the
particular "robustness" requirement (Def.\ref{t-ext}), which seems
reasonable from the both analytical and physical points of view.

Then we consider infinite-dimensional generalizations of several
entropic quantities having form (\ref{g-form+}), starting with the
analysis of continuity properties of the quantum mutual information
$I(A\!:\!B)_{\omega}$ defined by formula (\ref{mi-re}) as a function
of $\omega$ (Theorem \ref{main}). Then we describe extension
(\ref{c-ent-def+}) of the quantum conditional entropy  and its
applications.\smallskip

Special attention is paid to generalization of the conditional
mutual information
\begin{equation*}
    I(A\!:\!C|B)_{\omega}=H(\omega_{AB})+H(\omega_{BC})-H(\omega_{ABC})-H(\omega_{B})
\end{equation*}
having in mind its numerous applications.  It is shown that
$I(A\!:\!C|B)_{\omega}$ has an unique lower semicontinuous extension
to the whole set of states of an infinite-dimensional tripartite
system $ABC$ possessing all the basic properties of the conditional
mutual information valid in finite dimensions (Theorem
\ref{cmi-th}).\smallskip

Infinite-dimensional generalizations of some other measures of
quantum correlations in multipartite systems (topological
entanglement entropy, unconditional and conditional secrecy monotone
$S_n$, etc.) are also considered.\smallskip

In the study of  measures of quantum correlations we pay special
attention to their continuity properties. In finite dimensions any
quantity (\ref{g-form+}) is obviously continuous on the whole set of
states.\footnote{There exist discontinuous measures of quantum
correlations in finite-dimensional multipartite systems and their
discontinuity has a physical meaning. Such measure (called
irreducible three-party correlation) is considered in
\cite{Chen&Co,W&Co}. Its discontinuity is a corollary of the
discontinuity of the maximal entropy inference -- an interesting
purely quantum effect discovered by Knauf and Weis \cite{K&W}.}  In
infinite dimensions global continuity is a very strong requirement,
but one can try to obtain conditions for local continuity, i.e.
continuity with respect to variation of a states within a particular
subset. Intuitively, local continuity of some measure of quantum
correlations can be understood as stability with respect to local
perturbations of a state (which are unavoidable due to finite
accuracy of the state preparation procedure).

It turns out that for all commonly used quantum correlation measures
(defined by formula (\ref{g-form+}) in finite dimensions) there
exist simple sufficient conditions for local continuity expressed
via local continuity of one or several marginal entropies (not
necessarily involved in (\ref{g-form+})). These conditions look
especially surprising when local continuity of only one marginal
entropy  implies local continuity of the whole linear combination
(\ref{g-form+}) consisting of $n$ summands (see, for instance,
Proposition \ref{tee-pr}).\smallskip

We also obtain general results concerning preserving local
continuity under conditioning and partial trace (Proposition
\ref{main-c-2} and Corollary \ref{main-c-3}).\smallskip

In the second part of the paper we consider applications of the
general results of its first part to the theory of
infinite-dimensional quantum channels and their capacities. In
particular, we show that the classical entanglement-assisted
capacity with the constraint defined by the linear inequality $\Tr
F\rho\leq E$ is \emph{continuous on the set of all channels}
equipped with the strong convergence topology provided the von
Neumann entropy is continuous on the set of states satisfying this
inequality (Proposition \ref{eac-cont-cond}).

By using the extended conditional mutual information we show the
existence of the Fawzi-Renner recovery channel exactly reproducing
marginal states for all tripartite states (including states with
infinite marginal entropies) starting with the corresponding
finite-dimensional result from \cite{F&R} (Proposition
\ref{FR-r-m}).
\smallskip

\section{Preliminaries}

Let $\mathcal{H}$ be a separable Hilbert space,
$\mathfrak{B}(\mathcal{H})$ and $\mathfrak{T}( \mathcal{H})$ --
Banach spaces of all bounded operators and of all trace-class
operators in $\mathcal{H}$, $\mathfrak{T}_{+}(\mathcal{H})$ -- the
cone of positive operators in $\mathfrak{T}( \mathcal{H})$,
$\mathfrak{S}(\mathcal{H})$ -- the set of quantum states (operators
in $\mathfrak{T}_{+}(\mathcal{H})$ with unit trace)
\cite{H-SCI,N&Ch}.

Trace class operators (not only states) will be denoted by the Greek
letters $\rho$, $\sigma$, $\omega$, ... All others linear operators
(in particular, unbounded operators) will be denoted by the Latin
letters $A$, $B$, $F$, $H$, ...

Denote by $I_{\mathcal{H}}$ the unit operator in a Hilbert space
$\mathcal{H}$ and by $\id_{\mathcal{\H}}$ the identity
transformation of the Banach space $\mathfrak{T}(\mathcal{H})$.

A \emph{quantum operation} $\,\Phi$ from a system $A$ to a system
$B$ is a completely positive trace non-increasing linear map
$\mathfrak{T}(\mathcal{H}_A)\rightarrow\mathfrak{T}(\mathcal{H}_B)$,
where $\mathcal{H}_A$ and $\mathcal{H}_B$ are Hilbert spaces
associated with the systems $A$ and $B$. A trace preserving quantum
operation is called \emph{quantum channel} \cite{H-SCI,N&Ch}.

The \emph{von Neumann entropy} $H(\rho)=\mathrm{Tr}\eta(\rho)$ of a
state $\rho\in\mathfrak{S}(\mathcal{H})$, where $\eta(x)=-x\log x$,
has the natural extension to the cone
$\mathfrak{T}_{+}(\mathcal{H})$ (cf.\cite{L-2})\footnote{Here and in
what follows $\log$ denotes the natural logarithm.}
\begin{equation}\label{ent-ext}
H(\rho)=\mathrm{Tr}\rho H\!\left(\frac{\rho}{\mathrm{Tr}
\rho}\right)=\mathrm{Tr}\eta(\rho)-\eta(\mathrm{Tr}\rho),\quad \rho
\in\mathfrak{T}_{+}(\mathcal{H}).
\end{equation}

Nonnegativity, concavity and lower semicontinuity of the von Neumann
entropy on the cone $\mathfrak{T}_{+}(\mathcal{H})$ follow from the
corresponding properties of this function on the set
$\mathfrak{S}(\mathcal{H})$ \cite{L-2,W}. By definition
\begin{equation}\label{H-fun-eq}
H(\lambda \rho)=\lambda H(\rho),\quad \lambda\geq 0.
\end{equation}

The concavity of the von Neumann entropy is supplemented by the
inequality
\begin{equation}\label{w-k-ineq}
H\left(\lambda\rho+(1-\lambda)\sigma\right)\leq \lambda
H(\rho)+(1-\lambda)H(\sigma)+\max\{\Tr\rho,\Tr\sigma\}\shs
h_2(\lambda),
\end{equation}
where $h_2(\lambda)=\eta(\lambda)+\eta(1-\lambda)$, valid for any
operators $\rho,\sigma\in\mathfrak{T}_{+}(\mathcal{H})$. \smallskip

The \emph{quantum relative entropy} for two operators $\rho$ and
$\sigma$ in $\mathfrak{T}_{+}(\mathcal{H})$ is defined as follows
(cf.\cite{L-2})
$$
H(\rho\,\|\shs\sigma)=\sum_{i=1}^{+\infty}\langle
i|\,\rho\log\rho-\rho\log\sigma+\sigma-\rho\,|i\rangle,
$$
where $\{|i\rangle\}_{i=1}^{+\infty}$ is the orthonormal basis of
eigenvectors of the operator $\rho$ and it is assumed that
$H(\rho\,\|\sigma)=+\infty$ if $\,\mathrm{supp}\rho$ is not
contained in $\mathrm{supp}\shs\sigma$. This definition implies
\begin{equation}\label{H-fun-eq+}
H(\lambda\rho\,\|\shs\lambda\sigma)=\lambda
H(\rho\,\|\shs\sigma),\quad \lambda\geq0.
\end{equation}

We will use the following result of the purification
theory.\vspace{5pt}
\begin{lemma}\label{p-lemma}
\textit{Let $\mathcal{H}$ and $\mathcal{K}$ be Hilbert spaces such
that $\,\dim\mathcal{H}=\dim\mathcal{K}$. For an arbitrary pure
state $\omega_{0}$ in
$\,\mathfrak{S}(\mathcal{H}\otimes\mathcal{K})$ and an arbitrary
sequence $\{\rho_{k}\}$ of states in $\,\mathfrak{S}(\mathcal{H})$
converging to the state
$\rho_{0}=\mathrm{Tr}_{\mathcal{K}}\omega_{0}$ there exists a
sequence $\{\omega_{k}\}$ of pure states in
$\,\mathfrak{S}(\mathcal{H}\otimes\mathcal{K})$ converging to the
state $\omega_{0}$ such that\break
$\rho_{k}=\mathrm{Tr}_{\mathcal{K}}\omega_{k}$ for all $\,k$.}
\end{lemma}\vspace{5pt}

The assertion of Lemma \ref{p-lemma} can be proved by noting that
the infimum in the definition of the Bures distance (or the supremum
in the definition of the Uhlmann fidelity) between two quantum
states can be taken only over all purifications of one state with
fixed purification of the another state and that the convergence of
a sequence of states in the trace norm distance implies its
convergence in the Bures distance \cite{F&R,H-SCI, N&Ch}.\smallskip

We will use the following property of the von Neumann
entropy.\smallskip

\begin{lemma}\label{t-ext-l} \emph{Let $\,\omega_{A_{1}...A_{n}}$ be
a state of $\,n$-partite system $A_{1}...A_{n}$ and
$\{P^k_{A_{1}}\}_k\subset\B(\H_{A_{1}})$,...,
$\{P^k_{A_{n}}\}_k\subset\B(\H_{A_{n}})$ sequences of projectors
strongly converging to the identity operators
$I_{A_{1}}$,...,$I_{A_{n}}$. Let
$\;\omega^k_{A_{1}...A_{n}}=\lambda_k^{-1}Q_k\omega_{A_{1}...A_{n}}Q_k$,
where $Q_k= P^k_{A_1}\otimes \ldots\otimes P^k_{A_n}$,
$\lambda_k=\Tr Q_k\omega_{A_{1}...A_{n}}$ and $\,A_{i_1}...A_{i_m}$,
$m\leq n$, be a subsystem of $A_{1}...A_{n}$. Then}
$$
\lim_{k\rightarrow\infty}H(\omega^k_{A_{i_1}\ldots
A_{i_m}})=H(\omega_{A_{i_1}\ldots A_{i_m}})\leq+\infty.
$$
\end{lemma}\smallskip

\emph{Proof.}   By noting that
\begin{equation*}
\lambda_k\omega^k_{A_{i_1}...A_{i_m}}\leq
P^k_{A_{i_1}}\otimes...\otimes P^k_{A_{i_m}}
\,\omega_{A_{i_1}...A_{i_m}}\,P^k_{A_{i_1}}\otimes...\otimes
P^k_{A_{i_m}}\quad \forall k,
\end{equation*}
this assertion can be proved by using Simon's convergence theorems
for the von Neumann entropy \cite[the Appendix]{Ruskai}.$\square$
\smallskip

\begin{remark}\label{cont}  Throughout  the paper  we will consider that \emph{continuous}
functions on a metric space are \emph{finite} on this space (in
contrast to lower (upper) semicontinuous functions which can take
infinite values). We will say that \emph{local continuity of a
function $f$ implies local continuity of a function $g\shs$} if
$$
\lim_{k\rightarrow\infty}f(x_k)=f(x_0)\neq\pm\infty\quad\Rightarrow\quad
\lim_{k\rightarrow\infty}g(x_k)=g(x_0)\neq\pm\infty
$$
for any sequence $\{x_k\}$ converging to $x_0$.
\end{remark}\smallskip

We will refer to the following simple fact. \smallskip

\begin{lemma}\label{vsl} \emph{Let $f_1,...f_n$ be nonnegative lower
semicontinuous functions on a metric space $X$. Then local
continuity of $\,\sum_{k=1}^n f_k$ implies local continuity of all
the functions $f_1,...f_n$.}
\end{lemma}\smallskip

\begin{remark}\label{nat-ext}  Throughout  the paper  we will
assume that any function $F$ on the set $\S(\H)$ of quantum states
is extended to the cone $\T_+(\H)$ by the formula
$$
F(\rho)=[\Tr\rho]F\left(\frac{\rho}{\Tr\rho}\right)
$$
Note that such extension of a function $F(\rho)=\sum_k \alpha_k
H(\Phi_k(\rho))$, where $\Phi_k$ are positive linear maps, is
determined by the same formula provided that extension
(\ref{ent-ext}) of the von Neumann entropy is used.
\end{remark}\smallskip

\section{On faithful extension of entropic quantities}

Quantum correlations of a state  of a finite-dimensional $n$-partite
system $A_{1}...A_{n}$ are described by different entropic
quantities (such as quantum mutual information, quantum conditional
entropy, conditional mutual information, topological entanglement
entropy, etc.), defined as a real linear combination of marginal
entropies, i.e. as a function
\begin{equation}\label{g-form}
  F(\omega_{A_{1}...A_{n}})=\sum_k \alpha_k H(\omega_{X_k})
\end{equation}
on the set of all states of the system, where $\omega_{X_k}$ is a
partial state of $\omega_{A_{1}...A_{n}}$ corresponding to a
subsystem $X_k$ of $A_{1}...A_{n}$.\smallskip

In infinite dimensions such entropic quantity is correctly defined
if all the marginal entropies $\,H(\omega_{X_k})\,$ involved in the
corresponding linear combination (\ref{g-form}) are finite (or at
least this linear combination does not contain the uncertainty
$"\infty-\infty"$). Nevertheless such narrow domain of definition
can be extended by using althernative expressions, which are more
stable under passage to infinite dimensions than the linear
combination in (\ref{g-form}). Examples of such extensions for
quantum mutual information and for quantum conditional entropy have
been mentioned in the Introduction.

These examples motivate questions about possible extensions of
linear combination (\ref{g-form}) to states at which it is not
correctly defined and about requirements to these extensions in
general settings.\smallskip

In general the function $F$ in  (\ref{g-form}) is not lower or upper
semicontinuous on the set of states at which it is well defined (the
linear combination does not contain the uncertainty
$"\infty-\infty"$), but Lemma \ref{t-ext-l} shows that for any such
state this function possesses the following property:
\begin{equation}\label{t-ext-p}
\lim_{k\rightarrow\infty}F(\omega^k_{A_{1}...A_{n}})=F(\omega_{A_{1}...A_{n}})\in[-\infty,+\infty]
\end{equation}
for \emph{any} sequence of "truncated" states
$$
\omega^k_{A_{1}...A_{n}}=\lambda_k^{-1}Q_k\omega_{A_{1}...A_{n}}Q_k,\quad
Q_k= P^k_{A_1}\otimes \ldots\otimes P^k_{A_n},\;\lambda_k=\Tr
Q_k\omega_{A_{1}...A_{n}},
$$
determined by sequences
$\{P^k_{A_{1}}\}_k\subset\B(\H_{A_{1}})$,...,
$\{P^k_{A_{n}}\}_k\subset\B(\H_{A_{n}})$ of  projectors strongly
converging to the identity operators
$I_{A_{1}}$,...,$I_{A_{n}}$.\footnote{If $F(\omega_{A_{1}...A_{n}})$
is well defined then $F(\omega^k_{A_{1}...A_{n}})$ is well defined
for all $k$ (since finiteness of $H(\rho)$ implies finiteness of
$H(P\rho P)$ for any projector $P$).}
\smallskip

This property can be treated as a self-consistency or stability of
$F$ with respect to state truncation. It seems to be a reasonable
requirement for any measure of quantum correlations from the both
analytical and physical points of view. This motivates the following
definition.\smallskip

\begin{definition}\label{f-fun} Let $\mathcal{A}$ be a subset of states of a multipartite system
$A_{1}...A_{n}$ such that
\begin{equation}\label{A-cond}
\omega_{A_{1}...A_{n}}\in\mathcal{A}\quad\Rightarrow\quad [\Tr
Q\shs\omega_{A_{1}...A_{n}}]^{-1}Q\shs\omega_{A_{1}...A_{n}}Q \in
\mathcal{A},
\end{equation}
where $Q=P_{A_1}\otimes \ldots\otimes P_{A_n}$, for any projectors
$P_{A_{i}}\in\B(\H_{A_{i}})$, $i=\overline{1,n}$. We will say that a
function $F$ taking values in $[-\infty,+\infty]$ is \emph{faithful}
on $\mathcal{A}$ if for any state
$\omega_{A_{1}...A_{n}}\in\mathcal{A}$ the above condition
(\ref{t-ext-p}) is valid.
\end{definition}\medskip

It is clear that continuity of $F$ implies faithfulness of $F$, but
the reverse is not true. The simplest example is given by the von
Neumann entropy of a marginal state
$F(\omega_{A_{1}...A_{n}})=H(\omega_{X})$, $X\subseteq
A_{1}...A_{n}$, which is a faithful function on the whole set of
states of $A_{1}...A_{n}$ by Lemma \ref{t-ext-l}.

The faithfulness property seems to be a reasonable replacement for
the continuity in infinite dimensions. Many characteristics of
states of bi- and multipartite infinite-dimensional systems are
globally faithful but not continuous. For example, the bi- and
multipartite quantum mutual information, the Entanglement of
Formation of a bipartite state (defined as a "continuous" convex
roof, see \cite{EM}), etc. The faithfulness of these characteristics
follows from their lower semicontinuity and monotonicity under local
operations.\smallskip

\begin{lemma}\label{f-s-cond}
\emph{If $F$ is a lower semicontinuous function on a subset
$\mathcal{A}$ of $\T_+(\H_{A_{1}...A_{n}})$ satisfying condition
(\ref{A-cond}) and  $F(\Phi_1\otimes...\otimes\Phi_n(\omega))\leq
F(\omega)$ for any $\,\omega\in\mathcal{A}$ and arbitrary  quantum
operations $\,\Phi_1\!:A_1\rightarrow
\!A_1,..,\Phi_n\!:A_n\rightarrow \!A_n$ then $F$ is faithful on
$\mathcal{A}\cap\S(\H_{A_{1}...A_{n}})$.}
\end{lemma}\smallskip

The faithfulness is a very convenient  property from the analytical
point of view. It is the faithfulness of the bipartite quantum
mutual information that was implicitly used in \cite{Kuz} to prove
concavity and other basic properties of quantum conditional entropy
for its extension (\ref{c-ent-def+}).

The above arguments show that a function $\widehat{F}$ can be
treated as an adequate extension of a particular information
quantity $F$ to a subset $\mathcal{A}$ of $\S(\H_{A_{1}...A_{n}})$
if it is faithful on this subset.
\smallskip
\begin{definition}\label{t-ext}
A function $\widehat{F}$ is called \emph{faithful extension}
(briefly, \emph{$\mathfrak{F}$-extension}) of the function $F$ in
(\ref{g-form}) to a subset $\mathcal{A}\in\S(\H_{A_{1}...A_{n}})$
satisfying condition (\ref{A-cond}) if it is faithful on
$\mathcal{A}$ and $\widehat{F}(\omega)=F(\omega)$ for any state
$\omega\in\mathcal{A}$  for which $F(\omega)$ is well defined.
\end{definition}\medskip

Definition \ref{t-ext} implies the following simple observations
used below.\smallskip

\begin{lemma}\label{t-ext-r}
\emph{If  $\mathfrak{F}$-extension of a linear combination of
marginal entropies to a particular subset exists then it is uniquely
defined. }\end{lemma}\smallskip

\begin{lemma}\label{t-ext-pr}
\emph{If $\widehat{F}_1$ and $\widehat{F}_2$ are
$\mathfrak{F}$-extensions of quantities $F_1$ and $F_2$ to a
particular set $\mathcal{A}$ such that
$c_1\widehat{F}_1+c_2\widehat{F}_2$, $c_1,c_2\in\mathbb{R}$, is well
defined on $\mathcal{A}$ then $c_1\widehat{F}_1+c_2\widehat{F}_2$ is
a $\mathfrak{F}$-extension of the quantity $c_1F_1+c_2F_2$ to the
set $\mathcal{A}$.}
\end{lemma}\medskip

We finish this section by a general result concerning the case when
linear combination (\ref{g-form}) is bounded on the set where it is
well defined.\footnote{Examples corresponding to this case are
considered below (see Corollaries
\ref{main-c},\ref{FA-gen},\ref{cmi-pr-c},\ref{ent-ent-c}).} This
result (proved by Winter's modification of the Alicki-Fannes technic \cite{A&F,Winter}) shows that
boundedness of (\ref{g-form}) implies its uniform continuity and
gives estimates for its variation.

\smallskip

\begin{property}\label{FA-cont}
\emph{If a function $F(\omega_{A_{1}...A_{n}})=\sum_k \alpha_k
H(\omega_{X_k})$ is bounded on the set
$\;\S_{\mathrm{f}}=\{\shs\omega_{A_{1}...A_{n}}\shs|\shs\max_i\mathrm{rank}\shs\omega_{A_{i}}<+\infty\shs\}$
then it has a unique continuous extension $\widehat{F}$ to the set
$\,\S(\H_{A_{1}...A_{n}})$ such that
\begin{equation}\label{FA-ineq}
\!    |\widehat{F}(\omega^1)-\widehat{F}(\omega^2)|\leq
    \,\varepsilon\sup_{\omega,\omega'\in\S_{\mathrm{f}}}|F(\omega)-F(\omega')|+(1+\varepsilon)h_2\!\left(\frac{\varepsilon}{1+\varepsilon}\right)\sum_k|\alpha_k|
\end{equation}
for any $\,\omega^1,\omega^2\in\S(\H_{A_{1}...A_{n}})$ such that
$\;\varepsilon=\frac{1}{2}\|\shs\omega^1-\omega^2\|_1<1\,$, where
$\,h_2(\cdot)\,$ is the binary entropy.}\smallskip

\emph{If $\,F$ is a concave (corresp., convex) function on $\,\S_{\mathrm{f}}$ then $\,\widehat{F}$ is  a concave (corresp., convex) function on $\,\S(\H_{A_{1}...A_{n}})$ and $\,\sum_k|\alpha_k|$ in (\ref{FA-ineq}) can be replaced by
$\,\sum_{k:\alpha_k>0}|\alpha_k|$ (corresp., by $\,\sum_{k:\alpha_k<0}|\alpha_k|$).}
\end{property}
\smallskip

\emph{Proof.} It suffices to show that (\ref{FA-ineq}) holds with
$\widehat{F}=F$ for any states
$\,\omega^1,\omega^2\in\S_{\mathrm{f}}$, since this would imply that
$F$ is a uniformly continuous function on the dense subset
$\S_{\mathrm{f}}$ of $\S(\H_{A_{1}...A_{n}})$  uniquely extended, by
a standard way, to a continuous function $\widehat{F}$ on the set
$\,\S(\H_{A_{1}...A_{n}})$ satisfying (\ref{FA-ineq}).\smallskip

The concavity of the von Neumann entropy and inequality
(\ref{w-k-ineq}) imply
\begin{equation}\label{F-c-b}
\begin{array}{cc}
\lambda
F\left(\omega^1\right)+(1-\lambda)F\left(\omega^2\right)+C_-h_2(\lambda)\\\\\leq
F\left(\lambda\omega^1+(1-\lambda)\omega^2\right)\leq\\\\ \lambda
F\left(\omega^1\right)+(1-\lambda)F\left(\omega^2\right)+C_+h_2(\lambda)
\end{array}
\end{equation}
for any states $\omega^1,\omega^2\in\S_{\mathrm{f}}$, where
$\,C_-=\sum_{k:\alpha_k<0}\alpha_k\,$  and
$\,C_+=\sum_{k:\alpha_k>0}\alpha_k$.

\smallskip

Following \cite{Winter} introduce the state
$\,\omega^{*}=(1+\varepsilon)^{-1}(\omega^1+[\shs\omega^2-\omega^1]_+)$. Then
\begin{equation}\label{omega-star}
\frac{1}{1+\varepsilon}\,\omega^1+\frac{\varepsilon}{1+\varepsilon}\,\tilde{\omega}^1=\omega^{*}=
\frac{1}{1+\varepsilon}\,\omega^2+\frac{\varepsilon}{1+\varepsilon}\,\tilde{\omega}^2,
\end{equation}
where $\,\tilde{\omega}^1=\varepsilon^{-1}[\shs\omega^2-\omega^1]_+\,$
and
$\,\tilde{\omega}^2=\varepsilon^{-1}((1+\varepsilon)\omega^{*}-\omega^{2})$
are states in $\S_{\mathrm{f}}$. By applying (\ref{F-c-b}) to the
above convex decompositions of $\,\omega^{*}$ we obtain
$$
\frac{1}{1+\varepsilon}\,(F(\omega^1)-F(\omega^2))\leq\frac{\varepsilon}{1+\varepsilon}\,(F(\tilde{\omega}^2)-F(\tilde{\omega}^1))+(C_+-C_-)\shs
h_2\!\left(\frac{\varepsilon}{1+\varepsilon}\right)
$$
and
$$
\frac{1}{1+\varepsilon}\,(F(\omega^2)-F(\omega^1))\leq\frac{\varepsilon}{1+\varepsilon}\,(F(\tilde{\omega}^1)-F(\tilde{\omega}^2))+(C_+-C_-)\shs h_2\!
\left(\frac{\varepsilon}{1+\varepsilon}\right).
$$
These inequalities imply (\ref{FA-ineq}) with $\widehat{F}=F$, since
$\,C_+-C_-=\sum_k|\alpha_k|$. \smallskip

The last assertion of the proposition follows from the above proof. $\;\square$
\smallskip

If $\,d_A\doteq\dim\H_A<+\infty\,$ then Proposition \ref{FA-cont} imply the following continuity bounds for the von Neumann entropy and for the conditional entropy
\begin{equation}\label{H-c-b}
   |H(\omega^1)-H(\omega^2)|\leq
    \,\varepsilon\log d_A+(1+\varepsilon)h_2\!\left(\frac{\varepsilon}{1+\varepsilon}\right)
\end{equation}
and
\begin{equation}\label{CH-c-b}
  |H(A|B)_{\omega^1}-H(A|B)_{\omega^2}|\leq
    2\varepsilon\log d_A+(1+\varepsilon)h_2\!\left(\frac{\varepsilon}{1+\varepsilon}\right)
\end{equation}
(the concavity of the both these quantities was taken into account).

Continuity bound (\ref{H-c-b}) is close (and asymptotically equivalent) to the sharpest continuity bound for the von Neumann entropy
\begin{equation*}
   |H(\omega^1)-H(\omega^2)|\leq
    \varepsilon\log (d_A-1)+h_2(\varepsilon)
\end{equation*}
obtained by Audenaert \cite{Aud}. Continuity bound (\ref{CH-c-b}) coincides with the tight continuity bound
for the conditional entropy proved by Winter using (\ref{omega-star}) and the special representation for the conditional entropy \cite[Lemma 2]{Winter}. So, we have some reasons to expect that Proposition \ref{FA-cont} gives quite sharp continuity bounds despite its universality.

\section{Quantum mutual information and its use}

\subsection{Bipartite case}

Quantum correlations of a state $\,\omega_{AB}\,$ of a
finite-dimensional bipartite quantum system $AB$ are characterized
by the value
\begin{equation}\label{mi-d-1}
I(A\!:\!B)_{\omega}=H(\omega_{A})+H(\omega_{B})-H(\omega_{AB})
\end{equation}
called quantum mutual information of this state
\cite{MI-B,N&Ch,Tucci}.

In infinite dimensions the linear combination of marginal entropies
in (\ref{mi-d-1}) may contain the uncertainty $"\infty-\infty"$, but
it can be correctly defined for any state $\omega_{AB}$ (as a value
in $[0,+\infty]$) by the expression
\begin{equation}\label{mi-d-2}
I(A\!:\!B)_{\omega}=H(\omega_{AB}\shs\Vert\shs\omega_{A}\otimes
\omega_{B})
\end{equation}
coinciding with (\ref{mi-d-1}) for any state $\omega_{AB}$ with
finite marginal entropies.
\medskip

In infinite dimensions the right hand side of (\ref{mi-d-2}) plays a
role of "building block" in construction of many characteristics of
quantum systems and channels. It is used in extension
(\ref{c-ent-def+}) of quantum conditional entropy, in the
definitions of quantum mutual and coherent informations of an
infinite-dimensional quantum channel (see Sect.8.1,8.2 below), etc.
\smallskip

In some applications quantity (\ref{mi-d-1}) is used with
arbitrarily positive trace class operators $\omega_{AB}$ (not only
states). In this case (\ref{H-fun-eq}) and (\ref{H-fun-eq+}) imply
\begin{equation*}
I(A\!:\!B)_{\omega}=H\left(\omega_{AB}\,\Vert\;\frac{\omega_{A}\otimes\omega_{B}}{\Tr\omega_{AB}}\right).
\end{equation*}

Basic properties of the relative entropy show that $\omega\mapsto
I(A\!:\!B)_{\omega}$ is a lower semicontinuous function on the cone
$\T_{+}(\H_{AB})$ possessing the following properties:
\begin{enumerate}[\textrm{A}1)]
    \item $I(A\!:\!B)_{\omega}\geq0$ for any operator $\omega_{AB}$ and $I(A\!:\!B)_{\omega}=0$ if and only if $\omega_{AB}$ is a product operator, i.e.
    $[\Tr\omega_{AB}]\shs\omega_{AB}=\omega_{A}\otimes\omega_{B}$;
    \item monotonicity under local reduction: $I(A\!:\!BC)_{\omega}\geq
    I(A\!:\!B)_{\omega}$;
    \item monotonicity under local operations: $I(A\!:\!B)_{\omega}\geq
    I(A'\!:\!B')_{\Phi_{\!A}\otimes\Phi_{B}(\omega)}$ for arbitrary
    quantum operations $\Phi_A:A\rightarrow A'$ and $\Phi_B:B\rightarrow
    B'$; \footnote{If $\Phi_A$ and $\Phi_B$ are quantum channels then this property directly follows from the monotonicity of the relative entropy,
    if $\Phi_A$ and $\Phi_B$ are trace-non-preserving quantum operations then additional arguments are required, for example,
    formula (\ref{d-f}) in the Appendix can be used.}
    \item additivity: $I(AA'\!:\!BB')_{\omega\otimes\omega'}=
    I(A\!:\!B)_{\omega}+I(A'\!:\!B')_{\omega'}$.
\end{enumerate}\smallskip

By Lemma \ref{f-s-cond} the lower semicontinuity of
$I(A\!:\!B)_{\omega}$ and property A3 show that
$I(A\!:\!B)_{\omega}$ defined by (\ref{mi-d-2}) is the
$\mathfrak{F}$-extension of (\ref{mi-d-1}) to the set $\S(\H_{AB})$.
\smallskip

We will use the following upper bound
\begin{equation}\label{mi-u-b}
I(A\!:\!B)_{\omega}\leq 2\min\{H(\omega_A),H(\omega_B)\}
\end{equation}
mentioned in \cite{MI-B}. It directly follows from identity
(\ref{sp-ident}) in the Appendix.\smallskip

The following theorem gives local continuity conditions for the
bipartite quantum mutual information (as a function on the cone
$\T_{+}(\H_{AB})$), which will be essentially used below.\smallskip

\begin{theorem}\label{main} A) \emph{The limit relation
\begin{equation}\label{mi-l-r}
\lim_{k\rightarrow\infty}I(A\!:\!B)_{\omega^k}=I(A\!:\!B)_{\omega^0}
\end{equation}
holds for a sequence $\{\omega^k\}\subset\T_{+}(\H_{AB})$ converging
to an operator $\omega^0$ if one of the following conditions is
valid:}
\begin{enumerate}[a)]
    \item \emph{$\lim_{k\rightarrow\infty}H(\omega_X^k)=H(\omega_X^0)<+\infty$, where $X$ is \textbf{one} of the systems  $A$,$B$ (in
    this case the limit in (\ref{mi-l-r}) is finite according to (\ref{mi-u-b}));}
    \item \emph{$\lambda_k\omega^k\leq\Phi_A^k\otimes\Phi_B^k(\omega^0)$ for some
    sequences  $\{\Phi_A^k\}$ and $\{\Phi_B^k\}$ of local quantum operations and some sequence
    $\{\lambda_k\}\subset[0,1]\shs$ converging to $1$.}
\end{enumerate}

B) \emph{If (\ref{mi-l-r}) holds for a sequence
$\{\omega^k\}\subset\T_{+}(\H_{AB})$ as a finite limit then
\begin{equation}\label{mi-l-r+}
\lim_{k\rightarrow\infty}I(A'\!:\!B')_{\Phi_A\otimes\Phi_B(\omega^k)}=I(A'\!:\!B')_{\Phi_A\otimes\Phi_B(\omega^0)}<+\infty
\end{equation}
for arbitrary quantum operations $\,\Phi_A:A\rightarrow A'$ and
$\,\Phi_B:B\rightarrow B'$.}
\end{theorem}\medskip

Theorem \ref{main}B states, briefly speaking, that \emph{local
continuity of quantum mutual information is preserved by local
operations}.\smallskip

Theorem \ref{main} (proved in the Appendix) shows, in particular,
that the quantum mutual information $\,I(A\!:\!B)_{\omega}$ is
continuous on the set $\,\S(\H_{AB})$ if and only if either $A$ or
$B$ is a finite-dimensional system. Proposition \ref{FA-cont} and
upper bound (\ref{mi-u-b}) give the continuity bound for
$\,I(A\!:\!B)_{\omega}$ in this case.
\smallskip
\begin{corollary}\label{main-c}
\emph{If one of the systems $A$ and $B$, say $A$, is
finite-dimensional then $\,I(A\!:\!B)_{\omega}$ is a continuous
bounded function on the set $\,\S(\H_{AB})$ and
$$
|I(A\!:\!B)_{\omega^1}-I(A\!:\!B)_{\omega^2}|\leq 2\varepsilon
\log\dim\H_A+3(1+\varepsilon)h_2\!\left(\frac{\varepsilon}{1+\varepsilon}\right)
$$
for any $\,\omega^1,\omega^2\in\S(\H_{AB})$ such that
$\;\varepsilon=\frac{1}{2}\|\shs\omega^1-\omega^2\|_1<1\,$, where
$\,h_2(\cdot)\,$ is the binary entropy.}
\end{corollary}\medskip

\begin{remark}\label{main-r+} The continuity conditions for $\,I(A\!:\!B)_{\omega}$ in Theorem
\ref{main}A coincides with the continuity condition for the
Entanglement of Formation $\,E_F(\omega)$ of a state of an
infinite-dimensional bipartite system  \cite[Sect.6]{EM}.
\end{remark}\smallskip

The following example shows that condition a) in Theorem \ref{main}A
is not necessary for existence of a finite limit in (\ref{mi-l-r}).
\begin{example}\label{main-r}
Consider a sequence $\{\rho_k\}\subset\S(\H_A)$ converging to a
state $\rho_0$ such that $\lim_{k\rightarrow\infty} H(\rho_k)\neq
H(\rho_0)$. By Lemma \ref{p-lemma} there exists  a sequence
$\{\omega^k\}\subset\S(\H_{AB})$, $\H_B\cong\H_A$, converging to a
state $\omega^0\in\S(\H_{AB})$ such that $\omega^k_A=\rho_k$ for all
$k\geq0$. Let $\sigma_k=\omega^k_B$ and
$\,\tilde{\omega}^k=p_k\omega^k+(1-p_k)\rho_k\otimes\sigma_k$, where
$\{p_k\}$ is a sequence of positive numbers such that
$\lim_{k\rightarrow\infty} p_kH(\rho_k)=0$. The sequence
$\{\tilde{\omega}^k\}$ converges to the state
$\tilde{\omega}^0=\rho_0\otimes\sigma_0$. By using the convexity of
the relative entropy it is easy to see that
$$
\lim_{k\rightarrow\infty}\,I(A\!:\!B)_{\tilde{\omega}^k}=0=I(A\!:\!B)_{\tilde{\omega}^0}
$$
while
$$
\lim_{k\rightarrow\infty}H(\tilde{\omega}_A^k)\neq
H(\tilde{\omega}_A^0)\quad \textrm{and}\quad
\lim_{k\rightarrow\infty}H(\tilde{\omega}_B^k)\neq
H(\tilde{\omega}_B^0),
$$
since $\,\tilde{\omega}_A^k=\rho_k\,$ and
$\,\tilde{\omega}_B^k=\sigma_k\,$ for all $k\geq0$.\smallskip
\end{example}

Condition b) in Theorem \ref{main}A implies the following
observation used below.\smallskip
\begin{corollary}\label{main-c-n+} \emph{Let  $\,X$ and $Y$ be disjoint subsystems of $A_{1}...A_{n}$. The function
$\,\omega_{A_{1}...A_{n}}\mapsto I(X\!:\!Y)_{\omega}$ is faithful on
$\,\S(\H_{A_{1}...A_{n}})$.}\footnote{See Def.\ref{f-fun}. In
general faithfulness of the function $\,\omega_{XY}\mapsto
F(\omega_{XY})$ does not imply faithfulness of the function
$\,\omega_{A_{1}...A_{n}}\mapsto F(\omega_{XY})$.}\smallskip
\end{corollary}
\smallskip
\emph{Proof.}  We may assume that $X=A_1$ and $Y=A_2$. The validity
of relation (\ref{t-ext-p}) for the function
$\omega_{A_{1}...A_{n}}\mapsto I(A_1\!:\!A_2)_{\omega}$ follows from
the inequality
\begin{equation*}
\lambda_k\omega^k_{A_1A_2}\leq P^k_{A_1}\otimes P^k_{A_2}
\,\omega_{A_1A_2}\,P^k_{A_1}\otimes P^k_{A_2}\quad \forall k,
\end{equation*}
(where $\lambda_k$ is defined in (\ref{t-ext-p})) and condition b)
in Theorem \ref{main}A.$\square$\smallskip

\begin{remark}\label{main-c-n+r} Corollary \ref{main-c-n+}  and Lemma \ref{t-ext-l} show, by Lemma
\ref{t-ext-pr}, that the function
\begin{equation}\label{g-l-c}
    \omega_{A_{1}...A_{n}}\mapsto\sum_k \alpha_k I(X_{k}\shs:\shs
    Y_k)_{\omega}+\sum_k \beta_k H(\omega_{Z_k}),
\end{equation}
where $X_k,Y_k,Z_k$ are subsystems of $A_{1}...A_{n}$, is faithful
on the set of all states at which it is well defined (the linear
combination in (\ref{g-l-c}) does not contain the uncertainty
$"\infty-\infty"$).
\end{remark}

\subsection{Multipartite case}

Quantum mutual information of a state $\,\omega_{A_{1}...A_{n}}\,$
of a finite-dimensional $n$\nobreakdash-\hspace{0pt}partite quantum
system $A_{1}...A_{n}$ is defined as follows (cf.
\cite{SecMon,Herbut,N&Ch,3H+})
\begin{equation}\label{mi-d-3}
I(A_{1}\shs:\ldots:\shs A_{n})_{\omega}=\sum_{i=1}^n
H(\omega_{A_i})-H(\omega_{A_{1}...A_{n}}).
\end{equation}
In infinite dimensions it can be correctly defined for any state
$\omega_{A_{1}...A_{n}}$ (as a value in $[0,+\infty]$) by the
expression
\begin{equation}\label{mi-d-4}
I(A_{1}\shs:\ldots:\shs A_{n})_{\omega}=H(\shs\omega_{A_1\ldots
A_n}\shs\Vert\shs\omega_{A_1}\otimes\ldots\otimes \omega_{A_n}\shs)
\end{equation}
coinciding with (\ref{mi-d-3}) for any state
$\omega_{A_{1}...A_{n}}$ with finite marginal entropies. \medskip

Quantity (\ref{mi-d-3}) can be extended to the cone
$\T_{+}(\H_{A_1\ldots A_n})$ by the formula
$$
I(A_{1}\shs:\ldots:\shs A_{n})_{\omega}=
H\left(\shs\omega_{A_1\ldots
A_n}\,\Vert\;\frac{\omega_{A_1}\otimes\ldots\otimes
\omega_{A_n}}{[\Tr\omega_{A_1\ldots A_n}]^{n-1}}\shs\right).
$$

Basic properties of the relative entropy show that
$I(A_{1}\shs:\ldots:\shs A_{n})_{\omega}$ is a lower semicontinuous
function on the cone $\T_{+}(\H_{A_{1}...A_{n}})$ taking values in
$[0,+\infty]$  and possessing the analogs of the above properties
A1-A4. This can be shown by using the identity
$$
I(A_{1}\shs:\ldots:\shs A_{n})_{\omega}=I(A_{2}\shs:\shs
A_{1})_{\omega}+I(A_{3}\shs:\shs
A_{1}A_{2})_{\omega}+\ldots+I(A_{n}\shs:\shs
A_{1}...A_{n-1})_{\omega}
$$
which is directly verified for a state $\omega$ with finite marginal
entropies and can be extended to arbitrary states by approximation
using Corollary \ref{main-c-n+}. It follows from Lemma
\ref{f-s-cond} that $I(A_{1}\shs:\ldots:\shs A_{n})_{\omega}$
defined by (\ref{mi-d-4}) is a $\mathfrak{F}$-extension of
(\ref{mi-d-3}) to the set $\S(\H_{A_1\ldots A_n})$. \smallskip

The above identity makes possible to obtain from (\ref{mi-u-b}) the
upper bound
\begin{equation}\label{n-mi-u-b}
I(A_1\!:\ldots:\!A_n)_{\omega}\leq 2\min_{1\leq j\leq n}\sum_{i\neq
j}H(\omega_{A_{i}})
\end{equation}
and to derive from Theorem \ref{main} its $n$-partite version.
\smallskip

\begin{corollary}\label{main-c-n} A) \emph{The limit relation
\begin{equation}\label{mi-l-r-n}
\lim_{k\rightarrow\infty} I(A_{1}\shs:...:\shs
A_{n})_{\omega^k}=I(A_{1}\shs:...:\shs A_{n})_{\omega^0}
\end{equation}
holds for a sequence $\{\omega^k\}\subset\T_{+}(\H_{A_1...A_n})$
converging to an operator $\,\omega^0$ if one of the following
conditions is valid:}
\begin{enumerate}[a)]
    \item \emph{$\lim_{k\rightarrow\infty}H(\omega_{A_{i}}^k)=H(\omega_{A_{i}}^0)<+\infty\,$ for at least $\,n\!-\!1\,$ values of
    $\;i$ (in
    this case the limit in (\ref{mi-l-r-n}) is finite according to (\ref{n-mi-u-b}));} \footnote{It is easy to see that validity of this
    relation for $\,n\!-\!2\,$ values of $\,i$ does not imply
    (\ref{mi-l-r-n}).}
    \item \emph{$\lambda_k\omega^k\leq\Phi_{A_1}^k\otimes...\otimes\Phi_{A_n}^k(\omega^0)$ for some
    sequences  $\{\Phi_{A_1}^k\}$,...,$\{\Phi_{A_n}^k\}$ of local quantum operations and some sequence
    $\{\lambda_k\}\subset[0,1]\shs$ converging to $1$.}
\end{enumerate}

B) \emph{If (\ref{mi-l-r-n}) holds for a sequence
$\{\omega^k\}\subset\T_{+}(\H_{A_1...A_n})$ as a finite limit then
\begin{equation*}
\lim_{k\rightarrow\infty} I(A'_{1}\shs:...:\shs
A'_{n})_{\Phi_{A_1}\otimes...\otimes\Phi_{A_n}(\omega^k)}=I(A'_{1}\shs:...:\shs
A'_{n})_{\Phi_{A_1}\otimes...\otimes\Phi_{A_n}(\omega^0)}
\end{equation*}
for arbitrary quantum operations $\Phi_{A_1}:A_1\rightarrow
A'_1,...,\Phi_{A_n}:A_n\rightarrow A'_n$.}
\end{corollary}\medskip

Corollary \ref{main-c-n}B  states that local operations do not
destroy continuity of the multipartite quantum mutual information
(similar to the bipartite case).\smallskip

By using the condition b) in Corollary \ref{main-c-n}A one can
generalize Corollary \ref{main-c-n+}, i.e. to show global
faithfulness of the function $\omega_{A_{1}...A_{n}}\mapsto
I(A_{i_1}\shs:...:\shs A_{i_m})_{\omega}$ for any subsystems
$\,A_{i_1},...,A_{i_m}$ of $A_{1}...A_{n}$.
\smallskip

Proposition \ref{FA-cont} and upper bound (\ref{n-mi-u-b}) imply the
following result.
\smallskip
\begin{corollary}\label{main-c+}
\emph{If $\,n-1\,$ subsystems, say $A_1,...,A_{n-1}$, are
finite-dimensional then $I(A_1\!:\!...\!:\!A_n)_{\omega}$ is a
continuous bounded function on the set $\,\S(\H_{A_1...A_n})$ and
$$
|I(A_1\!:\!...\!:\!A_n)_{\omega^1}-I(A_1\!:\!...\!:\!A_n)_{\omega^2}|\leq
2\varepsilon C+(n+1)(1+\varepsilon)h_2\!\left(\frac{\varepsilon}{1+\varepsilon}\right)
$$
for any $\,\omega^1,\omega^2\in\S(\H_{A_1...A_n})$ such that
$\;\varepsilon=\frac{1}{2}\|\shs\omega^1-\omega^2\|_1<1\,$, where
$\,h_2(\cdot)\,$ is the binary entropy and $\,C=\log\dim\H_{A_1...A_{n-1}}$.}
\end{corollary}\medskip

\subsection{General relations between conditional and unconditional quantities}

For a quantity $F(\omega_{A_{1}...A_{n}})=\sum_k \alpha_k
H(\omega_{X_k})$ introduce the corresponding conditional quantity
\begin{equation}\label{cond-q}
F_{\cdot|B}(\omega_{A_{1}...A_{n}B})=\sum_k \alpha_k
\!\left[H(\omega_{X_kB})-H(\omega_{B})\right].
\end{equation}
By considering $F$ as a function of $\,\omega_{A_{1}...A_{n}B}\,$ we
have
$$
\left[F_{\cdot|B}-F\right]\!(\omega_{A_{1}...A_{n}B})=-\sum_k
\alpha_k
\!\left[H(\omega_{X_k})+H(\omega_{B})-H(\omega_{X_kB})\right]
$$
for any state with finite marginal entropies. Hence, Remark
 \ref{main-c-n+r}, upper bound
(\ref{mi-u-b}) and Theorem \ref{main} imply the following
observation.\smallskip

\begin{property}\label{main-c-2}
\emph{The difference $\left[F_{\cdot|B}-F\right]$ has the finite
$\mathfrak{F}$-extension
\begin{equation}\label{c-dif}
\left[F_{\cdot|B}-F\right]\!(\omega_{A_{1}...A_{n}B})=-\sum_{k}
\alpha_k I(X_k\!:\!B)_{\omega}
\end{equation}
to the set
$\,\left\{\shs\omega_{A_{1}...A_{n}B}\,|\,\min\left\{H(\omega_{B}),\sum_kH(\omega_{X_k})\right\}<+\infty\shs\right\}$
possessing the properties:}
\begin{enumerate}[1)]
    \item \emph{$\left|\!\left[F_{\cdot|B}-F\right]\!(\omega_{A_{1}...A_{n}B})\right|\leq
    \min\{H(\omega_{B})\!\left[|\sum_k\! \alpha_k|+\!\sum_{k}\!|\alpha_k|\right]\!,2\sum_{k}\!|\alpha_k|H(\omega_{X_k})\}$;}
    \item \emph{the function $\,\omega_{A_{1}...A_{n}B}\mapsto\left[F_{\cdot|B}-F\right]\!(\omega_{A_{1}...A_{n}B})$ is continuous on a subset
    $\mathcal{A}\subset\T_+(\H_{A_{1}...A_{n}B})$ if one the following conditions holds:}
\begin{enumerate}[a)]
\item \emph{the function $\,\omega_{A_{1}...A_{n}B}\mapsto H(\omega_{B})$
is continuous on $\mathcal{A}$;}
\item \emph{the functions $\,\omega_{A_{1}...A_{n}B}\mapsto H(\omega_{X_k})$
is continuous on $\mathcal{A}$ for all $\,k$.}
\end{enumerate}
\end{enumerate}
\end{property}

Proposition \ref{main-c-2} with condition a) shows, roughly
speaking, that conditioning upon a system with finite (corresp.
continuous) entropy \emph{does not destroy} finiteness  (corresp.
continuity) of a quantity $F(\omega_{A_{1}...A_{n}})=\sum_k \alpha_k
H(\omega_{X_k})$.\smallskip

Proposition \ref{main-c-2} with condition b) shows that finiteness
(corresp. continuity) of all marginal entropies in a linear
combination $F(\omega_{A_{1}...A_{n}})=\sum_k \alpha_k
H(\omega_{X_k})$ guarantees finiteness (corresp. continuity) of the
conditional quantity $F_{\cdot|B}$ \emph{regardless} of the system
$B$.\smallskip

By Proposition \ref{main-c-2} the formula
\begin{equation}\label{c-dif+}
F_{\cdot|B}(\omega_{A_{1}...A_{n}B})=\sum_k \alpha_k
\left[H(\omega_{X_k})-I(X_k\!:\!B)_{\omega}\right]
\end{equation}
defines  the $\mathfrak{F}$-extension of (\ref{cond-q}) to the set
$\,\left\{\shs\omega_{A_{1}...A_{n}B}\,|\,\max_k
H(\omega_{X_k})<+\infty\right\}$.\smallskip

\textbf{Note:} Proposition \ref{main-c-2} does not assert the
existence of $\mathfrak{F}$-extension of quantity (\ref{cond-q}) to
the set
$\,\{\shs\omega_{A_{1}...A_{n}B}\,|\,H(\omega_{B})<+\infty\shs\}$.\smallskip

\section{Extended quantum conditional entropy}

The quantum conditional entropy
\begin{equation}\label{c-e-d}
H(A|B)_{\omega}=H(\omega_{AB})-H(\omega_B)
\end{equation}
of a state of a finite-dimensional bipartite system $AB$ is
essentially used in analysis of quantum systems and channels despite
its possible negativity \cite{H-SCI,N&Ch}. It has the following
basic properties:
\begin{enumerate}[\textrm{B}1)]
    \item concavity: $H(A|B)_{p\omega_1+(1-p)\omega_2}\geq pH(A|B)_{\omega_1}+(1-p)H(A|B)_{\omega_2}$;
    \item monotonicity: $H(A|B)_{\omega}\geq H(A|BC)_{\omega},\;\,\omega=\omega_{ABC}$;
    \item subadditivity: $H(AA'|BB')_{\omega}\leq
    H(A|B)_{\omega}+H(A'|B')_{\omega},\;\,\omega=\omega_{AA'BB'}$.
\end{enumerate}\smallskip

In infinite dimensions formula (\ref{c-e-d}) defines a finite
quantity possessing  properties B1-B3 on the set
$\S_{\mathrm{f}}=\{\omega_{AB}\,|\,\max\{H(\omega_A),H(\omega_B)\}<+\infty\}$.
This narrow domain of definition of $H(A|B)_{\omega}$ is extended in
\cite{Kuz}, where it is shown that the quantity
\begin{equation}\label{c-ent-def++}
H_{\mathrm{e}}(A|B)_{\omega}=H(\omega_{A})-H(\omega_{AB}\shs\Vert\shs\omega_{A}\otimes
\omega_{B})=H(\omega_{A})-I(A\!:\!B)_{\omega}
\end{equation}
coinciding with (\ref{c-e-d}) on $\S_{\mathrm{f}}$ possesses
properties B1-B3 on the convex set
$\{\shs\omega_{AB}\,|\,H(\omega_{A})<+\infty\shs\}$ containing
states with $H(\omega_{AB})=H(\omega_B)=+\infty$. This extension
turns out to be very useful in the analysis of infinite-dimensional
quantum systems and channels \cite{Ren, EAC}.\footnote{This
extension of the quantum conditional entropy plays a basic role of
the proof of the generalized version of the
Bennett-Shor-Smolin-Thaplyal theorem given in \cite{EAC}.} Note that
(\ref{c-ent-def++}) is a partial case of (\ref{c-dif+}) for
$F(\omega_{A})=H(\omega_{A})$.
\smallskip

Proposition 1 in \cite{Kuz} and Proposition \ref{main-c-2}  imply
the following observations.\smallskip

\begin{property}\label{c-e-cont} A) \emph{The quantity $H_{\mathrm{e}}(A|B)_{\omega}$ defined by (\ref{c-ent-def++}) is a $\mathfrak{F}$-extension of the
quantum conditional entropy (\ref{c-e-d}) to the set
$\{\shs\omega_{AB}\,|\,H(\omega_{A})<+\infty\shs\}$ possessing
properties B1-B3 such that $|H_{\mathrm{e}}(A|B)_{\omega}|\leq
H(\omega_{A})$.}\smallskip

B) \emph{Local continuity of $H(\omega_{A})$ implies local
continuity of $H_{\mathrm{e}}(A|B)_{\omega}$.}
\end{property}\medskip

Continuity bound for the conditional entropy defined by formula (\ref{c-e-d}) under the condition $\,\dim\H_A<+\infty\,$ was originally obtained
by Alicki and Fannes in \cite{A&F} and then was strengthened by Winter in \cite{Winter}. Proposition \ref{FA-cont} gives the continuity bound for the extended quantum conditional entropy coinciding with Winter's continuity bound.\footnote{So, Corollary \ref{FA-gen} generalizes Winter's continuity bound to the case $\,H(\omega_B)=+\infty\,$. It can be directly derived from Lemma 2 in \cite{Winter} by approximation.}\smallskip

\begin{corollary}\label{FA-gen} \emph{If the system $A$ is finite-dimensional then
$\,H_{\mathrm{e}}(A|B)_{\omega}$ is a continuous bounded function on
the set $\,\S(\H_{AB})$ and
\begin{equation}\label{FA-gen-f}
|H_{\mathrm{e}}(A|B)_{\omega^1}-H_{\mathrm{e}}(A|B)_{\omega^2}|\leq2\varepsilon \log\dim\H_A+(1+\varepsilon)\shs h_2\!\left(\frac{\varepsilon}{1+\varepsilon}\right)
\end{equation}
for any $\,\omega^1,\omega^2\in\S(\H_{AB})$ such that
$\;\varepsilon=\frac{1}{2}\|\shs\omega^1-\omega^2\|_1<1\,$, where
$\,h_2(\cdot)\,$ is the binary entropy.}
\end{corollary}\medskip

Besides the above-mentioned direct applications the function
$H_{\mathrm{e}}(A|B)_{\omega}$ can be used for construction of
$\mathfrak{F}$-extensions for linear combinations of marginal
entropies of a special form. By Remark \ref{main-c-n+r} the function
\begin{equation}\label{g-l-c+}
    \omega_{A_{1}...A_{n}}\mapsto\sum_k \alpha_k H_{\mathrm{e}}(X_k|Y_k)_{\omega},
\end{equation}
where $X_k,Y_k$ are disjoint subsystems of $A_{1}...A_{n}$, is
faithful on the set of all states for which
$\,H(\omega_{X_k})<+\infty$ for all $k$. This and Proposition
\ref{c-e-cont} imply the following observation.
\smallskip
\begin{property}\label{main-c-1}
\emph{If a quantity $F(\omega_{A_{1}...A_{n}})=\sum_k \alpha_k
H(\omega_{X_k})$ can be represented as follows
\begin{equation}\label{sp-rep}
F(\omega_{A_{1}...A_{n}})=\beta_1 H(\omega_{A_{i_0}})+\sum_{k>1}
\beta_k\!\left[H(\omega_{A_{i_0}Y_k})-H(\omega_{Y_k})\right],
\end{equation}
where $Y_k$ is a particular subsystem of $A_{1}...A_{n}\backslash
A_{i_0}$ for all $\,k$, then  $F$ has the $\mathfrak{F}$-extension
$$
\widehat{F}(\omega_{A_{1}...A_{n}})=\beta_1
H(\omega_{A_{i_0}})+\sum_{k>1} \beta_k
H_{\mathrm{e}}(A_{i_0}|Y_k)_{\omega}
$$ to the set
$\,\{\shs\omega_{A_{1}...A_{n}}\,|\,H(\omega_{A_{i_0}})<+\infty\shs\}$
possessing the properties:}
\begin{enumerate}[1)]
    \item \emph{$|\widehat{F}(\omega_{A_{1}...A_{n}})|\leq
    H(\omega_{A_{i_0}})\sum_{k\geq1}
|\beta_k|$;}
    \item \emph{local continuity of the function $\;\omega_{A_{1}...A_{n}}\mapsto H(\omega_{A_{i_0}})$ implies local continuity of
    the function $\,\omega_{A_{1}...A_{n}}\mapsto\widehat{F}(\omega_{A_{1}...A_{n}})$.}
\end{enumerate}
\end{property}

\begin{remark}\label{main-r}
If a quantity $F(\omega_{A_{1}...A_{n}})=\sum_k \alpha_k
H(\omega_{X_k})$ has representation (\ref{sp-rep}) for some index
$i_0$ then the corresponding conditional quantity (\ref{cond-q})
also has representation (\ref{sp-rep}) for the same  index $i_0$ and
hence, by Proposition \ref{main-c-1}, it has the
$\mathfrak{F}$-extension to the set
$\{\shs\omega_{A_{1}...A_{n}B}\,|\,H(\omega_{A_{i_0}})<+\infty\shs\}$
possessing the above properties 1) and 2) (with
$\shs\omega_{A_{1}...A_{n}B}$ instead of
$\shs\omega_{A_{1}...A_{n}}$). $\square$
\end{remark}\medskip

For a quantity $F(\omega_{A_{1}...A_{n}})=\sum_k \alpha_k
H(\omega_{X_k})$ consider the "reduced" quantity
\begin{equation*}
F_{\cdot\backslash A_{i_0}}(\omega_{A_{1}...A_{n}})=\sum_k \alpha_k
H(\omega_{X_k\backslash A_{i_0}}),
\end{equation*}
where $\,\omega_{X_k\backslash A_{i_0}}=\Tr_{A_{i_0}}\omega_{X_k}$
if $A_{i_0}\subseteq X_k$ and $\,\omega_{X_k\backslash
A_{i_0}}=\omega_{X_k}$ otherwise. We have
$$
\left[F_{\cdot\backslash
A_{i_0}}-F\right]\!(\omega_{A_{1}...A_{n}})=\sum_{k\in K(i_0)}
\alpha_k \!\left[H(\omega_{X_k\backslash
A_{i_0}})-H(\omega_{X_k})\right]\!,
$$
where $K(i_0)\doteq\{\shs k\shs|\shs A_{i_0}\subseteq X_k\shs\}$,
i.e. the difference $\left[F_{\cdot\backslash A_{i_0}}-F\right]$ has
representation (\ref{sp-rep}). So, Proposition \ref{main-c-1}
implies the following observation used below.\smallskip

\begin{corollary}\label{main-c-3}
\emph{The difference $\left[F_{\cdot\backslash A_{i_0}}-F\right]$
has the finite $\mathfrak{F}$-extension
\begin{equation*}
\left[F_{\cdot\backslash
A_{i_0}}-F\right]\!(\omega_{A_{1}...A_{n}})=-\sum_{k\in K(i_0)}
\alpha_kH_{\mathrm{e}}(A_{i_0}|X_k\backslash A_{i_0})_{\omega}
\end{equation*} to the set
$\,\{\shs\omega_{A_{1}...A_{n}}\,|\,H(\omega_{A_{i_0}})<+\infty\shs\}$
possessing the properties:}
\begin{enumerate}[1)]
    \item \emph{$\left|\left[F_{\cdot\backslash
A_{i_0}}-F\right]\!(\omega_{A_{1}...A_{n}})\right|\leq
    H(\omega_{A_{i_0}})\sum_{k\in K(i_0)}|\alpha_k|$;}
    \item \emph{local continuity of the function $\;\omega_{A_{1}...A_{n}}\mapsto H(\omega_{A_{i_0}})$ implies local continuity of
    the function $\;\omega_{A_{1}...A_{n}}\mapsto\left[F_{\cdot\backslash
A_{i_0}}-F\right]\!(\omega_{A_{1}...A_{n}})$.}
\end{enumerate}
\end{corollary}

Corollary \ref{main-c-3} shows, roughly speaking, that reducing a
system with finite (corresp. continuous) entropy \emph{does not
destroy} finiteness  (corresp. continuity) of any quantity $F$
defined as a linear combination of marginal entropies.\smallskip

\textbf{Note:} Corollary \ref{main-c-3} does not assert the
existence of $\mathfrak{F}$-extensions to the set
$\{\shs\omega_{A_{1}...A_{n}}\,|\,H(\omega_{A_{i_0}})<+\infty\shs\}$
for the quantities $F$ and $F_{\cdot\backslash A_{i_0}}$
separately.\smallskip

\section{Conditional mutual information}

\subsection{Tripartite system}

The conditional mutual information of a state $\omega_{ABC}$ of a
tripartite finite-dimensional system $ABC$ is defined as follows
\begin{equation}\label{cmi-d}
    I(A\!:\!C|B)_{\omega}\doteq
    H(\omega_{AB})+H(\omega_{BC})-H(\omega_{ABC})-H(\omega_{B}).
\end{equation}
This quantity plays important role in different branches of quantum
information theory \cite{SqE,D&J,F&R,Herbut,Tucci,3H+,Z}, it has the
following basic properties:
\begin{enumerate}[\textrm{C}1)]
    \item $I(A\!:\!C|B)_{\omega}\geq0$ for any state $\omega_{ABC}$ and $I(A\!:\!C|B)_{\omega}=0$ if and only if there is a channel $\Phi:B\rightarrow BC$
    such that $\omega_{ABC}=\id_A\otimes\Phi(\omega_{AB})$ \cite{H&Co+};
    \item monotonicity under local conditioning: $I(AB\!:\!C)_{\omega}\geq
    I(A\!:\!C|B)_{\omega}$;
    \item monotonicity under local operations: $I(A\!:\!C|B)_{\omega}\geq
    I(A'\!:\!C'|B)_{\Phi_A\otimes\id_{\!B}\otimes\Phi_C(\omega)}$ for arbitrary
    quantum operations $\Phi_A:A\rightarrow A'$ and $\Phi_C:C\rightarrow
    C'$;\footnote{If either $\Phi_A$ or $\Phi_C$ is a trace non-preserving operation then $I(A'\!:\!C'|B)_{\Phi_A\otimes\id_{\!B}\otimes\Phi_C(\omega)}$
    is defined by (\ref{cmi-d}), where $H$ is the extended von Neumann entropy (\ref{ent-ext}), see Remark \ref{nat-ext}.}
    \item additivity: $I(AA'\!:\!CC'|BB')_{\omega\otimes\omega'}=
    I(A\!:\!C|B)_{\omega}+I(A'\!:\!C'|B')_{\omega'}$;
    \item duality: $I(A\!:\!C|B)_{\omega}=
    I(A\!:\!C|D)_{\omega}$ for any pure state $\omega_{ABCD}$ \cite{D&J}.
\end{enumerate}\smallskip
The nonnegativity of $I(A\!:\!C|B)_{\omega}$ is a basic result of
quantum information theory well known as \emph{strong subadditivity
of von Neumann entropy} \cite{Ruskai}.\smallskip

The conditional mutual information (\ref{cmi-d}) can be represented
by one of the formulae
\begin{equation}\label{cmi-d+}
    I(A\!:\!C|B)_{\omega}=I(A\!:\!BC)_{\omega}-I(A\!:\!B)_{\omega},
\end{equation}
\begin{equation}\label{cmi-d++}
    I(A\!:\!C|B)_{\omega}=I(AB\!:\!C)_{\omega}-I(B\!:\!C)_{\omega}.
\end{equation}
By these representations, the nonnegativity of $I(A\!:\!C|B)$ is a
direct corollary of the monotonicity of the relative entropy under
partial trace.\footnote{The monotonicity of the quantum relative
entropy and the strong subadditivity of the von Neumann entropy are
globally equivalent \cite{H-SCI}.}\smallskip

Formula (\ref{c-dif}) in this case implies
\begin{equation}\label{cmi-d+++}
I(A\!:\!C|B)_{\omega}=I(A\!:\!C)_{\omega}-I(A\!:\!B)_{\omega}-I(C\!:\!B)_{\omega}+I(AC\!:\!B)_{\omega}.
\end{equation}
The quantity $I(A\!:\!C|B)_{\omega}$ defined in (\ref{cmi-d}) can be
also represented as follows
\begin{equation}\label{cmi-d++++}
\begin{array}{c}
I(A\!:\!C|B)_{\omega}=I(A\!:\!C)_{\omega}+I(AB\!:\!D)_{\tilde{\omega}}+I(BC\!:\!D)_{\tilde{\omega}}
\\\\+I(AC\!:\!D)_{\tilde{\omega}}-4H(\omega_{ABC}),
\end{array}
\end{equation}
where $\tilde{\omega}=\tilde{\omega}_{ABCD}$ is any purification of
the state $\omega_{ABC}$.\smallskip

In infinite dimensions  the quantity $I(A\!:\!C|B)_{\omega}$ is well
defined by formula (\ref{cmi-d}) as a faithful function
(Def.\ref{f-fun}) on the set
$$
\S_0=\left\{\shs\omega_{ABC}\,|\, H(\omega_{ABC})<+\infty,
H(\omega_{B})<+\infty\right\}.
$$
By Remark \ref{main-c-n+r} formulae (\ref{cmi-d+}), (\ref{cmi-d++}),
(\ref{cmi-d+++}), (\ref{cmi-d++++}) define $\mathfrak{F}$-extensions
of (\ref{cmi-d}) respectively to the sets
$$
\begin{array}{ll}
  \S_1=\left\{\shs\omega_{ABC}\,|\,
I(A\!:\!B)_{\omega}<+\infty\right\}, &
\quad\S_2=\left\{\shs\omega_{ABC}\,|\,
I(B\!:\!C)_{\omega}<+\infty\right\}, \\\\
  \S_3=\left\{\shs\omega_{ABC}\,|\,
H(\omega_{B})<+\infty\right\}, &
\quad\S_4=\left\{\shs\omega_{ABC}\,|\,
H(\omega_{ABC})<+\infty\right\}.
\end{array}
$$
The following theorem shows that formulae
(\ref{cmi-d})-(\ref{cmi-d++++}) agree with each other (coincide on
the sets $\S_i\cap\S_j$) and can be extended to a unique lower
semicontinuous function on the set $\,\S(\H_{ABC})$ possessing basis
properties of the conditional mutual information. \smallskip

\begin{theorem}\label{cmi-th}
\emph{There exists  a unique  lower semicontinuous function\break
$I_\mathrm{e}(A\!:\!C|B)_{\omega}$ on the set $\,\S(\H_{ABC})$ such
that:
\begin{itemize}
   \item $I_\mathrm{e}(A\!:\!C|B)_{\omega}$ coincides with $I(A\!:\!C|B)_{\omega}$ given by
    (\ref{cmi-d}),(\ref{cmi-d+}), (\ref{cmi-d++}),
(\ref{cmi-d+++}), (\ref{cmi-d++++}) respectively on the sets
$\,\S_0,\S_1,\S_2,\S_3,\S_4$;
   \item $I_\mathrm{e}(A\!:\!C|B)_{\omega}$ possesses the above-stated
properties C1-C5 of conditional mutual information.
\end{itemize}
This function can be defined by one of the equivalent
expressions\footnote{According to Remark \ref{nat-ext} we consider
the mutual information $I(X\!:\!Y)_{\omega}$ as a function on the
cone $\T_{+}(\H_{XY})$, so that $I(X\!:\!Y)_{Q\shs\omega Q}=[\Tr
Q\shs\omega]I(X\!:\!Y)_{\frac{Q\shs\omega Q}{\Tr Q\shs\omega}}$.}
\begin{equation}\label{cmi-e+}
I_\mathrm{e}(A\!:\!C|B)_{\omega}=\sup_{P_A}\left[\shs
I(A\!:\!BC)_{Q\omega Q}-I(A\!:\!B)_{Q\omega
Q}\shs\right]\!,\;\,Q=P_A\otimes I_B\otimes I_C,
\end{equation}
\begin{equation}\label{cmi-e++}
I_\mathrm{e}(A\!:\!C|B)_{\omega}=\sup_{P_C}\left[\shs
I(AB\!:\!C)_{Q\omega Q}-I(B\!:\!C)_{Q\omega
Q}\shs\right]\!,\;\,Q=I_A\otimes I_B\otimes P_C,
\end{equation}
where the suprema are over all finite rank projectors
$P_X\in\B(\H_X),\, X\!=\!A,C.$}\medskip

\emph{The function $I_\mathrm{e}(A\!:\!C|B)_{\omega}$ satisfies
$\mathfrak{F}$-extension condition (\ref{t-ext-p}) for any state
$\omega_{ABC}$ such that  $\;\min\{I(A\!:\!B)_{\omega},
I(B\!:\!C)_{\omega},H(\omega_{ABC}),H(\omega_{B})\}<+\infty$.\footnote{It
follows from (\ref{mi-u-b}) that this condition can be replaced by
the stronger but more explicit condition
$\;\min\{H(\omega_{A}),H(\omega_{B}),H(\omega_{C}), H(\omega_{AB}),
H(\omega_{BC}),H(\omega_{ABC})\}<+\infty$.} For an arbitrary state
$\omega\in\S(\H_{ABC})$ the following weaker property is valid:
\begin{equation}\label{t-ext-p+}
I_\mathrm{e}(A\!:\!C|B)_{\omega}=\lim_{k\rightarrow\infty}\lim_{l\rightarrow\infty}I_\mathrm{e}(A\!:\!C|B)_{\omega^{kl}},
\end{equation}
where
$$
\omega^{kl}=\lambda_{kl}^{-1}Q_{kl}\shs\omega\shs Q_{kl},\quad
 Q_{kl}= P^k_{A}\otimes
P^l_{B}\otimes P^k_{C},\;\,\lambda_{kl}=\Tr Q_{kl}\shs\omega,
$$
$\{P^k_{A}\}_k\subset\B(\H_A)$, $\{P^l_B\}_l\subset\B(\H_{B})$,
$\{P^k_C\}_k\subset\B(\H_C)$ are sequences of projectors strongly
converging to the identity operators $I_A$,$I_B$,$I_C$ such that
$\,\min\{\mathrm{rank}P^k_{A},\mathrm{rank}P^k_{C}\}<+\infty\,$ for
all $\,k$.}
\end{theorem}\medskip

Theorem \ref{cmi-th} (proved in the Appendix) shows that the
function $I_\mathrm{e}(A\!:\!C|B)_{\omega}$ can be considered as an
extension of the conditional mutual information to the set of all
states of infinite-dimensional tripartite system. So, in what
follows we will denote it by $I(A\!:\!C|B)_{\omega}$ (omitting the
subscript $\mathrm{e}$).\smallskip

\begin{remark}\label{cmi-u-b}
By using formulae (\ref{cmi-d+})-(\ref{cmi-d++++}), the upper bound
(\ref{mi-u-b}) and Proposition \ref{main-c-2} it is easy to show
that $\,\frac{1}{2}I(A\!:\!C|B)_{\omega}$ is upper bounded by each
of the quantities
\begin{equation*}
    H(\omega_{A}),\, H(\omega_{C}),\, H(\omega_{AB}),\, H(\omega_{BC}),\,
    H(\omega_{B})+\textstyle\frac{1}{2}I(A\!:\!C),\,
    H(\omega_{ABC})+\textstyle\frac{1}{2}I(A\!:\!C).
\end{equation*}
\end{remark}
The following corollary shows, in particular, that local continuity
of at least one of these upper bounds implies local continuity of
$\,I(A\!:\!C|B)_{\omega}$.
\smallskip

\begin{corollary}\label{cmi-th-c-1} A) \emph{Local continuity of \textbf{one} of the marginal entropies
$H(\omega_{A})$, $H(\omega_{C})$, $H(\omega_{AB})$, $H(\omega_{BC})$
implies local continuity of $I(A\!:\!C|B)_{\omega}$.\footnote{This
means that
$$
\lim_{k\rightarrow\infty}H(\omega^k_{X})=H(\omega^0_{X})<+\infty
\quad\Rightarrow\quad\lim_{k\rightarrow\infty}I(A\!:\!C|B)_{\omega^k}=I(A\!:\!C|B)_{\omega^0}<+\infty
$$
for a sequence $\{\omega^k\}$ converging to a state $\omega^0$,
where $X$ is one of the systems $A,C,AB,BC$.} Local continuity of
\textbf{one} of the marginal entropies $H(\omega_{B})$,
$H(\omega_{ABC})$ implies local continuity of the difference
$\,I(A\!:\!C|B)_{\omega}-I(A\!:\!C)_{\omega}$.}\smallskip

B) \emph{If $\{\omega^k\}$ is a sequence of states in $\S(\H_{ABC})$
converging to a state $\omega^0$ such that
$\lambda_k\omega^k\leq\Phi_A^k\otimes\id_B\otimes\Phi_C^k(\omega^0)$
for some sequences  $\{\Phi_A^k\}$ and $\{\Phi_C^k\}$ of local
quantum operations and some sequence $\{\lambda_k\}\subset[0,1]\shs$
converging to $1$ then}
$\lim_{k\rightarrow\infty}I(A\!:\!C|B)_{\omega^k}=I(A\!:\!C|B)_{\omega^0}\leq+\infty$
\end{corollary}\medskip

\emph{Proof.} A) If either $H(\omega_{A})$ or $H(\omega_{C})$ is
continuous on a subset $\mathcal{A}$ of $\S(\H_{ABC})$ then
continuity of $I(A\!:\!C|B)_{\omega}$ on $\mathcal{A}$ follows from
Theorem \ref{main}A and formulae (\ref{cmi-d+}) and (\ref{cmi-d++})
correspondingly.\smallskip

If either $H(\omega_{AB})$ or $H(\omega_{BC})$ is continuous on a
subset $\mathcal{A}$ then Theorem \ref{main}A implies continuity on
$\mathcal{A}$ of the first term in (\ref{cmi-d++}) and in
(\ref{cmi-d+}) correspondingly. By Lemma \ref{vsl} the continuity of
$I(A\!:\!C|B)_{\omega}$ on $\mathcal{A}$ follows from its lower
semicontinuity (Theorem \ref{cmi-th}) and from the lower
semicontinuity of $I(A\!:\!B)_{\omega}$ and of
$I(B\!:\!C)_{\omega}$.\smallskip

If either $H(\omega_{B})$ or $H(\omega_{ABC})$ is continuous on a
subset $\mathcal{A}$ then the continuity of the difference
$\,I(A\!:\!C|B)_{\omega}-I(A\!:\!C)_{\omega}$ on $\mathcal{A}$
follows from Theorem \ref{main}A and formulae (\ref{cmi-d+++}) and
(\ref{cmi-d++++}) correspondingly (in the second case Lemma
\ref{p-lemma} and the equality $H(\omega_{ABC})=H(\omega_{D})$ are
used).
\smallskip

B) We will use the inequality
\begin{equation}\label{sp-ineq+}
 \lambda I(A\!:\!C|B)_{\rho}+(1-\lambda)I(A\!:\!C|B)_{\sigma}\leq
 I(A\!:\!C|B)_{\lambda\rho+(1-\lambda)\sigma}+2h_2(\lambda),
\end{equation}
where $h_2(\lambda)$ is the binary entropy, valid for any operators
$\rho,\sigma\in\T_+(\H_{ABC})$ such that
$\max\{\Tr\rho,\Tr\sigma\}\leq 1$. If all the marginal entropies of
the operators $\rho$ and $\sigma$ are finite then (\ref{sp-ineq+})
follows from (\ref{cmi-d}) and (\ref{w-k-ineq}). In general case
(\ref{sp-ineq+}) can be proved by approximating the operators $\rho$
and $\sigma$ by the double sequences of operators
$$
\rho_{kl} =Q_{kl}\shs\rho\shs Q_{kl}\quad \textrm{and} \quad
\sigma_{kl} =Q_{kl}\shs\sigma\shs Q_{kl},\quad Q_{kl}=P^A_k\otimes
P^B_l\otimes P^C_k,
$$
where $\{P^k_{A}\}\subset\B(\H_A)$, $\{P^l_{B}\}\subset\B(\H_B)$,
$\{P^k_C\}\subset\B(\H_{C})$ are sequences of finite rank projectors
strongly converging to the identity operators $I_A$,$I_B$,$I_C$.
Since (\ref{sp-ineq+}) holds for the operators $\rho_{kl}$ and
$\sigma_{kl}$ for all $k$ and $l$, validity of (\ref{sp-ineq+}) for
the operators $\rho$ and $\sigma$ can be shown by using property
(\ref{t-ext-p+}).

Inequality (\ref{sp-ineq+}), nonnegativity and monotonicity of the
conditional mutual information under local operations show that
$$
\lambda_k I(A\!:\!C|B)_{\omega^k}\leq
I(A\!:\!C|B)_{\Phi_A^k\otimes\id_B\otimes\Phi_C^k(\omega^0)}+2h_2(\lambda'_k)\leq
I(A\!:\!C|B)_{\omega^0}+2h_2(\lambda'_k),
$$
where
$\lambda'_k=\lambda_k[\Tr\shs\Phi_A^k\otimes\id_B\otimes\Phi_C^k(\omega^0)]^{-1}
\geq\lambda_k$. This inequality and the lower semicontinuity of the
conditional mutual information (Theorem \ref{cmi-th}) imply the
required limit relation. $\square$
\medskip

Corollary \ref{cmi-th-c-1} shows, in particular, that the
conditional mutual information $I(A\!:\!C|B)_{\omega}$ is continuous
on the set $\,\S(\H_{ABC})$ if either $A$ or $C$ is a
finite-dimensional system. Proposition \ref{FA-cont} and Remark
\ref{cmi-u-b} give the continuity bound for
$\,I(A\!:\!C|B)_{\omega}$ in this case.
\smallskip
\begin{corollary}\label{cmi-pr-c}
\emph{If one of the systems $A$ and $C$, say $A$, is
finite-dimensional then $\,I(A\!:\!C|B)_{\omega}$ is a continuous
bounded function on the set $\,\S(\H_{ABC})$ and
$$
|I(A\!:\!C|B)_{\omega^1}-I(A\!:\!C|B)_{\omega^2}|\leq
2\varepsilon \log\dim\H_A+4(1+\varepsilon)h_2\!\left(\frac{\varepsilon}{1+\varepsilon}\right)
$$
for any $\,\omega^1,\omega^2\in\S(\H_{ABC})$  such that
$\;\varepsilon=\frac{1}{2}\|\shs\omega^1-\omega^2\|_1<1\,$, where
$\,h_2(\cdot)\,$ is the binary entropy. If the system $B$
is finite-dimensional then the difference
$I(A\!:\!C|B)_{\omega}-I(A\!:\!C)_{\omega}$ is uniformly continuous
and bounded on the set $\,\S(\H_{ABC})$.}
\end{corollary}\medskip

We will use the following analog of Corollary
\ref{main-c-n+}.\smallskip
\begin{corollary}\label{main-c-n++} \emph{Let  $\,X$, $Y$ and $Z$ be disjoint subsystems of $A_{1}...A_{n}$ and $R=A_{1}...A_{n}\setminus XYZ$.
The function $\,\omega_{A_{1}...A_{n}}\mapsto I(X\!:\!Z|Y)_{\omega}$
satisfies $\mathfrak{F}$\nobreakdash-\hspace{0pt}exten-sion
condition (\ref{t-ext-p}) for any state $\omega_{A_{1}...A_{n}}$
such that
\begin{equation}\label{s-f-cond}
\min\{H(\omega_{X}),H(\omega_{Y}),H(\omega_{Z}),H(\omega_{XY}),H(\omega_{YZ}),H(\omega_{XYZ})\}<+\infty.
\end{equation}
For an arbitrary state $\,\omega\in\S(\H_{A_{1}...A_{n}})$ the
following weaker property is valid:
\begin{equation}\label{t-ext-p++}
I(X\!:\!Z|Y)_{\omega}=\lim_{k\rightarrow\infty}\lim_{l\rightarrow\infty}I(X\!:\!Z|Y)_{\omega^{klt}},
\end{equation}
for $\,t=k$ and for $\,t=l$, where
$$
\omega^{klt}=\lambda_{klt}^{-1}Q_{klt}\shs\omega\shs Q_{klt},\quad
 Q_{klt}= P^k_{X}\otimes
P^l_{Y}\otimes P^k_{Z}\otimes P^t_{R},\;\,\lambda_{klt}=\Tr
Q_{klt}\shs\omega,
$$
$\{P^k_{X}\}_k\subset\B(\H_X)$, $\{P^l_Y\}_l\subset\B(\H_{Y})$,
$\{P^k_Z\}_k\subset\B(\H_Z)$, $\{P^t_R\}_t\subset\B(\H_R)$,  are
sequences of projectors strongly converging to the identity
operators $I_X$,$I_Y$,$I_Z$,$I_R$ such that
$\,\min\{\mathrm{rank}P^k_{X},\mathrm{rank}P^k_{Z}\}<+\infty\,$ for
all $\,k$.}
\end{corollary}
\smallskip

\emph{Proof.} We will consider the cases $\,t=k$ and $\,t=l$
simultaneously.  Since
\begin{equation}\label{l-u-ineq}
\lambda_{klt}\omega^{klt}_{XYZ}\leq \tilde{\omega}^{kl}_{XYZ}\doteq
P^k_{X}\otimes P^l_{Y}\otimes P^k_{Z}\cdot \omega_{XYZ}\cdot
P^k_{X}\otimes P^l_{Y}\otimes P^k_{Z},
\end{equation}
inequality (\ref{sp-ineq+}) and nonnegativity of $\,I(X\!:\!Z|Y)$
imply
\begin{equation}\label{cmi-t-b}
\lambda_{klt}I(X\!:\!Z|Y)_{\omega^{klt}}\leq
I(X\!:\!Z|Y)_{\tilde{\omega}^{kl}}+2h_2\left(\lambda_{klt}[\Tr\tilde{\omega}^{kl}]^{-1}\right).
\end{equation}
Since Theorem \ref{cmi-th} shows that
$\lim_{k\rightarrow\infty}I(X\!:\!Z|Y)_{\tilde{\omega}^{kk}}=I(X\!:\!Z|Y)_{\omega}$
for any state $\omega_{A_{1}...A_{n}}$ satisfying (\ref{s-f-cond}),
the  first assertion  of the corollary follows from (\ref{cmi-t-b})
and the lower semicontinuity of the function
$\omega_{A_{1}...A_{n}}\mapsto I(X\!:\!Z|Y)_{\omega}$.\smallskip

To prove property (\ref{t-ext-p++}) denote
$\omega^{k*}=\lim_{l\rightarrow\infty}\omega^{klt}$ and
$\tilde{\omega}^{k*}=\lim_{l\rightarrow\infty}=\tilde{\omega}^{kl}$.
By the condition
$\,\min\{\mathrm{rank}P^k_{X},\mathrm{rank}P^k_{Z}\}<+\infty\,$
Corollary \ref{cmi-pr-c} implies
$$
\lim_{l\rightarrow\infty}I(X\!:\!Z|Y)_{\omega^{klt}}=I(X\!:\!Z|Y)_{\omega^{k*}}.
$$
The lower semicontinuity of the conditional mutual information and
its monotonicity under local operations show that
$$
\lim_{k\rightarrow\infty}I(X\!:\!Z|Y)_{\tilde{\omega}^{k*}}=I(X\!:\!Z|Y)_{\omega}.
$$
Since inequalities (\ref{l-u-ineq}) and (\ref{cmi-t-b}) hold with
$l\!=\!*$ if we set $P_Y^*=I_Y$ and
$\lambda_{k*}=\lim_{l\rightarrow\infty}\lambda_{klt}$, the above
limit relation  and the lower semicontinuity of the function
$\omega_{A_{1}...A_{n}}\mapsto I(X\!:\!Z|Y)_{\omega}$ imply
$\lim_{k\rightarrow\infty}I(X\!:\!Z|Y)_{\omega^{k*}}=I(X\!:\!Z|Y)_{\omega}$.
$\square$
\smallskip

The lower semicontinuity of $\,I(A\!:\!C|B)_{\omega}\,$ implies the
following observations concerning quantum mutual information.
\smallskip

\begin{corollary}\label{cmi-th-c-2} A) \emph{The function $\,\omega\mapsto
\left[\shs I(A\!:\!BC)_{\omega}-I(A\!:\!B)_{\omega}\right]$ is lower
semicontinuous on the set
$\,\{\shs\omega\in\T_{+}(\H_{ABC})\,|\,I(A\!:\!B)_{\omega}<+\infty\shs\}$.}\smallskip

B) \emph{Local continuity of the function $\,\omega_{ABC}\mapsto
I(A\!:\!BC)_{\omega}$ implies local continuity of the function
$\,\omega_{ABC}\mapsto I(A\!:\!B)_{\omega}$.}
\smallskip

C) \emph{Local continuity of the function $\,\omega_{ABC}\mapsto
H(\omega_{BC})$  implies local continuity of function
$\,\omega_{ABC}\mapsto I(A\!:\!B)_{\omega}$.}
\end{corollary}\smallskip

\emph{Proof.} Assertion A follows from Theorem \ref{cmi-th} and
formula (\ref{cmi-d+}). By Lemma \ref{vsl} assertion B follows from
A and the lower semicontinuity of quantum mutual information.
Assertion C follows from B and Theorem \ref{main}A.
$\square$.\smallskip

\subsection{Multipartite system}

The conditional mutual information of a state $\,\omega_{A_1 \ldots
A_nB}\,$ of a finite-dimensional multipartite system $A_1 \ldots
A_nB$ is defined as follows
\begin{equation}\label{cmi-mpd}
\begin{array}{cl}
     I(A_1\!:\ldots:\!A_n|B)_{\omega}&\doteq\displaystyle
    \sum_{i=1}^n H(A_i|B)_{\omega}-H(A_1 \ldots A_n|B)_{\omega}\\&
    =\displaystyle\sum_{i=1}^n H(\omega_{A_iB})-H(\omega_{A_1 \ldots A_nB})-(n-1)H(\omega_{B}).
\end{array}
\end{equation}
The analogs of the above-mentioned properties C1-C4 of the
tripartite conditional mutual information can be proved for
$I(A_1\!:\ldots:\!A_n|B)_{\omega}$ by using the representation (cf.
\cite{3H+})
\begin{equation}\label{cmi-mpd+}
\begin{array}{rl}
     I(A_1\!:\ldots:\!A_n|B)_{\omega}=I(A_2\!:\!A_1|B)_{\omega}\!\!& +\; I(A_3\!:\!A_1A_2|B)_{\omega}+...\\\\&+\;
     I(A_n\!:\!A_1...A_{n-1}|B)_{\omega}.
\end{array}
\end{equation}
and its modifications obtained by permuting indexes in the right
hand side.\smallskip

Formula (\ref{c-dif}) in this case has the form
\begin{equation}\label{cmi-dif}
I(A_1\!:\ldots:\!A_n|B)_{\omega}-I(A_1\!:\ldots:\!A_n)_{\omega}=I(A_1...A_n\!:\!B)_{\omega}-\sum_{i=1}^n
I(A_i\!:\!B)_{\omega}.
\end{equation}

Since (\ref{cmi-mpd+}) is valid with arbitrarily permuted indexes
$\,1,...,n\,$ in the right hand side,  Remark \ref{cmi-u-b} implies
that $I(A_1\!:\ldots:\!A_n|B)_{\omega}$ is upper bounded by the
value
\begin{equation}\label{cmi-n-u-b}
2\min_{1\leq j\leq n}\sum_{i\neq
j}\min\left\{H(\omega_{A_{i}}),H(\omega_{A_{i}B})\right\}.
\end{equation}

By using representation (\ref{cmi-mpd+}) and the extended
conditional mutual information described in Theorem \ref{cmi-th}
one can define $\,I(A_1\!:\ldots:\!A_n|B)_{\omega}$ for any state of
an infinite-dimensional system $A_1...A_nB$.
\smallskip

\begin{property}\label{cmi-pr}
A) \emph{The quantity $\,I(A_1\!:\!...\!:\!A_n|B)_{\omega}$ defined
in (\ref{cmi-mpd}) has a lower semicontinuous extension to the set
$\S(\H_{A_1 \ldots A_nB})$ possessing the analogs of above-stated
properties C1-C4 of conditional mutual information and upper bounded
by value (\ref{cmi-n-u-b}). This extension can be defined by formula
(\ref{cmi-mpd+}) in which each summand $I(X\!:\!Y|B)_{\omega}$
coincides with the function $I_{\mathrm{e}}(X\!:\!Y|B)_{\omega}$
described in Theorem \ref{cmi-th}.}\smallskip

B) \emph{This extension (also denoted
$\,I(A_1\!:\!...\!:\!A_n|B)_{\omega}$) satisfies
$\mathfrak{F}$-extension condition (\ref{t-ext-p}) for any state
$\omega_{A_1...A_nB}$ such that either (\ref{cmi-n-u-b}) or
$H(\omega_{B})$ is finite. For an arbitrary state
$\omega\in\S(\H_{A_1...A_nB})$ the following weaker property is
valid:
\begin{equation}\label{cmi-n-lr}
I(A_1\!:\ldots:\!A_n|B)_{\omega}=\lim_{k\rightarrow\infty}\lim_{l\rightarrow\infty}I(A_1\!:\ldots:\!A_n|B)_{\omega^{kl}},
\end{equation}
where
$$
\omega^{kl}=\lambda_{kl}^{-1}Q_{kl}\shs\omega\shs Q_{kl},\quad
Q_{kl}= P^k_{A_1}\otimes\ldots\otimes P^k_{A_n}\otimes P^l_{B},\;\,
\lambda_{kl}=\Tr Q_{kl}\shs\omega,
$$
$\{P^k_{A_i}\}_k\subset\B(\H_{A_i})$, $i=\overline{1,n}$, and
$\{P^l_B\}_l\subset\B(\H_{B})$ are sequences of projectors strongly
converging to the identity operators $I_{A_i}$ and $I_B$ such
that\break $\min_{1\leq j \leq n}\sum_{i\neq
j}\mathrm{rank}P^k_{A_i}<+\infty$ for all $\,k$.}\smallskip

C) \emph{Representation (\ref{cmi-mpd+}) is valid for
$\,I(A_1\!:\ldots:\!A_n|B)_{\omega}$ with arbitrarily permuted
indexes $\,1,...,n\,$ in the right hand side (provided each summand
$I(X\!:\!Y|Z)_{\omega}$ coincides with
$I_{\mathrm{e}}(X\!:\!Y|Z)_{\omega}$).}\smallskip

D) \emph{Local continuity of $\;n-1\,$ marginal entropies
$\,H(\omega_{X_{i_1}}),\ldots,H(\omega_{X_{i_{n-1}}})$, where
$X_{i_k}$ is either $A_{i_k}$ or $A_{i_k}B$, implies local
continuity of $\,I(A_1\!:...:\!A_n|B)_{\omega}$. Local continuity of
the marginal entropy $H(\omega_{B})$ implies local continuity of the
difference
$\,I(A_1\!:\!...\!:\!A_n|B)_{\omega}-I(A_1\!:\!...\!:\!A_n)_{\omega}$
having representation (\ref{cmi-dif}).}\smallskip

E) \emph{If $\;n-1\,$ subsystems, say $A_1,...,A_{n-1}$, are
finite-dimensional then $I(A_1\!:\!...\!:\!A_n|B)_{\omega}$ is a
continuous bounded function on the set $\,\S(\H_{A_1...A_nB})$ and
$$
|I(A_1\!:\!...\!:\!A_n|B)_{\omega^1}-I(A_1\!:\!...\!:\!A_n|B)_{\omega^2}|\leq 2\varepsilon
C+2n(1+\varepsilon)h_2\!\left(\frac{\varepsilon}{1+\varepsilon}\right)
$$
for any $\,\omega^1,\omega^2\in\S(\H_{A_1...A_nB})$ such that
$\;\varepsilon=\frac{1}{2}\|\shs\omega^1-\omega^2\|_1<1\,$, where
$\,h_2(\cdot)\,$ is the binary entropy and $\,C=\log\dim\H_{A_1...A_{n-1}}$ .}

\end{property}\medskip

\emph{Proof.} Assertion A  follows from Theorem \ref{cmi-th}, Remark
\ref{cmi-u-b} and assertion C proved below.\smallskip

Property (\ref{cmi-n-lr}) in the case
$\sum_{i=2}^n\mathrm{rank}P^k_{A_i}<+\infty$ for all $k$ follows
from the second assertion of Corollary \ref{main-c-n++}, i.e. limit
relation (\ref{t-ext-p++}), applied to each term in the right hand
side of (\ref{cmi-mpd+}). This property and (\ref{t-ext-p++}) make
possible to prove assertion C via approximation by noting that it is
valid for any state $\omega_{A_1...A_nB}$ such that
$\,\mathrm{rank}\shs\omega_{A_{i}}<+\infty$ for all $i$ and
$\,\mathrm{rank}\shs\omega_{B}<+\infty$.\smallskip

Now, by using assertion C, assertion B can be proved by applying
Corollary \ref{main-c-n++} to each term in the right hand side of
(\ref{cmi-mpd+}) with appropriately permuted indexes $\,1,...,n\,$
(and by using Lemmas  \ref{t-ext-r} and \ref{t-ext-pr}).

The first part of assertion D follows from Corollary
\ref{cmi-th-c-1}A applied to each term in the right hand side of
(\ref{cmi-mpd+}) with appropriately permuted indexes $\,1,...,n\,$.
The second part of assertion D follows from Proposition
\ref{main-c-2}.

Assertion E follows from Proposition \ref{FA-cont} and upper bound
(\ref{cmi-n-u-b}). $\square$

\section{Other information measures in multipartite systems}

\subsection{Quantum version of interaction information}

Consider the following characteristic of a state
$\omega_{A_1...A_n}$ of  $n$-partite quantum system
\begin{equation}\label{tee-d+}
\begin{array}{rl}
    I_n(\omega_{A_1...A_n})\!\!& \displaystyle =
    \sum_iH(\omega_{A_i})-\sum_{i<j}H(\omega_{A_iA_j})\\ & \displaystyle +\sum_{i<j<k}H(\omega_{A_iA_jA_k})-...-(-1)^n H(\omega_{A_1...A_n})
\end{array}
\end{equation}
which (up to the sign) can be considered as a noncommutative version
of the interaction information of a  $n$-partite classical system
\cite{J&B,McGill}.

Note that $I_1(\omega_A)$ is the von Neumann entropy of a
one-partite state $\omega_A$, $I_2(\omega_{AB})$ is the quantum
mutual information of a bipartite state $\omega_{AB}$, while
\begin{equation*}
\begin{array}{rl}
    I_3(\omega_{ABC})\!\!& =
    H(\omega_{A})+H(\omega_{B})+H(\omega_{C})\\\\ &-\,
    H(\omega_{AB})-H(\omega_{AC})-H(\omega_{BC})+H(\omega_{ABC})
\end{array}
\end{equation*}
is the topological entanglement entropy of a tripartite state
$\omega_{ABC}$ typically denoted $H_{\mathrm{topo}}(\omega_{ABC})$
(or $S_{\mathrm{topo}}(\omega_{ABC})$) \cite{Kitaev}. Despite
possible negativity the last quantity is also used as a special
measure of quantum correlations \cite{C&H,H&Co}.

An interesting feature of the linear combinations of marginal
entropies in (\ref{tee-d+}) consists in the fact that \textit{for
any} $n$ finiteness of \textit{only one} marginal entropy
$H(\omega_{A_i})$ "eliminates" all possible uncertainties
$"\infty-\infty"$ in (\ref{tee-d+}), while continuity of
$H(\omega_{A_i})$ guarantees continuity of
$I_n(\omega_{A_1...A_n})$.\smallskip
\begin{property}\label{tee-pr}
A) \emph{The quantity  $I_n(\omega_{A_1...A_n})$ defined in
(\ref{tee-d+}) has finite $\mathfrak{F}$-extension to the set
$\,\{\shs\omega_{A_1...A_n}\,|\,\exists\shs i:
H(\omega_{A_i})<+\infty\}\,$ such that} \footnote{We denote this
$\mathfrak{F}$-extension of $I_n(\omega_{A_1...A_n})$ by the same
symbol.}
\begin{equation*}
    |I_n(\omega_{A_1...A_n})|\leq 2^{n-1}\min\{H(\omega_{A_1}),\ldots,
    H(\omega_{A_n})\}.
\end{equation*}

B) \emph{If $H(\omega_{A_i})<+\infty$ then this
$\mathfrak{F}$-extension is given by the formula
\begin{equation}\label{tee-d++}
\begin{array}{c}
    I_n(\omega_{A_1...A_n})\displaystyle =H(\omega_{A_i})-
    \sum_jH_{\mathrm{e}}(A_i|A_j)_{\omega}+\sum_{j<k}H_{\mathrm{e}}(A_i|A_kA_j)_{\omega}\quad\qquad\\
    \displaystyle -\sum_{j<k<l}H_{\mathrm{e}}(A_i|A_jA_kA_l)_{\omega}
    +...+(-1)^{n-1} H_{\mathrm{e}}(A_i|A_1...A_{i-1}A_{i+1}...A_n)_{\omega},
\end{array}
\end{equation}
where all the indexes $k,j,l,...$ in each sum run over the set
$\{1,...,n\}\setminus\{i\}$ and $H_{\mathrm{e}}(X|Y)_{\omega}$ is
the extended quantum conditional entropy defined by
(\ref{c-ent-def++}).}\smallskip

C) \emph{Local continuity of \textbf{one} of the marginal entropies
$\,H(\omega_{A_1}),..., H(\omega_{A_n})$ implies local continuity of
$\,I_n(\omega_{A_1...A_n})$.}
\end{property}\medskip

\emph{Proof.} A) Assume that $H(\omega_{A_n})$ is finite. Consider
the quantity
$$
\begin{array}{c}
\displaystyle F(\omega_{A_1...A_n})=H(\omega_{A_n})+I_{n-1}(\omega_{A_1...A_{n-1}})-I_n(\omega_{A_1...A_n})=\sum_{i<n}H(\omega_{A_iA_n})\\\\
\displaystyle-\sum_{i<j<n}H(\omega_{A_iA_jA_n})
+\sum_{i<j<k<n}H(\omega_{A_iA_jA_kA_n})-...-(-1)^{n-1}
H(\omega_{A_1...A_n}).
\end{array}
$$
In terms of Corollary \ref{main-c-3} we have $F_{\cdot\backslash
A_{n}}=I_{n-1}$ and hence
\begin{equation}\label{tee-rep}
I_n(\omega_{A_1...A_n})=H(\omega_{A_n})+[F_{\cdot\backslash
A_{n}}-F\shs](\omega_{A_1...A_n}).
\end{equation}
So, Corollary \ref{main-c-3} implies the existence of
$\mathfrak{F}$-extension of the quantity  $I_n(\omega_{A_1...A_n})$
defined in (\ref{tee-d+}) to the set
$\,\{\shs\omega_{A_1...A_n}\,|\,H(\omega_{A_n})<+\infty\}\,$
determined by formula (\ref{tee-d++}) with $\,i=n\,$ such that
\begin{equation*}
    |I_n(\omega_{A_1...A_n})|\leq 2^{n-1}H(\omega_{A_n}).
\end{equation*}
To complete the proof of A it suffices to note that the
$\mathfrak{F}$-extensions of $I_n$ to the sets
$\,\{\shs\omega_{A_1...A_n}\,|\,H(\omega_{A_i})<+\infty\}\,$ and
$\,\{\shs\omega_{A_1...A_n}\,|\,H(\omega_{A_j})<+\infty\}\,$ agree
with each other by Lemma \ref{t-ext-r}.

B) If the function $\,\omega_{A_1...A_n}\mapsto H(\omega_{A_n})$ is
continuous on a set $\mathcal{A}$ then Corollary \ref{main-c-3} and
(\ref{tee-rep}) imply continuity of the function
$\,\omega_{A_1...A_n}\mapsto I_n(\omega_{A_1...A_n}) $ on
$\mathcal{A}$. $\square$ \medskip

\begin{remark}\label{tee-d++r}
For given $n$ general formula (\ref{tee-d++}) can be simplified. For
example, the $\mathfrak{F}$-extension of the topological
entanglement entropy $I_3(\omega_{A_1A_2A_3})$ to the set
$\{\shs\omega_{A_1A_2A_3}\,|\,H(\omega_{A_1})<+\infty\shs\}$ can be
expressed as follows
$$
I_{3}(\omega_{A_1A_2A_3})=I(A_1\!:\!A_2)_{\omega}-I(A_1\!:\!A_2|A_3)_{\omega},
$$
where $I(A_1\!:\!A_2|A_3)_{\omega}$ is the extended conditional
mutual information described in Theorem \ref{cmi-th}.
\end{remark}\smallskip

Propositions \ref{FA-cont} and \ref{tee-pr} imply the following
result. \smallskip

\begin{corollary}\label{ent-ent-c}
\emph{If one of the systems $A_1...A_n$, say $A_i$, is
finite-dimensional then $I_n(\omega_{A_1...A_n})$ is a continuous
bounded function on the set $\,\S(\H_{A_1...A_n})$ and
$$
|I_n(\omega^1)-I_n(\omega^2)|\leq 2^n\varepsilon
\log\dim\H_{A_i}+(2^n-1)(1+\varepsilon)h_2\!\left(\frac{\varepsilon}{1+\varepsilon}\right)
$$
for any $\,\omega^1,\omega^2\in\S(\H_{A_1...A_n})$  such that
$\;\varepsilon=\frac{1}{2}\|\shs\omega^1-\omega^2\|_1<1\,$, where
$\,h_2(\cdot)\,$ is the binary entropy.}
\end{corollary}\medskip

\subsection{Secrecy monotone $S_n$ (unconditional and conditional)}

Along with the quantum mutual information
$I(A_1\!:\!\ldots\!:\!A_n)$ the secrecy monotone
\begin{equation*}
     S_n(A_1\!:\ldots:\!A_n)_{\omega}=\sum_{i=1}^n H(\omega_{A_1...A_{i-1}A_{i+1}...A_{n}})-(n-1)H(\omega_{A_1\ldots A_n})
\end{equation*}
is proposed in \cite{SecMon} as a measure of quantum correlations of
a state $\omega_{A_1\ldots A_n}$ of a finite-dimensional $n$-partite
system.\footnote{The same quantity is independently proposed and
analyzed in \cite{Kum}, where it is called "operational quantum
mutual information".} Note that
$S_2(A_1\!:\!A_2)_{\omega}=I(A_1\!:\!A_2)_{\omega}$, so, the
quantity $S_n$ can be considered as a particular $n$-partite
generalization of the bipartite quantum mutual information. It can
be expressed as follows\smallskip
\begin{equation}\label{sec-m+}
\begin{array}{c}
     S_n(A_1\!:\ldots:\!A_n)_{\omega}=I(A_1\!:\!A_2...A_n)_{\omega} +\; I(A_2\!:\!A_3...A_n|A_1)_{\omega}\\\\+
     I(A_3\!:\!A_4...A_{n}|A_1A_2)_{\omega}+...+I(A_{n-1}\!:\!A_{n}|A_1...A_{n-2})_{\omega}.
\end{array}
\end{equation}

In applications the conditional version of $S_n$, i.e. the
characteristic
\begin{equation}\label{sec-m-c}
\begin{array}{rl}
     S_n(A_1\!:\ldots:\!A_n|B)_{\omega}\!&\displaystyle =\,\sum_{i=1}^n H(\omega_{A_1...A_{i-1}A_{i+1}...A_{n}B})\\\\&
     \displaystyle -\;(n-1)H(\omega_{A_1\ldots A_nB})-H(\omega_{B}).
\end{array}
\end{equation}
of a state $\,\omega_{A_1...A_nB}\,$ is also used \cite{3H+}. So, we
will consider infinite-dimensional generalization of the quantity
$\,S_n(A_1\!:\ldots:\!A_n|B)$, having in mind that
$\,S_n(A_1\!:\ldots:\!A_n)$ is a partial case of
$\,S_n(A_1\!:\ldots:\!A_n|B)$ for trivial $B$.

The conditional secrecy monotone $S_n$ can be represented by
conditioning (\ref{sec-m+}) as follows
\begin{equation}\label{sec-m-c+}
\begin{array}{c}
     S_n(A_1\!:\ldots:\!A_n|B)_{\omega}=I(A_1\!:\!A_2...A_n|B)_{\omega} +\; I(A_2\!:\!A_3...A_n|A_1B)_{\omega}\\\\+
     I(A_3\!:\!A_4...A_{n}|A_1A_2B)_{\omega}+...+I(A_{n-1}\!:\!A_{n}|A_1...A_{n-2}B)_{\omega}.
\end{array}
\end{equation}
Basic properties of the conditional mutual information show that the
quantity  $S_n(A_1\!:\ldots:\!A_n|B)_{\omega}$ is nonnegative and
does not increase under local operations $\Phi_{A_1}:A_1\rightarrow
A_1,...,\Phi_{A_n}:A_n\rightarrow A_n$, i.e.
\begin{equation}\label{s-n-m}
S_n(A_1\!:\ldots:\!A_n|B)_{\omega}\geq
S_n(A_1\!:\ldots:\!A_n|B)_{\Phi_{A_1}\otimes...\otimes\Phi_{A_n}\otimes\id_{B}(\omega)}.
\end{equation}

Formula (\ref{c-dif}) in this case has the form
\begin{equation}\label{s-n-dif}
\begin{array}{c}
S_n(A_1\!:\ldots:\!A_n|B)_{\omega}-S_n(A_1\!:\ldots:\!A_n)_{\omega}=\displaystyle(n-1)I(A_1...A_{n}\!:\!B)_{\omega}\\\displaystyle
-\sum_{i=1}^n I(A_1...A_{i-1}A_{i+1}...A_{n}\!:\!B)_{\omega}.
\end{array}
\end{equation}

Since (\ref{sec-m-c+}) is valid with arbitrarily permuted indexes
$\,1,...,n\,$ in the right hand side, Remark \ref{cmi-u-b} implies
that $S_n(A_1\!:\ldots:\!A_n|B)_{\omega}$ is upper bounded by the
value
\begin{equation}\label{s-n-u-b}
2\min_{1\leq j \leq n}\sum_{i\neq j}H(\omega_{A_{i}}).
\end{equation}

By using representation (\ref{sec-m-c+}) and the extended
conditional mutual information described in Sect.6. one can define
$\,S_n(A_1\!:\ldots:\!A_n|B)_{\omega}$ for any state of an
infinite-dimensional system $A_1...A_nB$.
\smallskip

\begin{property}\label{sec-m-pr}
A) \emph{The quantity $\,S_n(A_1\!:\ldots:\!A_n|B)_{\omega}$ defined
in (\ref{sec-m-c}) has a lower semicontinuous nonnegative extension
to the set $\S(\H_{A_1 \ldots A_nB})$ possessing property
(\ref{s-n-m}) and upper bounded by value (\ref{s-n-u-b}). This
extension can be defined by formula (\ref{sec-m-c+}) in which each
summand $I(X\!:\!Y|Z)_{\omega}$ coincides with the function
$I_{\mathrm{e}}(X\!:\!Y|Z)_{\omega}$ described in Theorem
\ref{cmi-th}.}\smallskip

B) \emph{This extension (also denoted
$\,S_n(A_1\!:\ldots:\!A_n|B)_{\omega}$) satisfies
$\mathfrak{F}$-extension condition (\ref{t-ext-p}) for any state
$\omega_{A_1 \ldots A_nB}$ such that (\ref{s-n-u-b}) is finite. For
an arbitrary state $\omega\in\S(\H_{A_1 \ldots A_nB})$ the following
weaker property is valid:
\begin{equation}\label{s-n-lr}
S_n(A_1\!:\ldots:\!A_n|B)_{\omega}=
\lim_{k_n\rightarrow\infty}...\lim_{k_1\rightarrow\infty}\lim_{l\rightarrow\infty}S_n(A_1\!:\ldots:\!A_n|B)_{\omega^{k_1...k_nl}},
\end{equation}
where
$$
\omega^{k_1...k_nl}=\lambda^{-1}Q\shs\omega\shs Q,\quad Q=
P^{k_1}_{A_1}\otimes\ldots\otimes P^{k_n}_{A_n}\otimes
P^l_{B},\quad\lambda=\Tr Q\shs\omega,
$$
$\{P^{k_i}_{A_i}\}\subset\B(\H_{A_i})$, $i=\overline{1,n}$, and
$\{P^l_B\}\subset\B(\H_{B})$ are any sequences of finite rank
projectors strongly converging to the identity operators $I_{A_i}$
and $I_B$.}\footnote{The limits over $k_1,...,k_n$ in (\ref{s-n-lr})
can be taken in arbitrary order. This follows from assertion C of
this proposition. The projectors in the sequence $\{P^l_B\}$ may be
arbitrary.}\smallskip

C) \emph{Representation (\ref{sec-m-c+}) is valid for
$\,S_n(A_1\!:\ldots:\!A_n|B)_{\omega}$ with arbitrarily permuted
indexes $\,1,...,n\,$ in the right hand side (provided each summand
$I(X\!:\!Y|Z)_{\omega}$ coincides with
$I_{\mathrm{e}}(X\!:\!Y|Z)_{\omega}$).}\smallskip

D) \emph{Local continuity of $\;n-1\,$ marginal entropies
$\,H(\omega_{A_{i_1}}),\ldots,H(\omega_{A_{i_{n-1}}})$ implies local
continuity of $\,S_n(A_1\!:\!...\!:\!A_n|B)_{\omega}$. Local
continuity of $H(\omega_{B})$ implies local continuity of the
difference
$S_n(A_1\!:...:\!A_n|B)_{\omega}-S_n(A_1\!:\!...\!:\!A_n)_{\omega}$
having representation (\ref{s-n-dif}).}\smallskip

E) \emph{If $\,n-1\,$ subsystems, say $A_1,...,A_{n-1}$, are
finite-dimensional then $S_n(A_1\!:...:\!A_n|B)_{\omega}$ is a
continuous bounded function on the set $\,\S(\H_{A_1...A_nB})$ and
$$
|S_n(A_1\!:...:\!A_n|B)_{\omega^1}-S_n(A_1\!:...:\!A_n|B)_{\omega^2}|\leq
2\varepsilon
C+2n(1+\varepsilon)h_2\!\left(\frac{\varepsilon}{1+\varepsilon}\right)
$$
for any $\,\omega^1,\omega^2\in\S(\H_{A_1...A_nB})$ such that
$\;\varepsilon=\frac{1}{2}\|\shs\omega^1-\omega^2\|_1<1\,$, where
$\,h_2(\cdot)\,$ is the binary entropy and $\,C=\log\dim\H_{A_1...A_{n-1}}$.}
\end{property}\medskip

\emph{Proof.}  Assertion A follows from Theorem \ref{cmi-th}, Remark
\ref{cmi-u-b} and assertion C proved below.\smallskip

Property (\ref{s-n-lr}) follows from Corollary\ref{cmi-pr-c} and the
second assertion of Corollary \ref{main-c-n++}, i.e. limit relation
(\ref{t-ext-p++}), applied to each term in the right hand side of
(\ref{sec-m-c+}). This property and (\ref{t-ext-p++}) make possible
to prove assertion C via approximation by noting that it is valid
for any state $\omega_{A_1...A_nB}$ such that
$\,\mathrm{rank}\shs\omega_{A_{i}}<+\infty$ for all $i$ and
$\,\mathrm{rank}\shs\omega_{B}<+\infty$.\smallskip

Now, by using assertion C, the first part of assertion B can be
proved by applying Corollary \ref{main-c-n++} to each term in the
right hand side of (\ref{sec-m-c+}) with appropriately permuted
indexes $\,1,...,n\,$ (and by using Lemmas \ref{t-ext-r} and
\ref{t-ext-pr}).

The first part of assertion D follows from Corollary
\ref{cmi-th-c-1}A applied to each term in the right hand side of
(\ref{sec-m-c+}) with appropriately permuted indexes $\,1,...,n\,$.
The second part of assertion D follows from Proposition
\ref{main-c-2}.

Assertion E follows from Proposition \ref{FA-cont} and upper bound
(\ref{s-n-u-b}). $\square$
\medskip

Other properties (monotonicity under local conditioning, additivity,
see \cite{3H+}) of the extension
$S_n(A_1\!:\ldots:\!A_n|B)_{\omega}$ defined in Proposition
\ref{sec-m-pr} can be derived from their validity in the
finite-dimensional settings by using approximation property
(\ref{s-n-lr}).

\subsection{The gap $I(A_1A'_1\!:\!...\!:\!A_nA'_n)-I(A'_1\!:\!...\!:\!A'_n)$ and the Wilde inequality}

In construction of entanglement measures in a multipartite
finite-dimensional quantum system $A_1...A_n$ the difference
\begin{equation}\label{inf-g-d}
\Delta I(\omega_{A_1A'_1...A_nA'_n})\doteq
I(A_1A'_1\!:\ldots:\!A_nA'_n)_{\omega}-I(A'_1\!:\ldots:\!A'_n)_{\omega}
\end{equation}
between mutual informations of a state of the extended system
$A_1A'_1...A_nA'_n$ is used \cite{YHW}. Basic properties of the
quantum mutual information show that the gap $\Delta
I(\omega_{A_1A'_1...A_nA'_n})$ is nonnegative and does not increase
under local operations $\Phi_{A_1}:A_1\rightarrow
A_1,...,\Phi_{A_n}:A_n\rightarrow A_n$, i.e.
\begin{equation}\label{inf-gap-m}
\Delta I(\omega_{A_1A'_1...A_nA'_n})\geq\Delta
I\left(\Phi_{A_1}\otimes...\otimes\Phi_{A_n}\otimes\id_{A_1'...A_n'}(\omega_{A_1A'_1...A_nA'_n})\right).
\end{equation}

Recently Wilde proved that
\begin{equation}\label{W-ineq}
\Delta
I(\omega_{A_1A'_1...A_nA'_n})\geq\frac{1}{4n^2}\left\|\omega_{A_1A'_1...A_nA'_n}-
\Phi_1\otimes...\otimes\Phi_n(\omega_{A'_1...A'_n})\right\|_1^2
\end{equation}
for particular recovery channels $\Phi_1:A'_1\rightarrow
A_1A'_1,...,\Phi_n:A'_n\rightarrow A_nA'_n$ \cite{Wilde}.\footnote{In the recent paper \cite{Wilde+} the version of this inequality not depending on $\,n\,$ is proved.} This
result shows that if the gap $\Delta I(\omega_{A_1A'_1...A_nA'_n})$
is close to zero then the state $\omega_{A_1A'_1...A_nA'_n}$ can be
approximately recovered from its marginal state
$\omega_{A'_1...A'_n}$ by the local recovery channels
$\Phi_1:A'_1\rightarrow A_1A'_1,...,\Phi_n:A'_n\rightarrow
A_nA'_n$.\smallskip

This result has several applications in quantum information theory
\cite{Wilde}. It can be generalized to all states of
infinite-dimensional quantum system $A_1A'_1...A_nA'_n$ by
constructing appropriate extension of the information gap $\Delta
I(\omega_{A_1A'_1...A_nA'_n})$ to the set
$\S(\H_{A_1A'_1...A_nA'_n})$ (see Corollary \ref{inf-g-pr-c}
below).\medskip

To construct this extension we will use the representation
\begin{equation}\label{inf-g-d+}
\begin{array}{rl}
\displaystyle\Delta I(\omega_{A_1A'_1...A_nA'_n})\!\!\! &
=I(A_1\!:\!A'_2...A'_n|A'_1)_{\omega}\\&\displaystyle+\sum_{i=2}^n
I(A_i\!:\!A_1...A_{i-1}A'_1...A'_{i-1}A'_{i+1}...A'_n|A'_i)_{\omega}
\end{array}
\end{equation}
valid for any state of a finite-dimensional system
$A_1A'_1..A_nA'_n$ \cite[Lemma 1]{Wilde} and the extended
conditional mutual information defined in Sect.6.
\smallskip

\begin{property}\label{inf-g-pr}
A) \emph{The quantity $\,\Delta I(\omega_{A_1A'_1...A_nA'_n})$
defined in (\ref{inf-g-d}) has a lower semicontinuous nonnegative
extension to the set $\,\S(\H_{A_1A'_1\ldots A_nA'_n})$ possessing
property (\ref{inf-gap-m}) and upper bounded by
$2\sum_{i=1}^{n}H(\omega_{A_{i}})$. This extension can be defined by
formula (\ref{inf-g-d+}) in which each summand
$I(X\!:\!Y|Z)_{\omega}$ coincides with the function
$I_{\mathrm{e}}(X\!:\!Y|Z)_{\omega}$ described in Theorem
\ref{cmi-th}.}\smallskip

B) \emph{This extension (also denoted $\,\Delta
I(\omega_{A_1A'_1...A_nA'_n})$) satisfies
$\mathfrak{F}$\nobreakdash-\hspace{0pt}extension condition
(\ref{t-ext-p}) for any state $\omega_{A_1A'_1\ldots A_nA'_n}$ such
that $\,H(\omega_{A_{i}})<+\infty$ for all $\,i$. For an arbitrary
state $\,\omega\in\S(\H_{A_1A'_1\ldots A_nA'_n})$ the following
weaker property is valid:
\begin{equation}\label{inf-gap-lr}
\Delta I(\omega)=
\lim_{k\rightarrow\infty}\lim_{k'\rightarrow\infty}\Delta
I(\omega^{kk'}),
\end{equation}
where
$$
\omega^{kk'}=\lambda^{-1}Q\shs\omega\shs Q,\quad
Q=P^{k}_{A_1}\otimes\ldots\otimes P^{k}_{A_n}\otimes
P^{k'}_{A'_1}\otimes\ldots\otimes P^{k'}_{A'_n},\quad\lambda=\Tr
Q\shs\omega,
$$
$\{P^{k}_{A_i}\}\subset\B(\H_{A_i})$ and
$\{P^{k'}_{A'_i}\}\subset\B(\H_{A'_i})$ are any sequences of finite
rank projectors strongly converging to the identity operators
$I_{A_i}$ and $I_{A'_i}$, $i=\overline{1,n}$.}\smallskip

C) \emph{Representation (\ref{inf-g-d+}) is valid for $\,\Delta
I(\omega_{A_1A'_1...A_nA'_n})$ with arbitrarily permuted indexes
$\,1,...,n\,$ in the right hand side (provided that each summand
$I(X\!:\!Y|Z)_{\omega}$ coincides with
$I_{\mathrm{e}}(X\!:\!Y|Z)_{\omega}$).}\smallskip

D) \emph{Local continuity of the marginal entropies
$\,H(\omega_{A_1}),\ldots,H(\omega_{A_n})$ implies local continuity
of $\,\Delta I(\omega_{A_1A'_1...A_nA'_n})$.}\smallskip

E) \emph{If the subsystems  $A_1,...,A_{n}$ are finite-dimensional
then $\,\Delta I(\omega_{A_1A'_1...A_nA'_n})$ is a continuous
bounded function on the set $\,\S(\H_{A_1A'_1\ldots A_nA'_n})$ and
$$
|\Delta I(\omega^1)-\Delta I(\omega^2)|\leq 2\varepsilon C+2(n+1)(1+\varepsilon)h_2\!\left(\frac{\varepsilon}{1+\varepsilon}\right)
$$
for any $\,\omega^1,\omega^2\in\S(\H_{A_1A'_1\ldots A_nA'_n})$ such
that $\;\varepsilon=\frac{1}{2}\|\shs\omega^1-\omega^2\|_1<1\,$, where
$\,h_2(\cdot)\,$ is the binary entropy and $\,C=
\log\dim\H_{A_1...A_{n}}$.}
\end{property}\medskip

\emph{Proof.} Assertion A is proved by using Theorem \ref{cmi-th}
and Remark \ref{cmi-u-b}.\smallskip

Assertion B is proved by applying Corollaries \ref{cmi-pr-c} and
\ref{main-c-n++} to each term in the right hand side of
(\ref{inf-g-d+}).

Assertion C is valid for any state $\omega_{A_1A'_1...A_nA'_n}$ such
that $\,\mathrm{rank}\shs\omega_{A_{i}}<+\infty$ and
$\,\mathrm{rank}\shs\omega_{A'_{i}}<+\infty$ for all $i$ by Lemma 1
in \cite{Wilde}. Its validity for arbitrary state can be shown via
approximation by using properties (\ref{t-ext-p++}) and
(\ref{inf-gap-lr}).\smallskip

Assertion D follows from  Corollary \ref{cmi-th-c-1}A applied to
each term in the right hand side of (\ref{inf-g-d+}).

Assertion E follows from Proposition \ref{FA-cont} and the upper
bound for the quantity $\Delta I(\omega_{A_1A'_1...A_nA'_n})$
mentioned in assertion A. $\square$
\medskip\smallskip

Inequality (\ref{W-ineq}) is proved in \cite{Wilde} by using the
following two facts:
\begin{itemize}
    \item the existence for any state $\omega_{ABC}$ of a finite-dimensional
tripartite system of a channel $\Phi:B\rightarrow BC$ (the
Fawzi-Renner recovery map) such that
\begin{equation*}
2^{-\frac{1}{2}I(A:C|B)_{\omega}}\leq F(\omega_{ABC},
\id_A\otimes\Phi(\omega_{AB})),
\end{equation*}
where $F(\rho,\sigma)\doteq\|\sqrt{\rho}\sqrt{\sigma}\|_1$ is the
quantum fidelity \cite{F&R};
    \item the validity of representation (\ref{inf-g-d+}) with arbitrarily permuted
indexes $\,1,...,n\,$ in the right hand side which shows that for
all $i=\overline{1,n}$ the following inequality holds
$$
\Delta I(\omega_{A_1A'_1...A_nA'_n})\geq
I(A_i\!:\!A_1..A_{i-1}A_{i+1}..A_nA'_1..A'_{i-1}A'_{i+1}..A'_n|A'_i)_{\omega}.
$$
\end{itemize}

The first fact is valid for all states of an infinite-dimensional
tripartite system provided
$I(A:C|B)_{\omega}=I_{\mathrm{e}}(A:C|B)_{\omega}$ -- the extended
conditional mutual information. For states with finite marginal
entropies this is proved in \cite{F&R}, for arbitrary states this
follows from Proposition \ref{FR-r-m} in Sect.8.4 below.

The second fact is  valid for all states of an infinite-dimensional
system $A_1A'_1...A_nA'_n$ by Proposition \ref{inf-g-pr}C. So, by
repeating the arguments from  \cite{Wilde} one can prove inequality
(\ref{W-ineq}) for all states of an infinite-dimensional system
$A_1A'_1...A_nA'_n$.
\smallskip

\begin{corollary}\label{inf-g-pr-c}
\emph{The Wilde inequality (\ref{W-ineq}) is valid for all states of
infinite-dimensional system $A_1A'_1...A_nA'_n$ provided $\,\Delta
I(\omega_{A_1A'_1...A_nA'_n})$ is the extension of the gap
$\,I(A_1A'_1\!:\!...\!:\!A_nA'_n)_{\omega}-I(A'_1\!:\!...\!:\!A'_n)_{\omega}$
described in Proposition \ref{inf-g-pr}. }
\end{corollary}

\section{Some applications}

\subsection{Triangle continuity relations for quantum channels}

Let $\,\Phi:A\rightarrow B$ be a quantum channel -- completely
positive trace preserving linear map
$\mathfrak{T}(\mathcal{H}_A)\rightarrow\mathfrak{T}(\mathcal{H}_B)$.
Stinespring's theorem implies the existence of a Hilbert space
$\mathcal{H}_E$ and of an isometry
$V:\mathcal{H}_A\rightarrow\mathcal{H}_B\otimes\mathcal{H}_E$ such
that
\begin{equation*}
\Phi(\rho)=\mathrm{Tr}_{E}V\rho V^{*},\quad
\rho\in\mathfrak{T}(\mathcal{H}_A).
\end{equation*}
The quantum  channel
\begin{equation}\label{c-channel}
\mathfrak{T}(\mathcal{H}_A)\ni
\rho\mapsto\widehat{\Phi}(\rho)=\mathrm{Tr}_{B}V\rho
V^{*}\in\mathfrak{T}(\mathcal{H}_E)
\end{equation}
is called \emph{complementary} to the channel $\Phi$
\cite[Ch.6]{H-SCI}.

It is well known that for a finite-dimensional channel $\,\Phi$ and
a state $\rho$ in $\mathfrak{S}(\mathcal{H}_A)$ the input entropy
$H(\rho)$, the output entropy $H(\Phi(\rho))$ and the \emph{entropy
exchange} $H(\Phi, \rho)\doteq H(\widehat{\Phi}(\rho))$ denoted
respectively $H_{\Phi}(\rho)$ and $H_{\widehat{\Phi}}(\rho)$ satisfy
the "triangle inequalities":
\begin{equation}\label{tr-e-1}
\begin{array}{c}
\left|H_{\Phi}(\rho)-H_{\widehat{\Phi}}(\rho)\right|\leq
H(\rho),\\\\
\left|H_{\Phi}(\rho)-H(\rho)\right|\leq
H_{\widehat{\Phi}}(\rho),\quad
\left|H_{\widehat{\Phi}}(\rho)-H(\rho)\right|\leq H_{\Phi}(\rho).
\end{array}
\end{equation}

The quantity $I_c(\Phi,\rho)\doteq
H_{\Phi}(\rho)-H_{\widehat{\Phi}}(\rho)$  called \emph{coherent
information} of a channel $\Phi$ at a state $\rho$ is an important
characteristic of a quantum channel related to its quantum capacity
\cite{H-SCI,N&Ch}.

The quantity $EG(\Phi,\rho)\doteq H_{\Phi}(\rho)-H(\rho)$  called
\emph{entropy gain}  of a channel $\Phi$ at a state $\rho$ is a
convex function of $\rho$ also used in analysis of information
properties of a quantum channel \cite{A,H-EG}.\smallskip

If $\Phi$ is an infinite-dimensional quantum  channel, then
inequalities (\ref{tr-e-1}) also hold provided all the terms are
finite.  Moreover, if we use the extension of the coherent
information $I_c(\Phi,\rho)=
H_{\Phi}(\rho)-H_{\widehat{\Phi}}(\rho)$ to the set
$\{\rho\in\S(\H_A)\,|\,H(\rho)<+\infty\}$ given by the
formula\footnote{The reasonability of this extension is shown in
\cite{H-Sh-4}.}
\begin{equation}\label{ext-1}
I_c(\Phi,\rho)= H\left(\Phi \otimes \mathrm{Id}_{R}
(|\varphi_{\rho}\rangle\langle\varphi_{\rho}|)\shs \|\shs \Phi
(\rho) \otimes \varrho\shs\right)-H(\rho),
\end{equation}
where $|\varphi_{\rho}\rangle$ is a purification of the state $\rho$
in $\mathcal{H}_A \otimes \mathcal{H}_R$ and
$\varrho=\mathrm{Tr}_{A}|\varphi_{\rho}\rangle\langle\varphi_{\rho}|$,
then upper bound (\ref{mi-u-b}) implies that the first inequality in
(\ref{tr-e-1}) becomes valid for arbitrary state $\rho$ (with
possible value $+\infty$ in the both sides).\smallskip

The quantities  $EG(\Phi,\rho)=H_{\Phi}(\rho)-H(\rho)$ and
$EG(\widehat{\Phi},\rho)=H_{\widehat{\Phi}}(\rho)-H(\rho)$ can be
extended respectively by the formulae
\begin{equation}\label{ext-2}
EG(\Phi,\rho)= H\left(V\rho
V^*\shs\|\shs\Phi(\rho)\otimes\widehat{\Phi}(\rho)\right)-H_{\widehat{\Phi}}(\rho)
\end{equation}
and
\begin{equation}\label{ext-3}
EG(\widehat{\Phi},\rho)= H\left(V\rho
V^*\shs\|\shs\Phi(\rho)\otimes\widehat{\Phi}(\rho)\right)-H_{\Phi}(\rho)
\end{equation}
to the sets $\{\rho\in\S(\H_A)\,|\,H_{\widehat{\Phi}}(\rho)<+\infty
\}$ and $\{\rho\in\S(\H_A)\,|\,H_{\Phi}(\rho)<+\infty \}$, where $V$
is any Stinespring isometry for $\Phi$ and $\widehat{\Phi}$ is the
version of complementary channel corresponding to this isometry via
(\ref{c-channel}). The right hand sides of (\ref{ext-2}) and
(\ref{ext-3}) can be written respectively as
$-H_{\mathrm{e}}(E|B)_{V\rho V^*}$ and $-H_{\mathrm{e}}(B|E)_{V\rho
V^*}$, where $H_{\mathrm{e}}(A|B)$ is the extended quantum
conditional entropy proposed in \cite{Kuz} and described in Sec.5.
So, the concavity of $H_{\mathrm{e}}(A|B)$ implies the convexity of
$EG(\Phi,\rho)$ and of $EG(\widehat{\Phi},\rho)$ defined by
(\ref{ext-2}) and (\ref{ext-3}) as functions of $\rho$. The upper
bound (\ref{mi-u-b}) shows that
$$
|EG(\Phi,\rho)|\leq
H_{\widehat{\Phi}}(\rho)\quad\textrm{and}\quad|EG(\widehat{\Phi},\rho)|\leq
H_{\Phi}(\rho).
$$
So, the second and the third inequalities in (\ref{tr-e-1}) are also
valid for arbitrary state $\rho$ (with possible value $+\infty$ in
the both sides) provided the values $H_{\Phi}(\rho)-H(\rho)$ and
$H_{\widehat{\Phi}}(\rho)-H(\rho)$ are defined respectively by
formulae (\ref{ext-2}) and (\ref{ext-3}).\smallskip

Theorem \ref{main}A makes possible to show that continuity of one of
the functions $H(\rho), H_{\Phi}(\rho), H_{\widehat{\Phi}}(\rho)$
implies continuity of the difference between two others.\smallskip

\begin{property}\label{triangle-cont-cond}
\textit{Let $\,\Phi:A\rightarrow B$ be a quantum channel and
$\;\widehat{\Phi}:A\rightarrow E\,$ its complementary channel. Then}
\begin{enumerate}[A)]
    \item \emph{local continuity of $\,H(\rho)$ implies local continuity
    of the coherent information
    $\,I_c(\Phi,\rho)=[H_{\Phi}(\rho)-H_{\widehat{\Phi}}(\rho)]\,$  defined by formula (\ref{ext-1});}
\item \emph{local continuity of $\,H_{\widehat{\Phi}}(\rho)$ implies local continuity of the entropy gain $\,EG(\Phi,\rho)=[H_{\Phi}(\rho)-H(\rho)]\,$
defined by formula (\ref{ext-2});}
\item \emph{local continuity of $\,H_{\Phi}(\rho)$ implies local continuity of the entropy gain
      $\,EG(\widehat{\Phi},\rho)=[H_{\widehat{\Phi}}(\rho)-H(\rho)]\,$ defined by formula (\ref{ext-3}).}
\end{enumerate}
\end{property}\vspace{5pt}

While triangle inequalities (\ref{tr-e-1}) show existence of a
triangle with sides of length $H(\rho), H_{\Phi}(\rho),
H_{\widehat{\Phi}}(\rho)$, Proposition \ref{triangle-cont-cond}
states that \emph{a small deformation of any side of this triangle
leads to a small deformation of the difference between the other
sides} (despite possible large deformation of each of these sides).
\vspace{5pt}

\emph{Proof.} Assertions B and C directly follow from Theorem
\ref{main}A. Assertion A is also  derived from Theorem \ref{main}A
by using Lemma \ref{p-lemma} and by noting that $H(\varrho)=H(\rho)$
for the state $\varrho$ in (\ref{ext-1}). $\square$ \medskip

Since $EG(\Phi,\rho)=-H_{\mathrm{e}}(E|B)_{V\rho V^*}$, by applying
Corollary \ref{FA-gen} one can strengthen assertion B of Proposition
\ref{triangle-cont-cond} in the case $\dim\H_E=k<+\infty$.\smallskip

\begin{corollary}\label{triangle-cont-cond-c}
\emph{If $\,\Phi:A\rightarrow B$ is a quantum channel with finite
Choi rank $\,k\,$ then the entropy gain (\ref{ext-2}) is a
continuous bounded function on the set $\,\S(\H_A)$ and
$$
|EG(\Phi,\rho_1)-EG(\Phi,\rho_2)|\leq 2\varepsilon \log
k+(1+\varepsilon)h_2\!\left(\frac{\varepsilon}{1+\varepsilon}\right)
$$
for any $\,\rho_1,\rho_2\in\S(\H_{A})$ such that
$\,\varepsilon=\frac{1}{2}\|\shs\rho_1-\rho_2\|_1<1$, where $h_2(\cdot)$
is the binary entropy.}
\end{corollary}\smallskip

By applying Corollary \ref{main-c} and by noting that
$\dim\H_R=\dim\H_A$ one can obtain continuity bounds for the
coherent information (\ref{ext-1}) and for the quantum mutual
information (\ref{mi-qc+}) in the case $\dim\H_A<+\infty$.

\subsection{Continuity conditions for the\\ functions $(\Phi,\rho)\mapsto I(\Phi,\rho)$ and $(\Phi,\rho)\mapsto I_c(\Phi,\rho)$}

The quantum mutual information is an important characteristic of a
quantum channel related to its  entanglement-assisted classical
capacity \cite{Adami,BSST,H-SCI,N&Ch}. For a finite-dimensional
channel $\Phi:A\rightarrow B$ it can be defined by the formula
\begin{equation}\label{mi-qc}
I(\Phi,\rho)=H(\rho)+H(\Phi(\rho))-H(\widehat{\Phi}(\rho)).
\end{equation}
In infinite dimensions this definition may contain the uncertainty
$"\infty-\infty"$, but it can be modified to avoid this problem as
follows
\begin{equation}\label{mi-qc+}
 I(\Phi,\rho) = H\left(\Phi \otimes \mathrm{Id}_{R}
(|\varphi_{\rho}\rangle\langle\varphi_{\rho}|)\shs \|\shs \Phi
(\rho) \otimes \varrho\shs\right),
\end{equation}
where $|\varphi_{\rho}\rangle$ is a purification of the state $\rho$
in $\mathcal{H}_A \otimes \mathcal{H}_R$ and
$\varrho=\mathrm{Tr}_{A}|\varphi_{\rho}\rangle\langle\varphi_{\rho}|$.
For an arbitrary quantum channel $\Phi$ the nonnegative function
$\rho\mapsto I( \Phi,\rho)$ defined by (\ref{mi-qc+}) is concave and
lower semicontinuous on the set $\mathfrak{S}(\mathcal{H}_A)$
\cite{H-Sh-4}. \smallskip

In study of the entanglement-assisted classical capacity of an
infinite-dimensional channel, in particular, in the analysis of its
continuity as a function of a channel (which can be interpreted as a
robustness or stability of the capacity with respect to perturbation
of a channel) it is necessary to explore continuity properties of
the quantum mutual information $I(\Phi,\rho)$ with respect to
simultaneous variation of $\Phi$ and $\rho$. This means that we have
to consider $I(\Phi,\rho)$ as a function of a pair $(\Phi,\rho)$,
i.e. as a function on the Cartesian product of the set
$\mathfrak{F}_{AB}$ of all quantum channels from $A$ to $B$ equipped
with appropriate topology (type of convergence) and the set
$\mathfrak{S}(\mathcal{H}_A)$ of input states.

Since a quantum channel $\Phi$ is a completely bounded map, the set
$\mathfrak{F}_{AB}$  is typically equipped with \emph{the norm of
complete boundedness} \cite{Paulsen}, which can be defined as the
upper bound of the operator norms of the maps
$\Phi\otimes\id_{\mathbb{C}^n}$, $n\in\mathbb{N}$. In infinite
dimensions along with the topology induced by the norm of complete
boundedness one can consider weaker topologies on the set of quantum
channels, in particular, the \textit{strong convergence topology}
generated by the strong operator topology on the set of all linear
bounded operators between the Banach spaces
$\mathfrak{T}(\mathcal{H}_A)$ and $\mathfrak{T}(\mathcal{H}_B)$
\cite{AQC}. The strong convergence of a sequence
$\{\Phi_{n}\}\subset\mathfrak{F}_{AB}$ to a quantum channel
$\Phi_{0}\in\mathfrak{F}_{AB}$ means that
$$
\lim_{n\rightarrow+\infty}\Phi_{n}(\rho)=\Phi_{0}(\rho)\quad
\forall\rho\in\mathfrak{S}(\mathcal{H}_A).
$$
The use of the strong convergence topology in infinite dimensions
seems preferable by the following reason. It is shown in \cite{Kr&W}
that closeness of two quantum channels in the norm of complete
boundedness means, roughly speaking, the operator norm closeness of
the corresponding Stinespring isometries. So, if we use the norm of
complete boundedness then we take into account only such
perturbations of a channel that corresponds to \emph{uniform
deformations} of the Stinespring isometry (i.e. deformations with
small operator norm). Physically, it seems reasonable to consider
the wider class of perturbations of a channel including the
perturbations corresponding to deformations of the Stinespring
isometry in the strong operator topology.

Thus, we will assume in what follows that $\mathfrak{F}_{AB}$ is the
set all channels $\,\Phi:A\rightarrow B$ equipped with the strong
convergence topology. By separability of the set
$\mathfrak{S}(\mathcal{H}_A)$ the strong convergence topology  on
the set $\mathfrak{F}_{AB}$ is metrizable (can be induced by some
metric). Note also that it is the strong convergence topology  that
makes the set $\mathfrak{F}_{AB}$ to be \emph{topologically}
isomorphic to a subset of states of a composite system (generalized
Choi-Jamiolkowski isomorphism) \cite[Proposition 3]{AQC}.

By using Theorem \ref{main} one can substantially amplify the
continuity condition for the function $\,(\Phi,\rho)\mapsto
I(\Phi,\rho)$  obtained in \cite[Proposition 5]{H-Sh-4}.
\smallskip

\begin{property}\label{mi-qc-cont}
A) \emph{Continuity of $\,H(\rho)$ on a set
$\mathcal{A}\subset\S(\H_A)$ implies continuity
    of the function  $\;(\Phi,\rho)\mapsto I(\Phi,\rho)\,$ on the set
    $\,\mathfrak{F}_{AB}\times\mathcal{A}$}.\smallskip

B) \emph{Local continuity of the function $\,(\Phi,\rho)\mapsto
H(\Phi(\rho))$ implies local continuity of the function
$\,(\Phi,\rho)\mapsto I(\Phi,\rho)$.}\smallskip

C) \emph{Local continuity of the function $\,(\Phi,\rho)\mapsto
I(\Phi,\rho)$ implies local continuity of the function
$\,(\Phi,\rho)\mapsto I(\Psi\circ\Phi,\rho)$ for any channel
$\,\Psi:B\rightarrow C$.}
\end{property}

\medskip
Assertion A of Proposition \ref{mi-qc-cont} states that
\begin{equation}\label{mi-c-impl}
\lim_{n\rightarrow\infty} H(\rho_n)=H(\rho_0)<+\infty\;\;
\Rightarrow\;\; \lim_{n\rightarrow\infty}
I(\Phi_n,\rho_n)=I(\Phi_0,\rho_0)<+\infty
\end{equation}
where $\rho_0=\lim_{n\rightarrow\infty}\rho_n$, for \emph{arbitrary}
sequence $\{\Phi_n\}$ of channels strongly converging to a channel
$\Phi_0$. In contrast to Proposition 5 in \cite{H-Sh-4}  the
existence of a sequence $\{\widehat{\Phi}_n\}$ converging to the
channel $\widehat{\Phi}_0$ is not required in (\ref{mi-c-impl}).
\smallskip

Assertions B and C can be formulated as the implications:
$$
\lim_{n\rightarrow\infty}
H(\Phi_n(\rho_n))=H(\Phi_0(\rho_0))<+\infty\; \Rightarrow\;
\lim_{n\rightarrow\infty} I(\Phi_n,\rho_n)=I(\Phi_0,\rho_0)<+\infty
$$
and
$$
\lim_{n\rightarrow\infty}\!I(\Phi_n,\rho_n)=\!I(\Phi_0,\rho_0)<+\infty\;
\Rightarrow\; \lim_{n\rightarrow\infty}\!
I(\Psi\circ\Phi_n,\rho_n)=\!I(\Psi\circ\Phi_0,\rho_0)<+\infty
$$
valid for sequences $\{\rho_n\}\subset\S(\H_A)$ and
$\{\Phi_n\}\subset\mathfrak{F}_{AB}$ converging respectively to a
state $\rho_0$ and to a channel $\Phi_0$ and arbitrary channel
$\Psi\in\mathfrak{F}_{BC}$.\smallskip

\emph{Proof.} Since the strong convergence of a sequence
$\{\Phi_n\}$ to a channel $\Phi_0$ implies the strong convergence of
the sequence $\{\Phi_n\otimes \id_{R}\}$ to the channel
$\Phi_0\otimes \id_{R}$ \cite{AQC}, assertions A and B follow from
Theorem \ref{main}A and Lemma \ref{p-lemma} (since
$H(\varrho)=H(\rho)$ for the state $\varrho$ in (\ref{mi-qc+}))
while assertion C follows from Theorem \ref{main}B. $\square$
\medskip

Proposition \ref{mi-qc-cont}A gives a continuity condition for the
coherent information $I_c(\Phi,\rho)$ defined by formula
(\ref{ext-1}) which means $\,I_c(\Phi,\rho)=I(\Phi,\rho)-H(\rho)$.
\smallskip

\begin{corollary}\label{c-inf-cont}
\emph{Continuity of $\,H(\rho)$ on a set
$\mathcal{A}\subset\S(\H_A)$ implies continuity
    of the function  $\;(\Phi,\rho)\mapsto I_c(\Phi,\rho)\,$ on the set
    $\,\mathfrak{F}_{AB}\times\mathcal{A}$}.
\end{corollary}
\smallskip

Corollary \ref{c-inf-cont} states that implication (\ref{mi-c-impl})
holds with $I_c(\Phi,\rho)$ instead of $I(\Phi,\rho)$, it can be
considered as a generalized version of Proposition
\ref{triangle-cont-cond}A.\smallskip

It is easy to see that the analog of Proposition \ref{mi-qc-cont}B
is not valid for the function $(\Phi,\rho)\mapsto I_c(\Phi,\rho)$.

\subsection{On continuity of the entanglement assisted classical
capacity as a function of channel}

In this section we substantially strengthen the conditions for
continuity of the entanglement-assisted classical capacity of an
infinite-dimensional quantum channel with linear constraint (as a
function of a channel) obtained in \cite{EAC}.\footnote{The
importance of studying continuity properties of quantum channel
capacities is discussed in \cite{L&S,AQC,EAC}. It is explained,
briefly speaking, by unavoidable perturbations of a channel used for
information transmission.}

\smallskip

A rate of transmission of classical information over a quantum
channel can be increased by using an entangled state as an
additional resource. A detailed description of the corresponding
protocol can be found in \cite{H-SCI,N&Ch}. The ultimate rate of
information transmission by this protocol is called
\emph{entanglement-assisted classical capacity} of a quantum
channel.

If $\Phi:A\rightarrow B$ is a finite-dimensional quantum channel
then the Bennett-Shor-Smolin-Thaplyal (BSST) theorem~\cite{BSST}
gives the following expression for its entanglement-assisted
classical capacity
\begin{equation*}
C_{\mathrm{ea}}(\Phi)=\sup_{\rho \in
\mathfrak{S}(\mathcal{H}_A)}I(\Phi,\rho),
\end{equation*}
where $I(\Phi,\rho)$ is the quantum mutual information defined by
(\ref{mi-qc}).

Continuity if this capacity as a function of a channel directly
follows from continuity of the function $(\Phi,\rho)\mapsto
I(\Phi,\rho)$ and compactness of the space of input states
\cite{L&S}. \smallskip

If $\Phi$ is an infinite-dimensional quantum channel then we have to
impose constraint on states used for coding information, typically
linear constraint determined by the inequality $\mathrm{Tr}F\rho
\leq E$, where $F$ is a positive operator and $E>0$. An operational
definition of the entanglement-assisted classical capacity
$C_{\mathrm{ea}}(\Phi ,F,E)$ of an infinite-dimensional quantum
channel $\Phi$ with the above linear constraint is given
in~\cite{H-c-w-c}, where the generalization of the BSST theorem is
proved under special restrictions on the channel $\Phi$ and on the
constraint operator $F$. A general version of the BSST theorem for
infinite-dimensional channel with linear constraints without any
simplifying restrictions is proved in \cite{EAC} (by using the
extended quantum conditional entropy $H_{\mathrm{e}}(A|B)$ proposed
in \cite{Kuz} and described in Sec.5), it states that
\begin{equation}\label{eaco}
C_{\mathrm{ea}}(\Phi ,F,E)=\sup_{\mathrm{Tr}F\rho \leq
E}I(\Phi,\rho)\leq+\infty
\end{equation}
for arbitrary  channel $\,\Phi $ and arbitrary constraint operator
$F$, where $I(\Phi,\rho)$ is the quantum mutual information defined
by formula (\ref{mi-qc+}).\smallskip

By noting that the function $(\Phi,\rho)\mapsto I(\Phi,\rho)$ is
lower semicontinuous on
$\mathfrak{F}_{AB}\times\mathfrak{S}(\mathcal{H}_A)$ (see
\cite{H-Sh-4}) it is easy to show that the function $\Phi \mapsto
C_{\mathrm{ea}}(\Phi ,F,E)$ is lower semicontinuous on
$\mathfrak{F}_{AB}$, i.e.
\begin{equation*}
\liminf_{n\rightarrow +\infty }C_{\mathrm{ea}}(\Phi _{n},F,E)\geq
C_{\mathrm{ea}}(\Phi_{0},F,E)\;\;(\leq +\infty )
\end{equation*}
for any sequence $\{\Phi _{n}\}$ of channels strongly converging to
a channel $\Phi_{0}$. But this function is not continuous in general
(see Example \ref{eac-discont} below).
\smallskip

Proposition \ref{mi-qc-cont}A implies the following sufficient
condition for global continuity of the function $\Phi\mapsto
C_{\mathrm{ea}}(\Phi,F,E)$.\smallskip

\begin{property}\label{eac-cont-cond} \emph{Let $\mathcal{K}_{F,E}$ be the set of input states $\rho$ such that $\,\Tr
F\rho\leq E$}\smallskip

\emph{If $\,H(\rho)$ is continuous on $\,\mathcal{K}_{F,E}$ then the
function $\,\Phi\mapsto C_{\mathrm{ea}}(\Phi,F,E)\,$ is continuous
on the set $\,\mathfrak{F}_{AB}$ of \textbf{all} channels (and upper
bounded by
$\displaystyle\sup_{\rho\in\mathcal{K}_{F,E}}\!\!2H(\rho)$).}
\end{property}
\medskip

The assumption of continuity of $\,H(\rho)$ on $\,\mathcal{K}_{F,E}$
holds for any $E>0$ if the operator $F$ satisfies the condition
$\Tr\exp(-\lambda F)<+\infty$ for all $\lambda>0$ \cite{W}. This
condition is valid if $F=R^{\top}\epsilon R$ -- Hamiltonian of a
many-mode Bosonic quantum system, where $\epsilon$ is a
nondegenerate energy matrix and $R$ are the canonical variables of
the system (see details in \cite[Ch.12]{H-SCI}). In this case
$C_{\mathrm{ea}}(\Phi,F,E)$ is the entanglement-assisted classical
capacity of a channel $\Phi$ under the condition that the mean
energy of states used for coding information is $\leq E$.
Proposition \ref{eac-cont-cond} implies the following
observation.\smallskip

\begin{corollary}\label{eac-cont-cond-c} \emph{Let $\,F=R^{\top}\epsilon R$ be a Hamiltonian of a many-mode Bosonic
quantum system $A$ and $E>0$.  The function $\,\Phi\mapsto
C_{\mathrm{ea}}(\Phi,F,E)\,$ is continuous on the set
$\,\mathfrak{F}_{AB}$ of \textbf{all} channels from the system A to
any system  B.}
\end{corollary}\smallskip

This corollary shows that the entanglement-assisted classical
capacity of a Bosonic Gaussian channel with energy constraint is
continuously varies under any perturbations of this
channel.\smallskip

\textit{Proof of Proposition \ref{eac-cont-cond}.} Since the set
$\mathcal{K}_{F,E}$ is closed and convex, the continuity of the
concave function $\,H(\rho)$ on $\mathcal{K}_{F,E}$ implies
boundedness of this function on $\mathcal{K}_{F,E}$.\footnote{If for
each $n$ there is a state $\rho_n\in\mathcal{K}_{F,E}$ such that
$H(\rho_{n})\geq 2^{n}$ then $
\sum_{n=1}^{+\infty}2^{-n}\rho_{n}\in\mathcal{K}_{F,E}$ and
$H\left(\sum_{n=1}^{+\infty}2^{-n}\rho_{n}\right)\geq\sum_{n=1}^{+\infty}2^{-n}H\left(\rho_{n}\right)=+\infty
$ contradicting to the continuity of $H$.} So, by Corollary 5 in
\cite{EC} the set $\mathcal{K}_{F,E}$ is compact.

Since the function $\,\Phi\mapsto C_{\mathrm{ea}}(\Phi,F,E)\,$ is
lower semicontinuous on the set $\,\mathfrak{F}_{AB}$ it suffices to
show that it is finite and upper semicontinuous.

Assume that there exists  a  sequence $\{\Phi_{n}\}$ of channels in
$\mathfrak{F}_{AB}$ strongly converging to a channel $\Phi_{0}$ such
that
\begin{equation}  \label{lim-exp+}
\lim_{n\rightarrow+\infty}C_{\mathrm{ea}}(\Phi_n,F,E)\geq
C_{\mathrm{ea} }(\Phi_0,F,E)+\varepsilon
\end{equation}
for some $\varepsilon>0$.  It follows from (\ref{eaco}) that for
each $n$ there is a state $\rho_n\in\mathcal{K}_{F,E}$ such that
$$
C_{\mathrm{ea}}(\Phi_n,F,E)<I(\Phi_n,\rho_n)+\varepsilon/2.
$$

By compactness of the set $\mathcal{K}_{F,E}$ we may assume (by
passing to a subsequence) that the sequence $\{\rho_n\}$ converges
to a particular state $ \rho_0\in\mathcal{K}_{F,E}$. Proposition
\ref{mi-qc-cont}A implies
\begin{equation*}
\lim_{n\rightarrow+\infty}I(\Phi_n,\rho_n)=I(\Phi_0,\rho_0).
\end{equation*}
 Since (\ref{eaco}) shows that $C_{\mathrm{ea}}(\Phi_0,F,E)\geq I(\Phi_0,\rho_0)$, this  contradicts
to (\ref{lim-exp+}). $\square$\smallskip

By using Proposition \ref{mi-qc-cont}B and the same arguments one
can prove the following "local" continuity condition. \smallskip

\begin{property}\label{eac-cont-cond+}  \emph{If $\,\mathcal{K}_{F,E}=\{\rho\in\mathfrak{S}(
\mathcal{H}_A)\,|\,\mathrm{Tr}F\rho \leq E\}$ is a compact set and
$\{\Phi_{n}\}$ is a sequence of channels strongly converging to a
channel $\Phi_{0}$ such that
$\displaystyle\lim_{n\rightarrow+\infty}H(\Phi_n(\rho_n))=H(\Phi_0(\rho_0))<+\infty$
for any  sequence $\{\rho_n\}\subset\mathcal{K}_{F,E}$ converging to
a state $\rho_0$ then}
\begin{equation}  \label{C-lim-exp}
\lim_{n\rightarrow+\infty}C_{\mathrm{ea}}(\Phi_n,F,E)=C_{\mathrm{ea}
}(\Phi_0,F,E)<+\infty.
\end{equation}
\end{property}

\begin{remark}\label{eac-cont-cond+r} The same  continuity condition holds for the Holevo capacity of an
infinite-dimensional channel with linear constraints
\cite[Proposition 7]{AQC}.
\end{remark}

\medskip

The following example shows that compactness of the set
$\mathcal{K}_{F,E}$ and continuity of the output entropies of all
the channels $\Phi_n$ do not imply (\ref{C-lim-exp}).\smallskip

\begin{example}\label{eac-discont}  Let $\{|k\rangle\}_{k\geq0}$ be an orthonormal basic in a separable Hilbert space
$\H_A=\H_B$ and $P_n=\sum_{k=1}^n |k\rangle\langle k|$ be a
projector of rank $n$. Consider the sequence of channels
$$
\Phi_n(\rho)=[\Tr(I_A-q_nP_n)\rho\shs]|0\rangle\langle 0|+q_nP_n\rho
P_n
$$
with finite-dimensional output space, where $\{q_n\}$ is a sequence
of positive numbers specified below.

Let $F=\sum_{k=0}^{+\infty}\log(\log (k+3)) |k\rangle\langle k|$ be
a positive operator. By the Lemma in \cite{H-c-w-c} the
corresponding set $\mathcal{K}_{F,E}$ is compact. Since
$\Tr\exp(-\lambda F)=+\infty$ for any $\lambda>0$, Proposition 1a in
\cite{EC} and its proof imply the existence of a sequence
$\{\rho_n\}\subset\mathcal{K}_{F,E}$ such that $\rho_n=P_n\rho_nP_n$
and $\lim_{n\rightarrow\infty}H(\rho_n)=+\infty$.

It is easy to show that $I(\Phi_n, \rho_n)\geq 2q_n
H(P_n\rho_nP_n)=2q_nH(\rho_n)$. Since
$\{\rho_n\}\subset\mathcal{K}_{F,E}$, this implies
$C_{\mathrm{ea}}(\Phi_n,F,E)\geq 2q_n H(\rho_n)$. So, for any
sequence $\{q_n\}$ such that $\,\lim_{n\rightarrow\infty}q_n=0\,$
and $\,\lim_{n\rightarrow\infty}q_n H(\rho_n)=C>0$ we have
$$
\liminf_{n\rightarrow+\infty}C_{\mathrm{ea}}(\Phi_n,F,E)\geq 2C,
$$
while the sequence $\{\Phi_{n}\}$ strongly converges to the
completely depolarizing channel
$\Phi_{0}(\rho)=[\Tr\rho]|0\rangle\langle 0|$ for which
$C_{\mathrm{ea}}(\Phi_0,F,E)=0$.
\end{example}

\subsection{On existence of the Fawzi-Renner recovery channel for arbitrary tripartite state}


The fundamental strong subadditivity property of the von Neumann
entropy, which means the nonnegativity of $\,I(A\!:\!C|B)_{\omega}$,
was recently specified by Fawzi and Renner who proved in \cite{F&R}
that for any state $\omega_{ABC}$ there exists a recovery channel
$\Phi:B\rightarrow BC$ such that
\begin{equation}\label{FR-ineq}
2^{-\frac{1}{2}I(A:C|B)_{\omega}}\leq F(\omega_{ABC},
\id_A\otimes\Phi(\omega_{AB}))
\end{equation}
where $F(\rho,\sigma)\doteq\|\sqrt{\rho}\sqrt{\sigma}\|_1$ is the
quantum fidelity between states $\rho$ and
$\sigma$.\footnote{Recently it was shown that the recovery channel
$\Phi:B\rightarrow BC$ satisfying (\ref{FR-ineq}) can be chosen
independently of $A$ \cite{SFR}.} This result can be considered as a
$\varepsilon$-version of the well-known characterization of a state
$\omega_{ABC}$ for which $I(A\!:\!C|B)_{\omega}=0$ as a Markov chain
(i.e. as a state reconstructed from its marginal state $\omega_{AB}$
by a channel $\id_A\otimes\Phi$). It has several important
applications in quantum information theory \cite{F&R,Wilde}.

The existence of a channel $\Phi$ satisfying (\ref{FR-ineq}) is
proved in \cite{F&R} in finite-dimensional settings by
quasi-explicit construction. Then, by using approximation technic,
this result is extended in \cite{F&R,SFR} to a state $\omega_{ABC}$
of infinite-dimensional system assuming that
$I(A\!:\!C|B)_{\omega}=H(A|B)_{\omega}-H(A|BC)_{\omega}$, i.e.
assuming that the marginal entropies of $\omega_{ABC}$ are finite.

It is also shown in Remark 5.3 in \cite{F&R} that in finite
dimensions a channel $\Phi:B\rightarrow BC$ satisfying
(\ref{FR-ineq}) can be chosen in such a way that
\begin{equation}\label{FR-cond}
[\Phi(\omega_{B})]_B=\omega_{B}\quad\textrm{and}\quad
[\Phi(\omega_{B})]_C=\omega_{C},
\end{equation}
i.e. a recovery channel $\Phi$ may exactly reproduce the marginal
states.\footnote{In Remark 5.3 in \cite{F&R}  the existence of a
quantum operation $\Phi'$ satisfying (\ref{FR-ineq}) such that
$[\Phi'(\omega_{B})]_B\leq\omega_{B}$ and
$[\Phi'(\omega_{B})]_C\leq\omega_{C}$ is shown. A quantum channels
$\Phi$ satisfying (\ref{FR-ineq}) and (\ref{FR-cond}) can be
obtained via $\Phi'$ as follows
$\Phi(\rho)=\Phi'(\rho)+[\Tr\rho-\Tr\Phi'(\rho)]\shs\sigma$, where
$\sigma$ is the normalized positive operator
$(\shs\omega_{B}-[\shs\Phi'(\omega_{B})]_B)\otimes(\shs\omega_{C}-[\shs\Phi'(\omega_{B})]_C)$.}
\smallskip

The properties of the extended conditional mutual information
(described in Section 6) make possible to show the existence of a
recovery channel satisfying (\ref{FR-cond}) for \emph{all} states of
infinite-dimensional tripartite system  starting with the
above-mentioned finite-dimensional result.
\smallskip

\begin{property}\label{FR-r-m} \textbf{(ID-version of Remark 5.3 in
\cite{F&R})} \emph{For arbitrary state $\omega_{ABC}$ of
infinite-dimensional tripartite system there exists a channel
$\,\Phi:B\rightarrow BC$ satisfying (\ref{FR-ineq}) and
(\ref{FR-cond}) provided $\,I(A:C|B)_{\omega}$ is the extended
conditional mutual information (described in Theorem \ref{cmi-th}).}
\end{property}\smallskip

We will use the following corollary of the compactness criterion for
families of quantum operations in the strong convergence topology
\cite[Corollary 2]{AQC}.\footnote{The strong convergence topology is
described in Sec.8.2}\smallskip

\begin{lemma}\label{cmi-conv}
\emph{Let $\,\rho_{A}$ be a full rank state in $\,\S(\H_A)$ and
$\,\{\Phi_n\}$ be a sequence of quantum operations from $A$ to $BC$
such that
$$
[\Phi_n(\rho_{A})]_B\leq\rho_{B}\quad \textit{and}\quad
[\Phi_n(\rho_{A})]_C\leq\rho_{C}\quad \forall n
$$
for some operators  $\,\rho_{B}\in\T_{+}(\H_B)$ and
$\,\rho_{C}\in\T_{+}(\H_C)$. Then the sequence $\{\Phi_n\}$ is
relatively compact in the strong convergence topology.}
\end{lemma}\smallskip

\emph{Proof.} It suffices to note that the set
$\{\omega\in\T_+(\H_{BC})\,|\,\omega_B\leq\rho_{B},
\omega_C\leq\rho_{C}\}$ is compact (see Corollary 6 in \cite{AQC})
and to apply Corollary 2 in \cite{AQC}. $\square$
\medskip

\emph{Proof of Proposition \ref{FR-r-m} .} By Remark 5.3 in
\cite{F&R} the assertion of the proposition is valid if $\omega_A$,
$\omega_B$ and $\omega_C$ are finite rank states. We will extend the
class of states for which this assertion is valid to the whole set
$\S(\H_{ABC})$ by several steps (1-4).

To simplify notations we will assume in each step that
$\H_X=\mathrm{supp}\shs\omega_X$ for $X=A,B,C,$  so that $\dim
A=\mathrm{rank}\shs\omega_A$, etc.

Throughout the proof we will assume that $P^n_X$ is the
\emph{spectral} projector of a state $\omega_{X}$ corresponding to
its $n$ maximal eigenvalues, $X=A,B,C$. Speaking about compactness
and convergence of a sequence of quantum operations we will have in
mind that the strong convergence topology is used. \smallskip

\emph{Step 1.} Assume that $\dim A\leq+\infty$ but $\dim B<+\infty$
and $\dim C<+\infty$. Let
$$
\omega^n_{ABC}=\lambda_n^{-1}Q_n\omega_{ABC}Q_n,\quad
Q_n=P^n_A\otimes I_B \otimes I_C,\;\lambda_n=\Tr Q_n\omega_{ABC}.
$$
By Remark 5.3 in \cite{F&R} for each $n$ there is a channel
$\Phi_n:B\rightarrow BC$ such that
\begin{equation}\label{FR-ineq-n}
2^{-\frac{1}{2}I(A:C|B)_{\omega^n}}\leq F(\omega^n_{ABC},
\id_A\otimes\Phi_n(\omega^n_{AB}))
\end{equation}
and
\begin{equation}\label{FR-cond-n}
[\Phi_n(\omega^n_{B})]_B=\omega^n_{B}\quad\textrm{and}\quad
[\Phi_n(\omega^n_{B})]_C=\omega^n_{C}
\end{equation}
Since $B$ and $BC$ are finite-dimensional systems, the sequence
$\{\Phi_n\}$ is relatively compact. So, we may assume that there
exists $\lim_{n\rightarrow\infty}\Phi_n=\Phi_*$. By Corollary
\ref{cmi-pr-c} we have
\begin{equation}\label{I-lr}
\lim_{n\rightarrow\infty}I(A\!:\!C|B)_{\omega^{n}}=I(A\!:\!C|B)_{\omega},
\end{equation}
while the continuity of the quantum fidelity (Lemma B.9 in
\cite{F&R}) implies
\begin{equation}\label{Fd-lr}
\lim_{n\rightarrow\infty}F(\omega^n_{ABC},
\id_A\otimes\Phi_n(\omega^n_{AB}))=F(\omega_{ABC},
\id_A\otimes\Phi_*(\omega_{AB})).
\end{equation}
It follows from (\ref{FR-ineq-n})-(\ref{Fd-lr}) that the channel
$\Phi_*$ satisfies (\ref{FR-ineq}) and (\ref{FR-cond}).\smallskip

\emph{Step 2.} Assume that $\dim A\leq+\infty$ and $\dim
B\leq+\infty$ but $\dim C<+\infty$. Let
$$
\omega^n_{ABC}=\lambda_n^{-1}Q_n\omega_{ABC}Q_n,\quad Q_n=I_A\otimes
P^n_B \otimes I_C,\;\lambda_n=\Tr Q_n\omega_{ABC}.
$$
By the previous step for each $n$ there is a channel
$\Phi_n:B_n\rightarrow B_nC$, where $B_n$ corresponds to the
subspace $P^n_B(\H_B)$, such that (\ref{FR-ineq-n}) and
(\ref{FR-cond-n}) hold. Consider the quantum operation
$\Psi_n=\Phi_n\circ\Pi_n$ from $B$ to $BC$, where
$\Pi_n(\cdot)=P^n_B(\cdot)P^n_B$. Since $P^n_B$ is a spectral
projector of $\omega_{B}$, it follows from (\ref{FR-cond-n}) that
\begin{equation}\label{p-s-f}
[\Psi_n(\omega_{B})]_B=\lambda_n\omega^n_{B}\leq\omega_{B}\quad\textrm{and}\quad
[\Psi_n(\omega_{B})]_C=\lambda_n\omega^n_{C}\leq\omega_{C}\quad
\forall n.
\end{equation}
Since $\omega_{B}$ is a full rank state, the sequence $\{\Psi_n\}$
is relatively compact by Lemma \ref{cmi-conv}. So, we may assume
that there exists $\lim_{n\rightarrow\infty}\Psi_n=\Phi_*$. It is
easy to see that $\Phi_*$ is a channel. It follows from
(\ref{p-s-f}) that the channel $\Phi_*$ satisfies condition
(\ref{FR-cond}). By noting that
$$
\id_A\otimes\Phi_*(\omega_{AB})=\lim_{n\rightarrow\infty}\id_A\otimes\Psi_n(\omega_{AB})=\lim_{n\rightarrow\infty}\id_A\otimes\Phi_n(\omega^n_{AB})
$$
and that relation (\ref{I-lr}) is also valid in this case by
Corollary \ref{cmi-pr-c} ($\dim C<+\infty$), we obtain from
(\ref{FR-ineq-n}),(\ref{I-lr}) and (\ref{Fd-lr}) that the channel
$\Phi_*$ satisfies condition (\ref{FR-ineq}).\smallskip

\emph{Step 3.} Assume that $\dim A=\dim B=\dim C=+\infty$ and
$\omega_{BC}$ is a full rank state. Let
$$
\omega^n_{ABC}=\lambda_n^{-1}Q_n\omega_{ABC}Q_n,\quad
 Q_n=I_A\otimes I_B \otimes
P^n_C,\;\lambda_n=\Tr Q_n\omega_{ABC}.
$$
By the previous step for each $n$ there is a channel
$\Phi_n:B\rightarrow BC$ such that (\ref{FR-ineq-n}) and
(\ref{FR-cond-n}) hold. Consider the  quantum operation
$\Psi_n=\Phi_n\circ\Theta\circ\Pi_n$ from $BC$ to $BC$, where
$\Theta(\cdot)=\Tr_C(\cdot)$  and $\Pi_n(\cdot)=I_B \otimes
P^n_C(\cdot)I_B \otimes P^n_C$. Since $P^n_C$ is a spectral
projector of $\omega_{C}$, it follows from (\ref{FR-cond-n}) that
\begin{equation*}
[\Psi_n(\omega_{BC})]_B=\lambda_n\omega^n_{B}\leq\omega_{B}\quad\textrm{and}\quad
[\Psi_n(\omega_{BC})]_C=\lambda_n\omega^n_{C}\leq\omega_{C}\quad
\forall n.
\end{equation*}
Since $\omega_{BC}$ is a full rank state, the sequence $\{\Psi_n\}$
is relatively compact by Lemma \ref{cmi-conv}. So, we may assume
that there exists $\lim_{n\rightarrow\infty}\Psi_n=\Psi_*$. It is
easy to see that $\Psi_*$ is a channel from $BC$ to $BC$.

Let $\Lambda(\rho)=\rho\otimes\sigma$ be a channel from $B$ to $BC$,
where $\sigma$ is a given state in $\S(\H_C)$. Consider the channel
$\Phi_*=\Psi_*\circ\Lambda$ from $B$ to $BC$. Since
$$
\Psi_*(\rho_{BC})=\lim_{n\rightarrow\infty}\Psi_n(\rho_{BC})=\lim_{n\rightarrow\infty}\Psi_n(\rho_{B}\otimes\sigma)=
\Psi_*(\rho_{B}\otimes\sigma)=\Phi_*(\rho_{B})
$$
for any state $\rho_{BC}\in\S(\H_{BC})$, we have
$\Psi_*=\Phi_*\circ\Theta$. By noting that
$$
\Phi_*(\omega_{B})=\lim_{n\rightarrow\infty}\Psi_n(\omega_{B}\otimes\sigma)=\lim_{n\rightarrow\infty}[\Tr
P^n_C\sigma]\shs\Phi_n(\omega_{B}),
$$
we obtain from (\ref{FR-cond-n}) that the channel $\Phi_*$ satisfies
condition (\ref{FR-cond}). Since
$$
\id_A\otimes\Phi_*(\omega_{AB})=\id_A\otimes\Psi_*(\omega_{ABC})=\lim_{n\rightarrow\infty}\id_A\otimes\Psi_n(\omega_{ABC})=\lim_{n\rightarrow\infty}
\id_A\otimes\Phi_n(\omega^n_{AB})
$$
it follows from (\ref{FR-ineq-n}),(\ref{I-lr}) and (\ref{Fd-lr})
that the channel $\Phi_*$ satisfies condition (\ref{FR-ineq}).  In
this case (\ref{I-lr}) follows from  the lower semicontinuity of
$I(A\!:\!C|B)_{\omega}$ and its monotonicity under local operations
(Theorem \ref{cmi-th}).\smallskip

\emph{Step 4.} To relax the full rank condition for $\omega_{BC}$
consider the sequence of states
$$
\omega^{n}_{ABC}=(1-\varepsilon_n)\omega_{ABC}+\varepsilon_n\omega_{AB}\otimes\omega_C,
$$
where $\varepsilon_n=1/n$. Since $\omega_{B}\otimes\omega_C$ is a
full rank state, $\omega^n_{BC}$ is a full rank state for each $n$.
By the previous step for each $n$ there is a channel
$\Phi_n:B\rightarrow BC$ such that (\ref{FR-ineq-n}) and
(\ref{FR-cond-n}) hold. Since $\omega^n_{B}=\omega_{B}$ and
$\omega^n_{C}=\omega_{C}$ for all $n$, it follows from
(\ref{FR-cond-n}) and Lemma \ref{cmi-conv} that the sequence
$\{\Phi_n\}$ is relatively compact. So, we may assume that there
exists $\lim_{n\rightarrow\infty}\Phi_n=\Phi_*$. It follows from
(\ref{FR-ineq-n})-(\ref{Fd-lr}) that the channel $\Phi_*$ satisfies
conditions (\ref{FR-ineq}) and (\ref{FR-cond}). In this case
(\ref{I-lr}) follows from Lemma \ref{cmi-conv+} below $\square$.
\smallskip

\begin{lemma}\label{cmi-conv+}
\emph{Let $\omega_{ABC}$ be an arbitrary state of a tripartite
system  and
$\omega^{\varepsilon}_{ABC}=(1-\varepsilon)\omega_{ABC}+\varepsilon\omega_{AB}\otimes\omega_C$,
where $\varepsilon\in(0,1)$, then}
$$
\lim_{\varepsilon\rightarrow+0}I(A\!:\!C|B)_{\omega^{\varepsilon}}=I(A\!:\!C|B)_{\omega}.
$$
\end{lemma}
\emph{Proof.} First assume that $I(A\!:\!C|B)_{\omega}$ is well
defined by formula (\ref{cmi-d+}), i.e.
\begin{equation*}
I(A\!:\!C|B)_{\omega}=H(\omega_{ABC}\shs\Vert\shs\omega_{A}\otimes
\omega_{BC})-H(\omega_{AB}\shs\Vert\shs\omega_{A}\otimes
\omega_{B}).
\end{equation*}
Since $\omega^{\varepsilon}_{X}=\omega_{X}$ for  $X=A,B,C,AB\,$ and
$\,\omega^{\varepsilon}_{BC}=(1-\varepsilon)\omega_{BC}+\varepsilon\omega_{B}\otimes\omega_C$,
the joint convexity of the quantum relative entropy implies
$$
\begin{array}{rl}
\!I(A\!:\!C|B)_{\omega^{\varepsilon}}\!\!\!&=H(\omega^{\varepsilon}_{ABC}\shs\Vert\shs\omega_{A}\otimes
\omega^{\varepsilon}_{BC})-H(\omega_{AB}\shs\Vert\shs\omega_{A}\otimes
\omega_{B})\\\\&\leq(1-\varepsilon)H(\omega_{ABC}\shs\Vert\shs\omega_{A}\otimes
\omega_{BC})+\varepsilon
H(\omega_{AB}\otimes\omega_C\shs\Vert\shs\omega_{A}\otimes
\omega_{B}\otimes\omega_C)\\\\&-\,H(\omega_{AB}\shs\Vert\shs\omega_{A}\otimes
\omega_{B})=(1-\varepsilon)\, I(A\!:\!C|B)_{\omega}
\end{array}
$$
By using approximation property (\ref{t-ext-p+}) it is easy to show
that the inequality
$$
I(A\!:\!C|B)_{\omega^{\varepsilon}}\leq(1-\varepsilon)\,
I(A\!:\!C|B)_{\omega},
$$
is valid for any state $\omega_{ABC}$. The assertion of the lemma
follows from this inequality and the lower semicontinuity of
$I(A\!:\!C|B)_{\omega}$ (Theorem \ref{cmi-th}). $\square$

\section*{Appendix: Proofs of Theorems \ref{main} and \ref{cmi-th}}

\emph{Proof Theorem \ref{main}.} A) Prove first that
\begin{equation}\label{mi-l-r-a}
\lim_{k\rightarrow\infty}I(A\!:\!B)_{\omega^k}=I(A\!:\!B)_{\omega^0}
\end{equation}
if condition b) is valid, i.e. if
$\lambda_k\omega^k\leq\Phi_A^k\otimes\Phi_B^k(\omega^0)$ for some
sequences  $\{\Phi_A^k\}$ and $\{\Phi_B^k\}$ of quantum operations
and some sequence $\{\lambda_k\}$ converging to $1$.

We will use the inequality
\begin{equation}\label{sp-ineq}
 \lambda I(A\!:\!B)_{\rho}+(1-\lambda)I(A\!:\!B)_{\sigma}\leq
 I(A\!:\!B)_{\lambda\rho+(1-\lambda)\sigma}+h_2(\lambda),
\end{equation}
where $h_2(\lambda)$ is the  binary entropy,  valid for arbitrary
operators $\rho,\sigma\in\T_+(\H_{AB})$ such that
$\max\{\Tr\rho,\Tr\sigma\}\leq1$. If all the marginal entropies of
the operators $\rho$ and $\sigma$ are finite then (\ref{sp-ineq})
directly follows from (\ref{mi-d-1}) and (\ref{w-k-ineq}). In
general case (\ref{sp-ineq}) can be proved by approximating the
operators $\rho$ and $\sigma$ by the sequences of operators
$$
\rho_k = P_A^k\otimes P_B^k \cdot \rho \cdot P_A^k\otimes P_B^k\quad
\textrm{and} \quad \sigma_k =P_A^k\otimes P_B^k \cdot \sigma \cdot
P_A^k\otimes P_B^k,
$$
where $\{P^k_{A}\}\subset\B(\H_A)$ and $\{P^k_B\}\subset\B(\H_{B})$
are sequences of finite rank projectors strongly converging to the
identity operators $I_A$ and $I_B$.

Since (\ref{sp-ineq}) holds for the operators $\rho_k$ and
$\sigma_k$ for all $k$, validity of (\ref{sp-ineq}) for the
operators $\rho$ and $\sigma$ follows from the relations
$$
\lim_{k\rightarrow+\infty}I(A\!:\!B)_{\varrho_k}=I(A\!:\!B)_{\varrho}\leq+\infty,
\quad \varrho=\rho,\,\sigma,\,\lambda\rho+(1-\lambda)\sigma,
$$
which directly follow from the lower semicontinuity of the quantum
mutual information and its monotonicity under local operations.

Inequality (\ref{sp-ineq}), nonnegativity and monotonicity of the
quantum mutual information under local operations show that
$$
\lambda_k I(A\!:\!B)_{\omega^k}\leq
I(A\!:\!B)_{\Phi_A^k\otimes\Phi_B^k(\omega^0)}+\gamma_kh_2(\lambda'_k)\leq
I(A\!:\!B)_{\omega^0}+\gamma_kh_2(\lambda'_k),
$$
where
$\gamma_k=\Tr\shs\Phi_A^k\otimes\Phi_B^k(\omega^0)\leq\Tr\omega^0$
and
$\lambda'_k=\gamma^{-1}_k\lambda_k\Tr\omega_k\geq\lambda_k\Tr\omega_k/\Tr\omega^0$.
This inequality and the lower semicontinuity of
$I(A\!:\!B)_{\omega}$ imply (\ref{mi-l-r-a}).\smallskip

In the next part of the proof we will use the identity
\begin{equation}\label{sp-ident}
    I(A\!:\!B)_{\omega}+I(B\!:\!C)_{\omega}=2H(\omega_{B})
\end{equation}
valid for any 1-rank operator $\omega\in\T_+(\H_{ABC})$ (with
possible value $+\infty$ in the both sides). If $H(\omega_{A})$,
$H(\omega_{B})$ and $H(\omega_{C})$ are finite then (\ref{sp-ident})
is easily verified by noting that $H(\omega_{A})=H(\omega_{BC})$,
$H(\omega_{B})=H(\omega_{AC})$ and $H(\omega_{C})=H(\omega_{AB})$.
In general case (\ref{sp-ident}) can be proved by approximating the
operator $\omega$ by the sequence of operators
$$
\omega^k=P_A^k\otimes P_B^k \otimes P_C^k\cdot\omega\cdot
P_A^k\otimes P_B^k \otimes P_C^k,
$$
where $\{P^k_{X}\}\subset\B(\H_X)$ is a  sequence of finite rank
projectors strongly converging to the identity operator $I_X$,
$X=A,B,C$. Since identity (\ref{sp-ident}) holds for each operator
$\omega^k$, its validity for the operator $\omega$ follows from the
relations
\begin{equation}\label{1-lr}
\lim_{k\rightarrow+\infty}I(X\!:\!Y)_{\omega^k}=I(X\!:\!Y)_{\omega}\leq+\infty,\quad
XY=AB,BC,
\end{equation}
and
\begin{equation}\label{2-lr}
\lim_{k\rightarrow+\infty}H(\omega^k_{B})=H(\omega_{B})\leq+\infty
\end{equation}
Since $\omega^k_{XY}\leq P_X^k \otimes P_Y^k \,\omega_{XY} P_X^k
\otimes P_Y^k$, relations (\ref{1-lr}) follow from the continuity
condition b) proved before. Relation (\ref{2-lr}) follows from Lemma
\ref{t-ext-l}.

To prove condition a) it suffices, by symmetry,  to prove
(\ref{mi-l-r-a}) assuming that there exists
$$
\lim_{k\rightarrow+\infty}H(\omega_B^k)=H(\omega^0_B)<+\infty.
$$
By Lemma \ref{p-lemma} there is a sequence $\{\tilde{\omega}^k\}$ of
1-rank operators in $\T_{+}(\H_{ABC})$ converging to an operator
$\tilde{\omega}^0$ such that $\tilde{\omega}^k_{AB}=\omega^k$ for
all $k\geq0$. Since the sequence $\{\tilde{\omega}^k_{BC}\}$
converges to the state $\tilde{\omega}^0_{BC}$, the lower
semicontinuity of the quantum mutual information shows that
\begin{equation*}
\liminf_{k\rightarrow+\infty}I(A\!:B)_{\tilde{\omega}^k}\geq
I(A\!:B)_{\tilde{\omega}^0},\quad
\liminf_{k\rightarrow+\infty}I(B\!:C)_{\tilde{\omega}^k}\geq
I(B\!:C)_{\tilde{\omega}^0}
\end{equation*}
while
$\,\lim_{k\rightarrow+\infty}H(\tilde{\omega}^k_{B})=H(\tilde{\omega}^0_{B})<+\infty\,$
by the assumption ($\tilde{\omega}^k_{B}=\omega^k_{B}$). So,
identity (\ref{sp-ident}) and Lemma \ref{vsl} imply
(\ref{mi-l-r-a}). \medskip

B) It suffices to assume that $\Phi_A$ is an arbitrary operation and
$\Phi_B=\id_B$. We have to show that continuity of the function
$\,I(A\!:\!B)_{\omega}$ on a subset
$\mathcal{A}\subset\T_+(\H_{AB})$ implies continuity of the function
$I(A'\!:\!B)_{\Phi_A\otimes\id_B(\omega)}$ on $\mathcal{A}$.

If $\shs\Phi_A$ is a quantum channels then this implication directly
follows from Corollary \ref{cmi-th-c-2}B, since by the Stinespring
representation $\Phi_A$ is isomorphic to a subchannel of a partial
trace.

If $\shs\Phi_A$ is a trace non-preserving  operation then consider
the channel $\Psi_A=\Phi_A\oplus\Delta$ from $A$ to $A''=A'\oplus
A^\mathrm{c}$, where $\Delta(\rho)=[\Tr\rho-\Tr\Phi_A(\rho)]\sigma$
is a quantum operation from $A$ to $A^\mathrm{c}$ determined by a
fixed state $\sigma\in\S(\H_{A^\mathrm{c}})$. We have\footnote{Here
we use the following property of the relative entropy
$H(\rho_1+\rho_2\shs\|\shs \sigma_1+\sigma_2)=H(\rho_1\shs\|\shs
\sigma_1)+H(\rho_2\shs\|\shs \sigma_2)\,$ if
$\,\rho_1\rho_2=\sigma_1\sigma_2=\rho_1\sigma_2=\sigma_1\rho_2=0$
\cite{L-2}.}
\begin{equation}\label{d-f}
\begin{array}{rl}
I(A''\!:\!B)_{\Psi_A\otimes\id_B(\omega_{AB})}\!\!& \doteq
H\left(\Psi_A\otimes\id_B(\omega_{AB})\shs\Vert\shs\Psi_A(\omega_{A})\otimes
\omega_{B}\right)\\\\& =
H\left(\Phi_A\otimes\id_B(\omega_{AB})\shs\Vert\shs\Phi_A(\omega_{A})\otimes
\omega_{B}\right)\\\\&+\,H\left(\Delta\otimes\id_B(\omega_{AB})\shs\Vert\shs\Delta(\omega_{A})\otimes
\omega_{B}\right)\\\\&= I(A'\!:\!B)_{\tilde{\omega}}+
H\left(\tilde{\omega}_{B}\shs\Vert\shs\lambda\omega_{B}\right)\\\\&+\,H\left(\Delta\otimes\id_B(\omega_{AB})\shs\Vert\shs\Delta(\omega_{A})\otimes
\omega_{B}\right),
\end{array}
\end{equation}
where $\tilde{\omega}_{A'B}=\Phi_A\otimes\id_B(\omega_{AB})$ and
$\lambda=\Tr\shs\tilde{\omega}_{A'B}$. Since $\Psi_A$ is a channel,
the continuity of $I(A\!:\!B)_{\omega}$ on $\mathcal{A}$ implies, by
the above remark, continuity of the left hand side of (\ref{d-f}) on
$\mathcal{A}$. Since all the summands in the right hand side of
(\ref{d-f}) are lower semicontinuous functions, Lemma \ref{vsl}
shows that all these summands are continuous on $\mathcal{A}$.
$\square$
\medskip

\emph{Proof of Theorem \ref{cmi-th}.} To show the uniqueness of a
function with the stated properties it suffices to assume that
$F(\omega)$ is a lower semicontinuous function on the set
$\,\S(\H_{ABC})$ possessing property C3 and coinciding with
$I(A\!:\!C|B)_{\omega}$ given by formula (\ref{cmi-d+}) on the set
of states with finite $I(A\!:\!B)_{\omega}$.

Chose any sequence of channels
$\Phi_A^k:\T(\H_A)\rightarrow\T(\H_A)$ with finite output entropy
such that $\,\lim_{k\rightarrow\infty}\Phi_A^k(\rho)=\rho\,$ for any
$\,\rho\in\S(\H_A)$. Then the lower semicontinuity of $F$ and
property C3 imply
$$
F(\omega)=\lim_{k\rightarrow\infty}F(\Phi_A^k\otimes\id_{BC}(\omega))=\lim_{k\rightarrow\infty}I(A\!:\!C|B)_{\Phi_A^k\otimes\id_{BC}(\omega)},\quad
$$
for any state $\omega\in\S(\H_{ABC})$, where
$I(A\!:\!C|B)_{\Phi_A^k\otimes\id_{BC}(\omega)}$ is defined by
formula (\ref{cmi-d+}), since upper bound (\ref{mi-u-b}) shows that
$I(A\!:\!B)_{\Phi_A^k\otimes\id_{BC}(\omega)}<+\infty$ for all $k$.
So, $F(\omega)$ is uniquely determined.\smallskip

By Remark \ref{main-c-n+r}  formulas (\ref{cmi-d+}),
(\ref{cmi-d++}), (\ref{cmi-d+++}) and (\ref{cmi-d++++}) determine
$\mathfrak{F}$-extensions of the quantity $I(A\!:\!C|B)_{\omega}$
defined in (\ref{cmi-d}) respectively to the sets
$$
\begin{array}{ll}
  \S_1=\left\{\shs\omega_{ABC}\,|\,
I(A\!:\!B)_{\omega}<+\infty\right\}, &
\quad\S_2=\left\{\shs\omega_{ABC}\,|\,
I(B\!:\!C)_{\omega}<+\infty\right\}, \\\\
  \S_3=\left\{\shs\omega_{ABC}\,|\,
H(\omega_{B})<+\infty\right\}, &
\quad\S_4=\left\{\shs\omega_{ABC}\,|\,
H(\omega_{ABC})<+\infty\right\}.
\end{array}
$$

By uniqueness of $\mathfrak{F}$-extension (Lemma \ref{t-ext-r}) any
pair of formulae (\ref{cmi-d+})-(\ref{cmi-d++++}) coincide on the
set $\S_i\cap\S_j$ where both of them are well defined. So, formulae
(\ref{cmi-d+})-(\ref{cmi-d++++}) correctly determine
$\mathfrak{F}$-extension of the quantity $I(A\!:\!C|B)$ defined in
(\ref{cmi-d}) to the set
\begin{equation*}
\S_*=\bigcup_{i=1}^4\S_i=\left\{\shs\omega_{ABC}\,|\,\min\{I(A\!:\!B)_{\omega},
I(B\!:\!C)_{\omega},H(\omega_{ABC}),H(\omega_{B})\}<+\infty\right\}
\end{equation*}
of states for which at least one of these formulae is well defined.
It follows, in particular, that formulae (\ref{cmi-d+}) and
(\ref{cmi-d++}) coincide on the set
\begin{equation}\label{T-0}
\T_0=\left\{\shs\omega\in\T_+(\H_{ABC})\,|\,
I(A\!:\!B)_{\omega}<+\infty, I(B\!:\!C)_{\omega}<+\infty\right\}
\end{equation}
containing the set
$\,\T_\mathrm{f}=\left\{\shs\omega\in\T_+(\H_{ABC})\,|\,
\rank\shs\omega_A<+\infty,\rank\shs\omega_C<+\infty\right\}$.\smallskip

First we will prove the stated properties of
$I_\mathrm{e}(A\!:\!C|B)_{\omega}$ for the function
\begin{equation}\label{cmi-e+++}
F(\omega)=\sup_{P_A,P_C}I(A\!:\!C|B)_{Q\shs \omega
Q},\;\,Q=P_A\otimes I_B\otimes P_C,
\end{equation}
on the cone $\,\T_{+}(\H_{ABC})$, where
$$
I(A\!:\!C|B)_{Q\shs \omega Q}=I(A\!:\!BC)_{Q\shs \omega
Q}-I(A\!:\!B)_{Q\shs \omega Q}=I(AB\!:\!C)_{Q\shs \omega
Q}-I(B\!:\!C)_{Q\shs \omega Q}
$$
(since $Q\shs \omega Q\in\T_{\mathrm{f}}$) and the supremum is over
all finite rank projectors $P_A$ in $\B(\H_A)$ and $P_C$ in
$\B(\H_C)$. Then we will show that $F(\omega)$ coincides with the
function $I_{\mathrm{e}}(A\!:\!C|B)_{\omega}$ defined by
(\ref{cmi-e+}). By symmetry this would imply the coincidence of
(\ref{cmi-e+}) and (\ref{cmi-e++}).

\smallskip

By Corollary \ref{main-c} the function $\,\omega\mapsto
\!I(A\!:\!C|B)_{Q\shs\omega Q}$, $Q=P_A\otimes I_B\otimes P_C$,  is
continuous on the cone $\,\T_{+}(\H_{ABC})$ for any finite rank
projectors $P_A$ and $P_C$. Hence $F(\omega)$ is a lower
semicontinuous function on  $\,\T_{+}(\H_{ABC})$.

The below Lemma \ref{T-0-l} shows, by symmetry, that the set $\T_0$
defined in (\ref{T-0}) is invariant under local operations
$\Phi_A:A\rightarrow A$ and $\Phi_C:C\rightarrow C$ and that
\begin{equation}\label{pmr}
I(A\!:\!C|B)_{\omega}\geq
I(A\!:\!C|B)_{\Phi_{\!A}\otimes\id_{B}\otimes\Phi_{\!C}(\omega)}
\quad \textrm{for any}\;\, \omega\in\T_{0}.
\end{equation}
Hence
$$
I(A\!:\!C|B)_{\omega}\geq I(A\!:\!C|B)_{Q\omega Q}\quad \textrm{for
any}\;\, \omega\in\T_{0},
$$
where $Q=P_A\otimes I_B\otimes P_C$, for any finite rank projectors
$P_A$ and $P_C$. It follows that
\begin{equation}\label{F-c}
F(\omega)=I(A\!:\!C|B)_{\omega}\quad\textrm{for any}\;\,
\omega\in\T_{\mathrm{f}},
\end{equation}
where $I(A\!:\!C|B)_{\omega}$ is given by the both formulae
(\ref{cmi-d+}) and (\ref{cmi-d++}).\smallskip

The lower semicontinuity of $F$ implies
\begin{equation}\label{F-lr}
F(\omega)=\lim_{k\rightarrow\infty} F(Q_k\omega Q_k),\quad
Q_k=P^k_A\otimes I_B\otimes P^k_C,
\end{equation}
for arbitrary operator $\omega\in\T_+(\H_{ABC})$ and any sequences
$\{P^k_A\}\subset\B(\H_A)$, $\{P^k_C\}\subset\B(\H_C)$ of finite
rank projectors strongly converging to the identity operators $I_A$,
$I_C$, since (\ref{cmi-e+++}) and (\ref{F-c}) show that
$$
F(\omega)\geq I(A\!:\!C|B)_{Q_k\omega Q_k}=F(Q_k\omega Q_k)\;\;
\textrm{for all}\;\; k.
$$
By definition of $\mathfrak{F}$-extension (\ref{F-lr}) implies
$F(\omega)=I(A\!:\!C|B)_{\omega}$ for any state $\omega$ in the
above-defined set $\S_*$.\smallskip

The nonnegativity of the function $F$ (the first part of C1) follows
from its definition (by the monotonicity of the relative entropy
under partial trace).

To prove C2 note that for any state $\omega$ formula (\ref{cmi-d++})
implies validity of C2 for all the operators $Q_k\omega Q_k$ in
(\ref{F-lr}). So, by using faithfulness of the quantum mutual
information and (\ref{F-lr}) we obtain
$$
I(AB:C)_{\omega}=\lim_{k\rightarrow\infty}I(AB:C)_{Q_k\omega
Q_k}\geq\lim_{k\rightarrow\infty} F(Q_k\omega Q_k)=F(\omega).
$$

To prove C3 note that for any state $\omega$ all the operators
$Q_k\omega Q_k$ in (\ref{F-lr}) belong to the set
$\T_\mathrm{f}\subset\T_0$. So, by using (\ref{pmr}), (\ref{F-c}),
(\ref{F-lr}) and the lower semicontinuity of $F$, we obtain
$$
F(\omega)=\lim_{k\rightarrow\infty} F(Q_k\omega Q_k)\geq
\liminf_{k\rightarrow\infty} F(\Psi(Q_k\omega Q_k))\geq
F(\Psi(\omega)),
$$
where $\Psi=\Phi_A\otimes\id_B\otimes\Phi_C$.\footnote{To simplify
notation we consider quantum operations $A\rightarrow A$ and
$C\rightarrow C$. Generalization to quantum operations $A\rightarrow
A'$ and $C\rightarrow C'$ is obvious.}\smallskip

The validity of property (\ref{t-ext-p}) for any state in the set
$\S_*$ follows from the existence of $\mathfrak{F}$-extension of
$I(A\!:\!C|B)$ to this set and its coincidence with $F$ proved
before. Property (\ref{t-ext-p+}) means that
$$
\lim_{l\rightarrow\infty}
F(\omega^{kl})=F(\omega^{k})\quad\textrm{and}\quad
\lim_{k\rightarrow\infty} F(\omega^{k})=F(\omega),
$$
where $\omega^{k}=\lim_{l\rightarrow\infty}\omega^{kl}$.  Since the
state $\omega^{k}$ belong to the set $\S_*$ for each $k$, property
(\ref{t-ext-p}) holds for this state implying the first of these
limit relations. The second one follows from the lower
semicontinuity of $F$ and property C3 proved before.

Now we can show coincidence of the function
$I_{\mathrm{e}}(A\!:\!C|B)$ defined by (\ref{cmi-e+}) with the
function $F$ defined by (\ref{cmi-e+++}).

Let $Q_1=P_A\otimes I_B\otimes I_C$ and $Q_2=P_A\otimes I_B\otimes
P_C$. The proved properties of the function $F$ imply
$$
I(A\!:\!C|B)_{Q_1\omega Q_1}=F(Q_1\omega Q_1)\geq F(Q_2\omega
Q_2)=I(A\!:\!C|B)_{Q_2\omega Q_2}
$$
and
$$
I(A\!:\!C|B)_{Q_1\omega Q_1}=F(Q_1\omega Q_1)\leq F(\omega)
$$
for any state $\omega$. The first of these inequalities implies
$I_{\mathrm{e}}(A\!:\!C|B)_{\omega}\geq F(\omega)$, while the second
one shows that $I_{\mathrm{e}}(A\!:\!C|B)_{\omega}\leq F(\omega)$.

Properties C4-C5 of $I_{\mathrm{e}}(A\!:\!C|B)_{\omega}$ can be
derived by approximation from the same properties of the conditional
mutual information in the finite-dimensional settings. In the case
of C4 it suffices to use property  (\ref{t-ext-p+}). In the case of
C5  the second part of Corollary \ref{main-c-n++} (i.e. property
(\ref{t-ext-p++})) is necessary.\footnote{Property C5 of
$I_{\mathrm{e}}(A\!:\!C|B)_{\omega}$ is not used in the proof of
Corollary \ref{main-c-n++}.}
\smallskip

To prove the second part of C1 note that
$I_{\mathrm{e}}(A\!:\!C|B)_{\omega}=0$ implies, by Proposition
\ref{FR-r-m} in Sect.8.4. (proved independently), the existence of a
channel $\Phi:B\rightarrow BC$ such that
$\,\omega_{ABC}=\id_A\otimes\Phi(\omega_{AB})$. The converse
statement follows from definition (\ref{cmi-e+}) of
$I_{\mathrm{e}}(A\!:\!C|B)_{\omega}$. $\square$\smallskip

\begin{lemma}\label{T-0-l}
\emph{For arbitrary quantum operation $\,\Phi_A:A\rightarrow A$ the
set $\,\T_0$ is invariant under the map
$\,\Phi_{\!A}\otimes\id_{BC}$ and $\,I(A\!:\!C|B)_{\omega}\geq
I(A\!:\!C|B)_{\Phi_{\!A}\otimes\id_{BC}(\omega)}$ for any
$\,\omega\in\T_{0}$.}
\end{lemma}\smallskip

\emph{Proof.} If $\shs\Phi_A$ is a  channel then the assertion of
the lemma directly follows from monotonicity of the quantum mutual
information and formula (\ref{cmi-d++}), since in this case
$I(B\!:\!C)_{\Phi_{\!A}\otimes\id_{BC}(\omega)}=I(B\!:\!C)_{\omega}$
for any state $\omega_{ABC}$.

If $\shs\Phi_A$ is a trace non-preserving  operation then consider
the channel $\Psi_A=\Phi_A\oplus\Delta$ from $A$ to $A'=A\oplus
A^\mathrm{c}$, where $\Delta(\rho)=[\Tr\rho-\Tr\Phi_A(\rho)]\sigma$
is a quantum operation from $A$ to $A^\mathrm{c}$ determined by a
fixed state $\sigma\in\S(\H_{A^\mathrm{c}})$.

Let $\tilde{\omega}_{ABC}=\Phi_A\otimes\id_{BC}(\omega_{ABC})$,
$\omega^\mathrm{c}_{ABC}=\Delta\otimes\id_{BC}(\omega_{ABC})$ and
$\lambda=\Tr\shs\tilde{\omega}_{ABC}$. To prove the invariance of
$\,\T_0$ it suffices to note that  $I(A\!:\!B)_{\tilde{\omega}}\leq
I(A\!:\!B)_{\omega}$ by monotonicity of the quantum mutual
information and that inequality (\ref{sp-ineq}) implies
$I(B\!:\!C)_{\tilde{\omega}}\leq
I(B\!:\!C)_{\tilde{\omega}+\omega^\mathrm{c}}+h_2(\lambda)=I(B\!:\!C)_{\omega}+h_2(\lambda)$
(the last equality holds, since $\Psi_A$ is a channel).\smallskip

Similarly to (\ref{d-f}) we have
\begin{equation}\label{d-f+}
I(A'B\!:\!C)_{\tilde{\omega}+\omega^\mathrm{c}}=
I(AB\!:\!C)_{\tilde{\omega}}+
H\left(\tilde{\omega}_{C}\shs\Vert\shs\lambda\omega_{C}\right)+H\left(\omega^\mathrm{c}_{ABC}\shs\Vert\shs\omega^\mathrm{c}_{AB}\otimes
\omega_{C}\right),
\end{equation}
while the joint convexity of the relative entropy and relation
(\ref{H-fun-eq+}) imply
\begin{equation}\label{d-f++}
\begin{array}{rl}
I(B\!:\!C)_{\tilde{\omega}+\omega^\mathrm{c}}\!\!&=\,
H\left(\tilde{\omega}_{BC}+\omega^\mathrm{c}_{BC}\shs\Vert\shs(\tilde{\omega}_{B}+\omega^\mathrm{c}_{B})\otimes
\omega_{C}\right)\\\\&\leq\,
H\left(\tilde{\omega}_{BC}\shs\Vert\shs\tilde{\omega}_{B}\otimes
\omega_{C}\right)+H\left(\omega^\mathrm{c}_{BC}\shs\Vert\shs\omega^\mathrm{c}_{B}\otimes
\omega_{C}\right)\\\\&=\,I(B\!:\!C)_{\tilde{\omega}}+
H\left(\tilde{\omega}_{C}\shs\Vert\shs\lambda\omega_{C}\right)+H\left(\omega^\mathrm{c}_{BC}\shs\Vert\shs\omega^\mathrm{c}_{B}\otimes
\omega_{C}\right).
\end{array}
\end{equation}
We have $ I(A'B\!:\!C)_{\tilde{\omega}+\omega^\mathrm{c}}\leq
I(AB\!:\!C)_{\omega} $ and
$I(B\!:\!C)_{\tilde{\omega}+\omega^\mathrm{c}}=I(B\!:\!C)_{\omega}$
(since $\Psi_A$ is a channel). Hence it follows from (\ref{d-f+})
and (\ref{d-f++}) that
$$
I(AB\!:\!C)_{\tilde{\omega}}-I(B\!:\!C)_{\tilde{\omega}}\leq
I(AB\!:\!C)_{\omega}-I(B\!:\!C)_{\omega}-\delta
$$
where
$\,\delta=H\left(\omega^\mathrm{c}_{ABC}\shs\Vert\shs\omega^\mathrm{c}_{AB}\otimes
\omega_{C}\right)-H\left(\omega^\mathrm{c}_{BC}\shs\Vert\shs\omega^\mathrm{c}_{B}\otimes
\omega_{C}\right)\geq0\,$ by monotonicity of the relative entropy.
$\square$ \medskip

I am grateful to A.S.Holevo and to the participants of his seminar
"Quantum probability, statistic, information" (the Steklov
Mathematical Institute) for useful discussion. I am also grateful to
G.G.Amosov for the help in solving the particular question and to M.M.Wilde for useful comments.
Special thanks to A.Winter for his modification of the Alicki-Fannes technic used in new version of Proposition 1. 

The research is funded by the grant of Russian Science Foundation
(project No 14-21-00162).

\end{document}